\documentclass[aps, 10pt, twocolumn,prb,superscriptaddress,showpacs, showkeys]{revtex4-1}
\usepackage[colorlinks,bookmarks=false,citecolor=blue,linkcolor=red,urlcolor=blue]{hyperref}
\usepackage{graphicx,epstopdf,color}
\usepackage{amsfonts}
\usepackage{amsmath,amssymb,mathrsfs}
\usepackage{bm}
\usepackage{pst-grad}

\def\bc{\begin{center}}
\def\ec{\end{center}}

\newcommand{\ket}[1]{\left|#1\right\rangle}
\newcommand{\bra}[1]{\left\langle#1\right|}

\newcommand{\braopket}[3]{\langle#1|#2|#3\rangle}

\newcommand{\bea}{\begin{eqnarray}}
\newcommand{\eea}{\end{eqnarray}}
\newcommand{\ketu}{\ket{\uparrow}}
\newcommand{\ketd}{\ket{\downarrow}}
\newcommand{\ketuu}{\ket{\uparrow\uparrow}}
\newcommand{\ketdd}{\ket{\downarrow\downarrow}}
\newcommand{\ketud}{\ket{\uparrow\downarrow}}
\newcommand{\ketdu}{\ket{\downarrow\uparrow}}
\newcommand{\brau}{\bra{\uparrow}}

\def\ie{\emph{i.e.},\ }

\newcommand{\be}{\begin{equation}}
\newcommand{\ee}{\end{equation}}

\newcommand{\av}[1]{\langle#1\rangle}

\newcommand{\bfss}{{\boldsymbol{S}}}

\newcommand{\bfbb}{\boldsymbol{B}}

\def\bmx{\begin{pmatrix}}
\def\emx{\end{pmatrix}}

\begin{document}

\title{Non-perturbative stochastic method for driven spin-boson model}
%

\author{Peter P. Orth}
\affiliation{Institute for Theory of Condensed Matter, Karlsruhe Institute of Technology (KIT), 76131 Karlsruhe, Germany }
\author{Adilet Imambekov}
\affiliation{Department of Physics and Astronomy, Rice University, Houston, TX, 77005, USA}
\author{Karyn Le Hur}
\affiliation{Center for Theoretical Physics, Ecole Polytechnique, CNRS, 91128 Palaiseau Cedex, France}
\affiliation{Department of Physics, Yale University, New Haven, Connecticut 06520, USA}

\begin{abstract}
We introduce and apply a numerically exact method for investigating the real-time dissipative dynamics of quantum impurities embedded in a macroscopic environment beyond the weak-coupling limit. We focus on the spin-boson Hamiltonian that describes a two-level system interacting with a bosonic bath of harmonic oscillators. This model is archetypal for investigating dissipation in quantum systems and tunable experimental realizations exist in mesoscopic and cold-atom systems. It finds abundant applications in physics ranging from the study of decoherence in quantum computing and quantum optics to extended dynamical mean-field theory.
Starting from the real-time Feynman-Vernon path integral, we derive an exact stochastic Schr\"odinger equation that allows us to compute the full spin density matrix and spin-spin correlation functions beyond weak coupling. We greatly extend our earlier work (P.~P. Orth, A. Imambekov, and K. Le Hur, Phys. Rev. A {\bf 82},~032118~(2010)) by fleshing out the core concepts of the method and by presenting a number of interesting applications. 
Methodologically, we present an analogy between the dissipative dynamics of a quantum spin and that of a classical spin in a random magnetic field. This analogy is used to recover the well-known non-interacting-blip-approximation in the weak-coupling limit. We explain in detail how to compute spin-spin autocorrelation functions. 
As interesting applications of our method, we explore the non-Markovian effects of the initial spin-bath preparation on the dynamics of the coherence $\sigma^x(t)$ and of $\sigma^z(t)$ under a Landau-Zener sweep of the bias field. We also compute to a high precision the asymptotic long-time dynamics of $\sigma^z(t)$ without bias and demonstrate the wide applicability of our approach by calculating the spin dynamics at non-zero bias and different temperatures.
\end{abstract}

\maketitle

\section{Introduction}
\label{sec:introduction}
The coupling of a system to its environment leads to irreversible energy flow between system and environment, thus giving rise to the phenomenon of dissipation~\cite{QuantumTransportAndDissipation-Dittrich-Book,VanKampen-StochasticProcesses-Book,weissdissipation}. In addition, thermal and quantum fluctuations in the environmental bath cause fluctuations of the system degrees of freedom, which results in a Brownian motion~\cite{PhysRev.36.823,PTP.33.423, Grabert1988115, Caldeira1983587}. These features of bath-induced dissipation and fluctuations occur both in classical and quantum systems. A quantum system that becomes entangled with the bath exhibits decoherence, the suppression of coherence between different states in the system~\cite{RevModPhys.75.715,DecoherenceNATOProceedings-2004}.
The effect of decoherence is particularly crucial if one wants to implement a quantum computer~\cite{nielsen_chuang_qc,RevModPhys.73.357,Vion03052002,PhysRevA.76.042319,Schoelkopf_Nature_2008}, where phase coherence between the two qubit states $\ketu$ and $\ketd$ is used as a resource. 
In fact, virtually no system is completely isolated from its surrounding making dissipation and decoherence ubiquitous in physics, chemistry and biology~\cite{VanKampen-StochasticProcesses-Book, Nitzan_chem_dyn_in_condensed_phase_book, PhysRevLett.68.998, XuSchulten-ChemPhys1994, StockburgerMak-ChemPhys-1996, PhysRevLett.99.267202, PhysRevB.84.245109, Chin13082012}. 

An impurity spin embedded in a macroscopic environment also emerges as an effective model for strongly correlated materials within (extended) dynamical mean-field theory~\cite{RevModPhys.68.13,PhysRevLett.89.236402,PhysRevB.61.5184,PhysRevB.66.085120}. In this work, we consider the paradigmatic spin-boson model~\cite{weissdissipation,RevModPhys.59.1}, which is a variant of the Caldeira-Leggett model~\cite{PhysRevLett.46.211,caldeira_quantum_1983}. Here, the system consists of only two states and the environment is described by a bosonic bath of harmonic oscillators. 

The spin-boson model with an Ohmic bath is a particularly rich model as it exhibits a wealth of interesting phenomena such as a delocalization-localization quantum phase transition of the spin for sufficiently strong coupling to the environment~\cite{PhysRevLett.25.450, PhysRevLett.54.263, vojta_philmag_2006, lehur_entanglement_spinboson, Florens-SBBookChapter-2010, KarynLeHur-UnderstandingQPT-Article}. There exist exact mappings to the anisotropic Kondo model~\cite{PhysRevLett.49.681,PhysRevLett.49.1545,PhysRevB.32.4410, PhysRevB.1.4464}, to the interacting resonance level model and to the one-dimensional Ising model with $1/r^2$ interaction~\cite{SpohnDumcke1985,PhysRevB.1.4464, 0022-3719-4-5-011}. Various theoretical proposals have been made to experimentally implement the Ohmic spin-boson model in a controllable low-energy circuit. In particular, the recent progress in nanotechnology~\cite{CastellanosBeltran-Lehnert-ApplPhysLett_2007, Manucharyan10022009,chung:216803, Mebrahtu-QPTResLev-Nature-2012,Pop-PhaseSlips-NatPhys-2010,PhysRevLett.106.217005, PhysRevLett.109.137002} allows for great control on the dissipation strength of resistive (Ohmic) environments~\cite{PhysRevLett.84.346, kopp:220401, PhysRevLett.88.226404, PhysRevB.85.235104, PhysRevLett.108.233603, PhysRevLett.105.100505, PhysRevB.85.184302}. In principle, this development could lead to the realization of a tunable spin-boson model with microwave photons~\cite{LeClair-PhysLettA-1997, PhysRevB.58.1872, PhysRevA.82.063816, PhysRevB.85.140506, PhysRevLett.110.017002, Delsing-arXiv2012}. A tunable spin-boson Hamiltonian with Ohmic dissipation can also be realized using trapped ions~\cite{porras:010101} or in cold-atom systems~\cite{recati:040404, orth:051601, PhysRevB.82.144423, PhysRevB.79.241105, PhysRevA.75.013406,  PhysRevLett.104.200402, PhysRevLett.105.045303, PhysRevLett.107.145306, WeitenbergBlochKuhr-Nature-2011}, where sound modes of a one-dimensional Bose-Einstein condensate mimic the bosonic environment. 

Several methods have been devised to investigate the dissipative spin dynamics in this system, for example the well-known non-interacting blip approximation (NIBA)~\cite{RevModPhys.59.1,weissdissipation} and extensions to it,~\cite{0295-5075-80-4-40005,goerlich_low-temperature_1989,weiss_dynamics_1989}, (non)-Markovian master equations,~\cite{koch:230402,PhysRevE.61.R4687,PhysRevB.71.035318,BreuerPetruccione-Book,PhysRevLett.107.210402,PhysRevB.84.235140,PhysRevA.84.023416, PhysRevB.85.224301}, (iterative) path-integral sampling and related techniques~\cite{springerlink:10.1007/BF01320834,PhysRevB.50.15210, Makri-QUAPI-ChemPhysLett-1994,makri_numerical_1995,grifoni_driven_1998, 1367-2630-10-11-115005, nalbach:220401,PhysRevB.77.195316,PhysRevLett.102.150601, arXiv:1207.6995, PhysRevLett.110.010402}, Monte-Carlo methods on the Keldysh contour,~\cite{RevModPhys.83.349,PhysRevB.78.235110,PhysRevB.79.153302, PhysRevB.81.035108}, and various renormalization group approaches.~\cite{PhysRevB.74.245113,RevModPhys.80.395,PhysRevB.71.045122,PhysRevLett.95.086406,PhysRevLett.92.196804,PhysRevB.82.144423, PhysRevB.70.121302, PhysRevB.84.155110,PhysRevB.63.180302, epjst/e2009-00962-3, PhysRevLett.104.106801, andergassen_renormalization_2011, springerlink:10.1140/epjst/e2010-01219-x,PhysRevB.78.092303,0953-8984-21-1-015601, PhysRevB.85.085113, arXiv:1211.0293} The Feynman-Vernon real-time functional integral formalism~\cite{feynmanvernon}, which will be the starting point in the following, is particularly well suited to study this class of models since one can easily eliminate the environmental degrees of freedom. 

Here, our primary goal is to introduce and apply a non-perturbative and numerically exact method to investigate the dissipative dynamics of the Ohmic spin-boson model beyond the weak spin-bath coupling limit. Starting from the real-time functional integral description, we derive an exact non-perturbative stochastic Schr\"odinger equation (SSE).~\cite{imambek_jetp_02, imambekov:063606,ImambekovGritsevDemler-FundNoiseFermiProceed2006, Hofferberth-NatPhys-2008, PhysRevA.82.032118} Compared to earlier SSE approaches,~\cite{Kleinert1995224,PhysRevLett.80.2657,PhysRevLett.82.1801,PhysRevLett.88.170407,Stockburger2004159} our method allows exact consideration of the initial spin-bath correlations as we derive in the main text below. Our approach works both at zero and at finite temperatures, and we may easily consider a bias field $\epsilon(t)$ with arbitrary time-dependence in the Hamiltonian. We have previously applied this method to investigate the spin dynamics during a Landau-Zener sweep $\epsilon(t) = v t$ with velocity $v$.~\cite{PhysRevA.82.032118} In addition, it may also be applied to other many-body environments, and in particular to a fermionic environment, that can be represented in the form of a Coulomb gas such as the Kondo model.~\cite{imambekov:063606,ImambekovGritsevDemler-FundNoiseFermiProceed2006,0022-3719-4-5-011, Hofferberth-NatPhys-2008}

The main idea of our method is to recast the problem of finding the exact path-integral amplitudes into the form of a numerically solvable linear stochastic equation~\cite{imambek_jetp_02, imambekov:063606, PhysRevA.82.032118}. The quantum spin evolution is given by the average solution over different stochastic realizations. We explicitly derive an analogy to a classical spin in a random magnetic field. Compared to the NIBA, we treat the blip-blip interactions in an exact manner here, then solve the SSE numerically.

The starting point to study the influence of the environment on the dynamics of a quantum spin is the spin-boson Hamiltonian
\begin{align}
  \label{eq:1}
  H &= \frac{\Delta}{2} \sigma^x + \frac{\epsilon(t)}{2} \sigma^z + \frac{\sigma^z}{2} \sum_{k} \lambda_k (b^\dag_k + b_k) + \sum_k \omega_k b^\dag_k b_k \,.
\end{align}
The spin is described by Pauli matrices $\sigma^{\alpha}$, $\alpha = x,y,z$, and the operators $b_k$ describe bosonic oscillators with momentum $k$ and frequency $\omega_k$. They fulfill bosonic commutation relations $[b_k, b^\dag_q] = \delta_{k,q}$. In Eq.~\eqref{eq:1} we have set the reduced Planck constant $\hbar = 1$.  The spin part of the Hamiltonian contains a tunneling element $\Delta$, which induces transitions between eigenstates $\{\ket{\uparrow}, \ket{\downarrow}\}$ of $\sigma^z$. It also contains a bias field $\epsilon(t)$, which can be time-dependent and sets the energy difference between states $\ket{\uparrow}$ and $\ket{\downarrow}$. The spin couples to the bath mode $b_k$ via the $\sigma^z$ component and with coupling strength $\lambda_k$. 

It is well-known that the spin-bath coupling is uniquely characterized by the bath spectral function, which we assume to be of Ohmic form
\begin{align}
  \label{eq:2}
  J(\omega) &= \pi \sum_k \lambda_k^2 \delta(\omega - \omega_k) = 2 \pi \alpha \omega e^{- \omega/\omega_c} \,.
\end{align}
In the following, we take the bath cutoff frequency $\omega_c$ to be the largest energy scale in the system. The dimensionless parameter $\alpha \geq 0$ describes the dissipation strength. For $\alpha < 1$, the interaction with the bath renormalizes the characteristic tunneling frequency scale of the spin from its bare value $\Delta$ to 
\begin{align}
  \label{eq:3}
  \Delta_r &= \Delta \Bigl( \frac{\Delta}{\omega_c} \Bigr)^{\frac{\alpha}{1-\alpha}} \,.
\end{align}
This important energy scale governs the low-energy Fermi-liquid fixed point in the delocalized regime $\alpha < 1$, and the spin dynamics for $\alpha < 1/2$~\cite{RevModPhys.59.1, weissdissipation, lehur_entanglement_spinboson}. At $\alpha_c = 1$ there occurs a localization quantum phase transition where the bath completely suppresses tunneling between the two spin states, and formally $\Delta_r = 0$.  

In this article, we are interested to calculate the real-time dynamics of the spin $\av{\sigma^\alpha(t)}$ for different initial preparations of spin and bath. We also consider the dynamics of the spin-spin correlation function $C_z(t) = \av{\sigma^z(t) \sigma^z(0)}$. We focus on the regime of dissipation strength $0 < \alpha < 1/2$, where $\av{\sigma^z(t)}$ exhibits damped coherent oscillations. We emphasize that perturbative master-equation approaches fail for $\alpha \gtrsim 0.1$ and the regime is experimentally accessible~\cite{Manucharyan10022009,Mebrahtu-QPTResLev-Nature-2012,Pop-PhaseSlips-NatPhys-2010,PhysRevLett.106.217005}. Non-perturbative methods are thus required to reliably calculate the time-evolution of the spin. In this article, we will not discuss even stronger spin-bath coupling $\alpha > 1/2$, where the dynamics of $\av{\sigma^z(t)}$ becomes completely incoherent~\cite{egger_crossover_1997,PhysRevLett.80.4370}, before it is completely suppressed for $\alpha \geq 1$~\cite{RevModPhys.59.1,PhysRevLett.49.1545,PhysRevLett.95.196801}.

Our method allows us to investigate the non-Markovian effects of the initial spin-bath preparation on the dynamics of $\av{\sigma^\alpha(t)}$. We distinguish two different preparation schemes. In both cases the spin and bath are first brought into contact at a time $t_0$. At this time, we assume that the spin is in a given pure state, for example $\rho_S(t_0) = \ketu \brau$, and the bath is in canonical equilibrium at temperature $T$. The total state at time $t_0$ is thus given by the product state $\rho(t_0) = \rho_S(t_0) \otimes \rho_B(t_0)$ with $\rho_B(t_0) = \exp(-H_B/T)/\text{Tr} [\exp(-H_B/T)]$ where $H_B = \sum_{k} \omega_k b^\dag_k b_k$ and the Boltzmann constant is set to $k_B = 1$. We then hold the spin fixed in state $\ketu$ until time $t_I \geq t_0$. This can be achieved, for instance, by applying a large bias field $\epsilon(t) = \epsilon_0 \theta(-t+t_I)$ with $|\epsilon_0| \gg \Delta$. During the time interval $[t_0, t_I]$ spin and bath are in contact. The large-bias constraint is released for $t > t_I$, and the spin starts to evolve in time. The system thus starts out from a non-equilibrium state at time $t_I$, which is a spin-bath product state 
\begin{align}
  \label{eq:145}
  \rho(t_I) = \ketu \brau \otimes \rho_B(t_I) \,.
\end{align}
We now distinguish between two cases: either we send $t_0 \rightarrow -\infty$ or we set $t_0 = t_I$. This yields different bath states $\rho_B(t_I)$ at time $t_I$. We study the influence of the two different preparation schemes on the spin dynamics in detail below. 

We also discuss initial states which are not spin-bath product states, and consider a system starting out from its equilibrium state at time $t=0$. Computing the spin-spin correlation function $\av{\sigma^z(t) \sigma^z(0)}$, we demonstrate the effect of initial spin-bath correlations present at $t=0$.


The structure of the paper is as follows: after this introduction, we briefly develop the real-time functional integral description in Sec.~\ref{sec:real-time-functional} mainly to introduce our notation. In Sec.~\ref{sec:non-pert-stoch} we explain in detail our novel non-perturbative stochastic Schr\"odinger equation approach. We derive all important results that show how to exactly solve for the driven Ohmic spin-boson dynamics provided that $\omega_c \gg \Delta$. In Sec.~\ref{sec:anal-class-spin}, we make an analogy of the quantum spin evolution to the dynamics of a classical spin in a random magnetic field. We also expose the relation to the NIBA in a transparent manner. In Sec.~\ref{sec:spin-bath-prep} we discuss the influence of the spin-bath preparation on the dynamics of the spin. We provide different examples where the initial state of the bath has a pronounced effect on the time evolution of the spin. We show that such non-Markovian signatures can be seen in both $\av{\sigma^x(t)}$ and $\av{\sigma^z(t)}$. In Sec.~\ref{sec:correlation-function}, we describe how spin-spin correlation functions such as $C_z(t) = \av{\sigma^z(t) \sigma^z(0)}$ can be calculated within the SSE. In Sec.~\ref{sec:spin-dynam-avsigm}, we discuss the dynamics of the spin expectation value $\av{\sigma^z(t)}$ in various physically relevant situations. This component of the spin is most interesting since, in contrast to $\sigma^{x,y}(t)$, it exhibits \emph{universal dynamics} for large $\omega_c$, and is thus relevant to the dynamics of a Kondo spin. We close this article in Sec.~\ref{sec:spin-dynam-avsigm} with a summary and a discussion of open questions and current limitations of the SSE method. In the Appendices we provide for completeness all relevant formulas from the NIBA, corrections to the NIBA, and a rigorous Born approximation result of Ref.~\onlinecite{PhysRevB.71.035318}. 

\section{Real-time functional description}
\label{sec:real-time-functional}
To study the non-equilibrium dynamics of the spin-boson model in a non-perturbative way, we employ the real-time functional integral description.~\cite{RevModPhys.59.1,weissdissipation} In this section, we set the stage and introduce a few technical concepts following the seminal work of Leggett \emph{et al.} in Ref.~\onlinecite{RevModPhys.59.1}. Then, in Sec.~\ref{sec:non-pert-stoch}, in contrast to Ref.~\onlinecite{RevModPhys.59.1} we shall treat the blip-blip interaction exactly. This affects, for example, the long-time behavior of the spin dynamics. 

We are interested to calculate the spin reduced density matrix $\rho_S(t) = \text{Tr}_B \rho(t)$, where $\rho(t)$ is the full density matrix and $\text{Tr}_B$ denotes the trace over the bath. Its components can be expressed using real-time functional integrals as 
\begin{align}
  \label{eq:4}
  \bra{\sigma_f} \rho_S(t)\ket{\sigma'_f} = \int {\cal D}\sigma(\cdot) \int {\cal D}\sigma'(\cdot) {\cal A}[\sigma] {\cal A}^*[\sigma'] F[\sigma,\sigma']\,,
\end{align}
with $\sigma_f, \sigma_f' \in \{ \ketu, \ketd \}$. Here, ${\cal D}\sigma(\cdot)$ denotes integration over all real-time spin paths $\sigma(t)$ with fixed initial and final conditions. A spin path $\sigma(s)$ jumps back and forth between the two values $\sigma = \pm 1$. The initial conditions describe the preparation of the spin, while different final conditions $\sigma(t) = \sigma_f$, $\sigma'(t) = \sigma'_f$ yield different elements of the spin reduced density matrix at time $t$. 

The integrand in Eq.~\eqref{eq:4} contains ${\cal A}[\sigma]$ and ${\cal A}^*[\sigma']$, which denote the amplitude of the spin to follow a path in the absence of the bath. The effect of the environment on the spin is captured by the real-time \emph{influence functional}~\cite{feynmanvernon,RevModPhys.59.1}
\begin{align}
  \label{eq:5}
  F[\sigma,\sigma'] &= \exp \Bigl( - \frac{1}{\pi} \int_{t_0}^t ds \int_{t_0}^s ds' \bigl[ - i L_1(s-s') \xi(s) \eta(s') \nonumber \\ & \quad  + L_2(s-s') \xi(s) \xi(s') \bigr] \Bigr) \,,
\end{align}
where we have introduced symmetric and antisymmetric spin paths $\eta(s) = \frac12 [ \sigma(s) + \sigma'(s)]$ and $\xi(s) = \frac12 [ \sigma(s) - \sigma'(s)]$. It results from an exact integration over the bath degrees of freedom~\cite{feynmanvernon,weissdissipation}. It contains the real and imaginary parts of the force autocorrelation function of the environment $\pi \av{X(t) X(0)}_T = L_2(t) - i L_1(t)$ with $X = \sum_k \lambda_k (b^\dag_k + b_k)$ and 
\begin{align}
  \label{eq:6}
  L_1(t) &= \int_0^\infty d\omega J(\omega) \sin \omega t \\
\label{eq:7}
  L_2(t) &= \int_0^\infty d\omega J(\omega) \cos \omega t \coth \beta \omega/2\,,
\end{align}
where $\beta = 1/k_B T$ with temperature $T$. 

Next, one parametrizes a general (double) spin path and inserts it into the functional integral in Eq.~\eqref{eq:4}. Since the spin is held fixed at times $t < t_I$, the double spin path is constrained to one of the diagonal (or ``sojourn'') states $\{ \ketuu, \ketdd \}$. If we are interested in a diagonal element (population) of $\rho_s(t)$, we fix the final state of the spin path to be a ``sojourn'' state as well. To calculate an off-diagonal element (coherence), we let the spin path end at time $t$ in an off-diagonal (or ``blip'') state $\{ \ketud, \ketdu \}$. 

For a path that ends in a sojourn state and makes $2n$ transitions at time $t_I < t_1 < t_2 < \cdots < t_{2n} < t$ along the way, we write the spin paths as
\begin{align}
  \label{eq:8}
  \xi(t) &= \sum_{j=1}^{2n} \Xi_j \theta(t - t_j) \\
\label{eq:9}
  \eta(t) &= \sum_{j=0}^{2n} \Upsilon_j \theta(t - t_j)\,.
\end{align}
The variables $\{\Xi_1, \ldots, \Xi_{2n} \} = \{ \xi_1, - \xi_1, \ldots, - \xi_n\}$ with $\xi_j = \pm 1$ describe the $n$ off-diagonal or ``blip'' parts of the path spent in the states $\{\ketud, \ketdu\}$ during times $t_{2m-1} < t < t_{2m}$ ($m = 1, \ldots, n$), where $\xi(t) = \pm 1$ and $\eta(t) = 0$. The variables $\{\Upsilon_0, \ldots, \Upsilon_{2n} \} = \{ \eta_0, - \eta_0, \ldots, \eta_n\}$, on the other hand, characterize the $(n+1)$ diagonal or ``sojourn'' parts of the path during times $t_{2m} < t < t_{2m+1}$ ($m=0, \ldots, n$), where $\eta(t) = \pm 1$ and $\xi(t) = 0$. The beginning of the initial sojourn is either at $t_0 \rightarrow - \infty$ or at $t_0 = t_I$, depending on whether spin and bath are in contact at $t < t_I$. We discuss the influence of this initial preparation on the dynamics in detail later. Formally we have $t_{2n+1} \equiv t$, and the path's boundary conditions specify $\eta_0$ and $\eta_n$. Altogether, the two-spin path is completely characterized by the variables $ \{t_0, t_1, \ldots, t_{2n}; \xi_1, \ldots, \xi_n; \eta_0 = 1, \eta_1, \ldots, \eta_{n-1}, \eta_n\}$.
A spin path that ends in a blip state is written in an analogous way. 

Using this parametrization of the spin path in Eqs.~\eqref{eq:8} and \eqref{eq:9}, we may perform the time integrations in the influence functional in Eq.~\eqref{eq:5}, which yields
\begin{align}
  \label{eq:10}
  F_n\bigl[\{\Xi_j\}, \{\Upsilon_j\}, \{t_j\}\bigr] = {\cal Q}_1 {\cal Q}_2
\end{align}
where
\begin{align}
  \label{eq:11}
  {\cal Q}_1 &= \exp \biggl[ \frac{i}{\pi} \sum_{j > k \geq 0}^{2n} \Xi_j \Upsilon_k Q_1 (t_j - t_k) \biggr] \\
  \label{eq:12}
  {\cal Q}_2 &= \exp \biggl[ \frac{1}{\pi} \sum_{j> k \geq 1}^{2n} \Xi_j \Xi_k Q_2(t_j - t_k) \biggr]\,.
\end{align}
The bath functions $Q_{1,2}(t)$ are the second integrals of $L_{1,2}(t)$, \emph{i.e.} $\ddot{Q}_{1,2} = L_{1,2}$. Explicitly, they read for an Ohmic spectral density
\begin{align}
  \label{eq:13}
  Q_1(t) &= 2 \pi \alpha \tan^{-1}(\omega_c t) \\
\label{eq:14}
  Q_2(t) &= \pi \alpha \ln (1 + \omega_c^2 t^2) + 2 \pi \alpha \ln \Bigl(\frac{\beta}{\pi t} \sinh \frac{\pi t}{\beta} \Bigr) \,.
\end{align}
The influence functional is a product of two terms: $\mathcal{Q}_1$ and $\mathcal{Q}_2$. While $\mathcal{Q}_1$ describes a coupling between the blip and all previous sojourn parts of the path, the term $\mathcal{Q}_2$ contains the interaction between all blips (including a self-interaction). 

The environment induces a (long-range) interaction between the spin path at different times. 
The state of the spin at time $t$ depends on its state at earlier times, which leads to a non-Markovian Heisenberg equation of motion for the spin. The form of the interaction depends, of course, on the spectral density $J(\omega)$ and the temperature $T$. At zero temperature, for example, one finds that $L_2(t) = 2 \pi \alpha \omega_c^2 (1 - \omega_c^2 t^2)/(1 + \omega_c^2 t^2)^2$ only decays algebraically in time. Non-Markovian effects are thus pronounced, especially at long times. At high temperatures, on the other hand, the blip-blip interaction becomes short-ranged. In the white-noise limit at $T > \omega_c$, for example, one derives $L_2(t) = 2 \pi \alpha k_B T \delta(t)$ and the dynamics is Markovian.

The path integral of the reduced density matrix in Eq.~\eqref{eq:4} also depends on the free spin-path amplitudes ${\cal A}[\sigma]$ and ${\cal A}^*[\sigma']$. These amplitudes contribute a factor of $i \xi \eta \Delta/2$ for each transition between a sojourn state $\eta$ and a blip state $\xi$, as well as a bias-dependent phase factor 
\begin{align}
  \label{eq:15}
    H_n = \exp \biggl[ i \sum_{j=1}^{2n} \Xi_j h_\epsilon(t_j) \biggr]
\end{align}
with
\begin{align}
  \label{eq:16}
    h_\epsilon(t) = \int_{t_I}^t dt' \epsilon(t')\,. 
\end{align}
Altogether, the diagonal element of the density matrix describing the probability 
\begin{align}
  \label{eq:17}
  p(t) = \bra{\uparrow} \rho_S(t) \ketu
\end{align}
to find the system in state $\ketu$ at time $t$ is given by a series in the tunneling coupling $\Delta^2$
\begin{align}
  \label{eq:18}
  p(t) = 1 + \sum_{n=1}^\infty \Bigl( \frac{i \Delta}{2} \Bigr)^{2n} \int_{t_I}^t dt_{2n} \cdots \int_{t_I}^{t_2} dt_1 \sum_{\{\xi_j, \eta_j\}} F_n H_n\,.
\end{align}
The sum is only over even exponents of $\Delta^{2n}$, because we are calculating a diagonal element of $\rho_S(t)$. The spin expectation value $\av{\sigma^z(t)} \equiv P(t)$ can be expressed as
\begin{align}
  \label{eq:19}
  \av{\sigma^z(t)} \equiv P(t) = 2 p(t) - 1 \,.
\end{align}
In contrast, for an off-diagonal element of $\rho_S(t)$ the path ends in a blip state $\xi_{2n} = \pm 1$ and one finds 
\begin{align}
  \label{eq:20}
  \bra{\uparrow} \rho_S(t) \ketd &= \av{\sigma^+(t)} = i \xi_{2n} \sum_{n=1}^\infty \Bigl( \frac{i \Delta}{2} \Bigr)^{2n-1}  \nonumber \\ & \times \int_{t_I}^t dt_{2n-1} \cdots \int_{t_I}^{t_2} dt_1 \sum_{\{\xi_j, \eta_j\}} F_n H_n\,,
\end{align}
where $\xi_{2n} = 1$ for this off-diagonal element and $\sigma^+ = \frac12( \sigma^x + i \sigma^y)$. Note the presence of a boundary term in Eq.~\eqref{eq:20} at the final time $t$, since it now determines the end of the last blip, \emph{i.e.}, $t_{2n} =t$. 

The formal series expansions in Eqs.~\eqref{eq:18} and \eqref{eq:20} are exact. What makes these expressions complicated is the fact that the coupling between the spin paths in the influence functional $F$ is long-range in time. Hence, one must consider all terms coupling different blips and sojourns. Their analytical evaluation is only possible in special cases, \emph{e.g.} at $\alpha = 1/2$, or if one simplifies them using approximations. The most prominent so-called non-interacting blip approximation (NIBA) is discussed in detail in Appendix~\ref{sec:non-interacting-blip}. It simply neglects all interactions between different blips. In the next section, we introduce a novel method that allows us to take all terms in the influence functional exactly into account. In particular, unlike the NIBA, we exactly consider the long-range interactions between different blips. This is achieved by mapping the problem onto a linear stochastic equation that can be easily solved numerically. 


\section{Non-perturbative stochastic Schr\"odinger equation method}
\label{sec:non-pert-stoch}
We now present a method to evaluate the full spin reduced density matrix $\rho_S(t)$ in a numerically exact manner. Its element $\bra{i} \rho_S(t) \ket{j}$ with $i,j \in \{\uparrow, \downarrow\}$ is calculated by averaging over solutions of a non-perturbative stochastic Schr\"odinger equation (SSE). We explicitly derive the SSE from the expressions in Eqs.~\eqref{eq:18} and \eqref{eq:20}. This method works for all temperatures and, importantly, for an arbitrary time-dependent bias field $\epsilon(t)$. 

In contrast to other numerical approaches such as the real-time Monte-Carlo method,~\cite{springerlink:10.1007/BF01320834,PhysRevB.50.15210,PhysRevE.61.5961} or the quasi-adiabatic path integral scheme,~\cite{PhysRevB.77.195316,makri_numerical_1995,Makri-QUAPI-ChemPhysLett-1994, nalbach:220401} we will not directly evaluate the real-time path integral in Eqs.~\eqref{eq:18} and~\eqref{eq:20}. 
Instead, we first decouple the terms bilinear in the blip and sojourn variables in the influence functional $F_n = \mathcal{Q}_1 \mathcal{Q}_2$ in Eq.~\eqref{eq:10} using Hubbard-Stratonovich transformations. We then obtain $\bra{i} \rho_S(t) \ket{j}$ as a statistical average over solutions of a stochastic Schr\"odinger equation. 

We want to emphasize that our method takes all terms in the influence functional exactly into account. In particular, we fully account for all interactions between different blips. Although this method is quite powerful, it is so far restricted to the case of an Ohmic bath with $0 < \alpha < 1/2$ and a large cutoff frequency $\omega_c \gg \Delta$ (scaling limit). The reason for this limitation will become clear in the following, when we show that $\mathcal{Q}_1$ greatly simplifies in the Ohmic scaling limit. Although we can formally apply our approach also for sub-Ohmic and super-Ohmic bath spectral functions, it remains an open questions to sufficiently improve the numerical convergence properties to make it useful in practice. 


\subsection{Blip-blip interaction part $\boldsymbol{\mathcal{Q}_2}$ }
\label{sec:blip-blip-inter}
We first analyze the ${\cal Q}_2$ part of the influence functional $F_n$ in Eq.~\eqref{eq:10}, that describes the interactions between blips. Before we can apply a Hubbard-Stratonovich transformation, we must diagonalize the kernel $Q_2(t)$ and write it in a factorized form as~\cite{ImambekovGritsevDemler-FundNoiseFermiProceed2006,PhysRevA.82.032118} 
\begin{align}
  \label{eq:21}
  Q_2(t_j - t_k) = \pi \alpha \Bigl[ G_0 + \sum_{m=1}^{m_{\text{max}}} G_m \Psi_m(t_j) \Psi_m(t_k) \Bigr]\,.
\end{align}
We truncate the sum and keep $m_{\text{max}}$ terms, but always check that the final result is independent of $m_{\text{max}}$. 

To achieve this, we expand $Q_2(t)$ in a Fourier series. To obtain only negative Fourier coefficients, we rather expand $\tilde{Q}_2(\tau) = Q_2(\tau) - Q_2(2)$ and write 
\begin{align}
  \label{eq:22}
    Q_2(\tau) = Q_2(2) + \pi \alpha \Bigl[ g_0 + \sum_{m=1}^{m_{\text{max}/2}} g_m \cos \frac{m \pi \tau}{2} \Bigr]\,,
\end{align}
on the interval $\tau \in (-2,2)$. Here, $\tau = (t-t_I)/t_{\text{tot}}$ is a rescaled time which depends on the total length of our numerical simulation $t_{\text{tot}}$. The time $\tau = 0$ corresponds to the initial time $t_I$, when the large-bias constraint on the spin is turned-off. The time $\tau = 1$ corresponds to the final time $t_{\text{max}} = t_{\text{tot}} + t_I$ of our numerical simulation. Note that $Q_2(2)$ is a constant that depends on the length of the simulation $t_{\text{tot}}$. At $T=0$, it reads for example $Q_2(2) = \pi \alpha \ln [1 + 4 \omega_c^2 t_{\text{tot}}^2 ]$. Since we obtain the Fourier coefficients $g_m$ numerically, this approach is quite general and can be used for various forms of the bath correlation function $Q_2(t)$, as arise for different spectral densities $J(\omega)$.   

From the Fourier expansion, we identify the (trigonometric) eigenfunctions as $\Psi_{2k-1}(\tau) = \cos \frac{k \pi \tau}{2}$ as well as $\Psi_{2k}(\tau) = \sin \frac{k \pi \tau}{2}$. The coefficients in Eq.~\eqref{eq:21} read
\begin{align}
  \label{eq:23}
  G_0 &= g_0 + \frac{1}{\pi \alpha} Q_2(2) \\
\label{eq:24}
  G_{2k-1} &= G_{2k} = g_k < 0\,.
\end{align}
We can thus write $\mathcal{Q}_2$ in factorized form as 
\begin{align}
  \label{eq:25}
    {\cal Q}_2 &= \exp \Bigl\{-n \alpha \bigl[ \ln ( 1 + 4 \omega_c^2 t_{\text{tot}}^2 ) + G \bigr] \Bigr\} \nonumber \\ & \quad \times \prod_{m=1}^{m_{\text{max}}} \exp \biggl\{ \frac12 \Bigl[ \sqrt{\alpha G_m} \sum_{j = 1}^{2n} \Xi_j \Psi_m(t_j) \Bigr]^2 \biggr\}\,,
\end{align}
with constant $G = \sum_{m=0}^{m_{\text{max}}/2} g_m$. Note that $\lim_{m_{\text{max}} \rightarrow \infty} G = - \ln ( 1 + 4 \omega_c^2 t_{\text{tot}}^2 )$, so the prefactor in Eq.~\eqref{eq:25} approaches unity in this limit. To derive Eq.~\eqref{eq:25}, we have used that $\sum_{j>k \geq 1}^{2n} \Xi_j \Xi_k = - n$ and $\Xi_j^2 = 1$. Since $G_m < 0$ for $m \geq 1$ it is more appropriate to write $\sqrt{\alpha G_m} = i \sqrt{- \alpha G_m}$. Next, we decouple the blip variables $\{\Xi_j\}$ in the exponent in Eq.~\eqref{eq:25} using a total of $m_{\text{max}}$ Hubbard-Stratonovich transformations, resulting in 
\begin{align}
  \label{eq:26}
    {\cal Q}_2 &= \exp \Bigl\{-n \alpha \bigl[ \ln ( 1 + 4 \omega_c^2 t_{\text{tot}}^2 ) + G \bigr] \Bigr\} \nonumber \\ & \quad \times \int d{\cal S} \exp \Bigl\{ i \sum_{j=1}^{2n} \Xi_j h_s(\tau_j) \Bigr\}\,.
\end{align}
The integral over the Hubbard-Stratonovich variables $\{s_m\}$ reads
\begin{align}
  \label{eq:27}
    \int d{\cal S} &= \prod_{m=1}^{m_{\text{max}}} \int_{-\infty}^\infty \frac{d s_m}{\sqrt{2 \pi}} e^{- s_m^2/2}\,,
\end{align}
and we have introduced a real (height) function 
\begin{align}
  \label{eq:28}
    h_s(\tau) = \sum_{m=1}^{m_{\text{max}}} s_m \sqrt{- \alpha G_m} \Psi_m(\tau)\,.
\end{align}
The function $h_s(\tau)$ contains information about the environment via the eigenfunctions and eigenvalues of the bath correlation function $Q_2(t)$. It also depends on the Hubbard-Stratonovich variables $\{s_m\}$, which can be interpreted as Gaussian distributed random (noise) variables. One thus finds that $\av{h_s(t)}_{\mathcal{S}} = 0$ and $\av{h_s(t) h_s(s)}_{\mathcal{S} } = \alpha G_0 - Q_2(t-s)/\pi$, while all higher moments vanish. 

\subsection{Blip-sojourn interaction part $\boldsymbol{\mathcal{Q}_1}$}
\label{sec:blip-sojo-inter}
Let us now turn to the $\mathcal{Q}_1$ part of the influence functional $F_n = \mathcal{Q}_1 \mathcal{Q}_2$, that couples the blip and sojourn part of the spin path. It is important to distinguish between the first sojourn, which occurs during initial time $t_0 \leq t \leq t_I$ when the spin is polarized, and all other sojourns. We thus separate $\mathcal{Q}_1 = \mathcal{Q}_1^{(0)} \mathcal{Q}_1^{(1)}$. 

The contribution of the first sojourn $\mathcal{Q}_1^{(0)}$ encodes the initial preparation of the system. It is given by the terms where $k=0,1$ in Eq.~\eqref{eq:11} and reads $\mathcal{Q}_1^{(0)} = \exp[ \frac{i}{\pi} \sum_{j=1}^{2n} \Xi_j \{ Q_1(t_j - t_0) - Q_1(t_j - t_1) \} ]$. Within our method we can take it into account in an exact manner. This is described in Sec.~\ref{sec:spin-bath-prep}, and allows us to study the effect of the spin-bath preparation on the spin dynamics. It plays an important role, \emph{e.g.} for the dynamics of $\av{\sigma^x(t)}$ or if the bias $\epsilon(t)$ depends on time. The spin-bath preparation is less important for the dynamics of $\av{\sigma^z(t)}$ at constant bias~\cite{weissdissipation}. 

The contribution of all later sojourns $\mathcal{Q}_1^{(1)}$ is given by the terms with $k \geq 2$ in Eq.~\eqref{eq:11} and reads $\mathcal{Q}_1^{(1)} = \exp[ \frac{i}{\pi} \sum_{j > k \geq 2}^{2n} \Xi_j \Upsilon_k Q_1(t_j - t_k)]$. Fortunately, it takes a particularly simple form for an Ohmic bath and if $\Delta/\omega_c \ll 1$ and $\alpha < 1/2$ (scaling limit), where one may safely approximate~\cite{RevModPhys.59.1}
\begin{align}
  \label{eq:29}
  Q_1(t) &= 2 \pi \alpha \tan^{-1} (\omega_c t) \approx \alpha \pi^2 \theta(t)\,.
\end{align}
We thus find that 
\begin{align}
  \label{eq:30}
  \mathcal{Q}_1^{(1)} &= \exp \Bigl[ i \pi \alpha \sum_{j> k \geq 2}^{2n} \Xi_j \Upsilon_k \Bigr] = \exp \Bigl[ i \pi \alpha \sum_{k=1}^{n-1} \xi_{k+1} \eta_k \Bigr]\,.
\end{align}
This reflects the fact that the main contribution to the path integral stems from paths with spin-flip separations larger than $\omega_c^{-1}$. 

If we use the scaling form $Q_1(t) = \alpha \pi^2 \theta(t)$ for the first sojourn as well, this corresponds to the spin-bath preparation where $t_0 = t_I$. In this case, the complete blip-sojourn interaction term $\mathcal{Q}_1 = \mathcal{Q}_1^{(0)} \mathcal{Q}_1^{(1)}$ is given by
\begin{align}
  \label{eq:31}
      {\cal Q}_1 &=  \exp \Bigl[ i \pi \alpha \sum_{j> k \geq 0}^{2n} \Xi_j \Upsilon_k \Bigr] = \exp \Bigl[ i \pi \alpha \sum_{k=0}^{n-1} \xi_{k+1} \eta_k \Bigr]\,.
\end{align}
Let us finally note that we can, in principle, deal with the blip-sojourn interaction term ${\cal Q}_1$ in a similar way as with $\mathcal{Q}_2$. In this case, we must first separate the bath correlation function $Q_1(t)$ in the exponent into a symmetric part $Q_1(|t|)$ and an anti-symmetric part $Q_1(t)$, in order to extend the sum over the blip and sojourn variables to $j \leq k$. Then, we can diagonalize the kernels, complete the square in the exponent and linearize it using Hubbard-Stratonovich transformations. The resulting expression for the height function $h_s(\tau)$, however, is no longer purely real, but also contains an imaginary component. This leads to slow convergence properties, similar to the case of the sign problem known from Monte-Carlo sampling. This currently limits our SSE approach to an Ohmic bath with $\Delta/\omega_c \ll 1$ and $0 < \alpha < 1/2$ (see also Sec.~\ref{sec:open-quest-curr}), where $Q_1(t)= \alpha \pi^2 \theta(t)$.

\subsection{Stochastic Schr\"odinger Equation (SSE)}
\label{sec:stoch-schr-equat}
We now use the form of the influence functional $F_n = \mathcal{Q}_1 \mathcal{Q}_2$ that we have derived in the last two sections to obtain the spin reduced density matrix as a statistical average of solutions of a stochastic Schr\"odinger equation. Employing Eqs.~\eqref{eq:31},~\eqref{eq:26} and~\eqref{eq:15} a diagonal entry of $\rho_S(t)$ can be written as
\begin{align}
  \label{eq:32}
  p(\tau) &= 1 + \int d{\cal S} \sum_{n=1}^\infty \biggl( \frac{i \Delta t_{\text{tot}} e^{- (\alpha/2)\bigl[ \ln ( 1 + 4 \omega_c^2 t_{\text{tot}}^2 ) + G \bigr]}}{2} \biggr)^{2n} \nonumber \\ & \;\; \times \int_0^\tau d\tau_{2n} \cdots \int_0^{\tau_2} d\tau_1 \sum_{\{ \xi_j, \eta_j\}} \exp \Bigl[ i \pi \alpha \sum_{k=0}^{n-1} \eta_k \xi_{k+1} \Bigr] \nonumber \\ & \; \; \times \prod_{j=1}^{2n} \exp \bigl[ i \, \Xi_j h(\tau_j) \bigr]\,.
\end{align}
Here, we have defined the total height function 
\begin{align}
  \label{eq:33}
  h(\tau) = h_s(\tau) + h_\epsilon(\tau)\,.
\end{align}
It contains both the random height function $h_s(\tau)$ in Eq.~\eqref{eq:28} as well as the bias-dependent part which reads 
\begin{align}
  \label{eq:34}
 h_\epsilon(\tau) = \int_0^{t_{\text{tot}} \tau} d\tau' \epsilon(\tau') \,.
\end{align}
Without the summation over the blip and sojourn variables $\{\xi_j, \eta_j\}$ the expression in Eq.~\eqref{eq:32} has the form of a time-ordered exponential, averaged over the random noise variables $\{s_m\}$. This summation, however, can easily be incorporated into a product of matrices in the vector space of two-spin states $\{\ketuu, \ketud, \ketdu, \ketdd\}$, that read~\cite{imambek_jetp_02}
\begin{align}
  \label{eq:36}
  V &= V_0
  \begin{pmatrix} 
    0 & e^{- i h(\tau)} & - e^{i h(\tau)} & 0 \\ 
    e^{i \pi \alpha} e^{i h(\tau)} & 0 & 0 & - e^{- i \pi \alpha} e^{i h(\tau)} \\
    - e^{- i \pi \alpha} e^{- i h(\tau)} & 0 & 0 & e^{i \pi \alpha} e^{- i h(\tau)} \\
    0 & - e^{- i h(\tau)} & e^{i h(\tau)} & 0
\end{pmatrix} \,,
\end{align}
where
\begin{align}
  \label{eq:37}
  V_0 = \frac12 \Delta t_{\text{tot}} \exp\Bigl\{-(\alpha/2) \bigl[ \ln ( 1 + 4 \omega_c^2 t_{\text{tot}}^2 ) + G \bigr] \Bigr\}\,.
\end{align}
Note that $V_0 = \frac12 \Delta t_{\text{tot}}$ for $m_{\text{max}} \rightarrow \infty$. It is worth emphasizing that the two-spin basis states simply correspond to the four elements of the reduced density matrix $\bra{i} \rho_S \ket{j}$. The final two-spin state $\ket{ij}$ with $i,j \in \{\uparrow, \downarrow\}$ of the real-time spin path determines which density matrix element $\bra{i} \rho_S(t) \ket{j}$ is calculated. A product of matrices of the type in Eq.~\eqref{eq:36} automatically satisfies the requirement that transitions between two-spin states occur via single spin flips. A two-spin path consists of an alternating sequence of sojourn (diagonal) and blip (off-diagonal) parts. The different signs in Eq.~\eqref{eq:36} stem from the free-spin contribution $i \Delta \xi \eta/2$ for each spin flip between blip state $\xi$ and sojourn state $\eta$. 

We finally arrive at the central result of our work. With Eq.~\eqref{eq:36} we can express Eq.~\eqref{eq:32} as a time-ordered exponential
\begin{align}
  \label{eq:38}
  p(\tau) = \int d{\cal S} \braopket{\Phi_f}{ T e^{-i \int_0^\tau ds V(s) } }{\Phi_i}\,.
\end{align}
Here, $T$ is the usual time-ordering operator. The two-spin states $\ket{\Phi_i}$ and $\ket{\Phi_f}$ are the initial and final states of the spin path. When calculating the diagonal element $p(\tau)=\brau \rho_S(\tau) \ketu$, we thus have $\Phi_f = \ketuu$. Since we consider an initial polarization of the spin in state $\ketu$, it follows that $\ket{\Phi_i} = \ketuu$ as well. We can evaluate the amplitudes on the right-hand side of Eq.~\eqref{eq:38} by solving the stochastic Schr\"odinger equation
\begin{align}
  \label{eq:39}
  i \frac{\partial}{\partial \tau} \ket{\Phi(\tau)} = V(\tau) \ket{\Phi(\tau)}
\end{align}
with initial and final conditions $\Phi_{i,f} = (1,0,0,0)^T$. The vector $(1,0,0,0)^T$ corresponds to the basis state $\ketuu$. The integration $\int d\mathcal{S}$ over the Hubbard-Stratonovich variables is performed by averaging the result over $N$ different realizations of the noise variables $\{s_m\}$. One then obtains $p(\tau)$ by averaging over the different results 
\begin{align}
  \label{eq:40}
 p(\tau) = \frac{1}{N} \sum_{k=1}^N \Phi_1^{(k)}(\tau) = \av{\Phi_1(\tau) }_{\mathcal{S}} \,,
\end{align}
where $\Phi_1(\tau)$ is the first component of $\ket{\Phi(\tau)}$ and $\av{\cdot}_{\mathcal{S}}$ denotes the average over the Hubbard-Stratonovich random noise variables $\{s_m\}$. We note that $\av{\Phi_1(\tau) }_{\mathcal{S}}$ is purely real as required. The spin expectation value $\av{\sigma^z(t)}$ is given by $\av{\sigma^z(t)} \equiv P(t) = 2 p(t) - 1$. 

Other components of the spin reduced density matrix $\rho_S$ can be computed simply by using different final conditions $\Phi_{f}$. In order to calculate the off-diagonal element $\brau \rho_S(t)\ketd$ for instance, we must project onto the final state $\ketud$ which corresponds to $\Phi_f = (0,1,0,0)^T$. In this case we also need to consider a boundary term in the influence functional at $\tau=\tau_{2n}$ which arises if the spin path ends in a blip state~\cite{weissdissipation}. It appears as if the system steps back to a sojourn state at the final time $\tau$. This can be implemented by multiplying $\Phi_2(\tau)$ with $(\Delta_r t_{\text{tot}}/(\omega_c/\Delta))^\alpha$, where $\Delta_r = \Delta (\Delta/\omega_c)^{\alpha/(1 - \alpha)}$ is the renormalized tunneling element. 

The different spin expectation values are found from 
\begin{align}
  \label{eq:41}
  \av{\sigma^x(t)} &= 2 (\Delta_r t_{\text{tot}}\Delta/\omega_c)^\alpha \av{\Phi'_2(t_{\text{tot}} \tau)}_{\mathcal{S}} \\
\label{eq:42}
  \av{\sigma^y(t)} &= 2 (\Delta_r t_{\text{tot}}\Delta/\omega_c)^\alpha \av{\Phi''_2(t_{\text{tot}} \tau)}_{\mathcal{S}} \\
\label{eq:43}
\av{\sigma^z(t)} &= 2 \av{\Phi'_1(t_{\text{tot}} \tau)}_{\mathcal{S}} - 1\,,
\end{align}
with $\Phi'_\alpha = \text{Re}~\Phi_\alpha$, $\Phi''_{\alpha} = \text{Im}~\Phi_\alpha$ and $V_0$ given in Eq.~\eqref{eq:37}. Here, we have set $t_I = 0$ for notational clarity. 
Because of the boundary factor $(\Delta_r t_{\text{tot}}\Delta/\omega_c)^\alpha$, the spin expectation values $\av{\sigma^{x,y}(t)}$ are not universal functions and strictly vanish in the scaling limit $\Delta/\omega_c \rightarrow 0$. A universal function depends on the bath cut-off frequency $\omega_c$ only through the renormalized tunneling element $\Delta_r$, and is thus universal as a function of the dimensionless variable $y = \Delta_r t$. 

Apart from using the scaling form $Q_1(t) \approx \alpha \pi^2 \theta(t)$ [see Eq.~\eqref{eq:29}], the final expressions in Eq.~\eqref{eq:40}-\eqref{eq:43} are still exact in the limit $m_{\text{max}} \rightarrow \infty$ and $N \rightarrow \infty$. In practice, of course, we work with finite values of typically $m_{\text{max}} \approx 4000$ and $N \sim 10^6 - 10^7$. We always check that the final result for $p(t)$ and $\brau \rho_S(t)\ketd$ is independent of $m_{\text{max}}$ and $N$. As the numerical accuracy of our results scales with the number of noise realizations as $N^{-1/2}$, we are able to routinely calculate the spin expectation values $\av{\sigma^{x,y,z}(t)}$ up to an (absolute) uncertainty as small as $10^{-4}$.

\subsection{Symmetries of the stochastic equations}
\label{sec:symm-stoch-equat}
It is interesting to note that the differential equations in Eq.~\eqref{eq:39} with the initial condition $\Phi_i = (1,0,0,0)^T$ in fact obey the additional symmetries
\begin{align}
  \label{eq:44}
  \Phi^{\prime \prime}_1(\tau) &= 0 \\
\label{eq:45}
\Phi_3^*(\tau) &= \Phi_2(\tau) \\
\label{eq:46}
\Phi_4(\tau) &= 1 - \Phi_1(\tau) \,.
\end{align}
From the eight real variables $\{\Phi'_{1,2,3,4}, \Phi^{\prime \prime}_{1,2,3,4}\}$ in fact only three are independent. Choosing as independent variables $\{ \Phi'_1, \Phi'_2, \Phi^{\prime \prime}_2\}$ the differential equations~\eqref{eq:39} read explicitly
\begin{align}
  \label{eq:47}
  \partial_\tau \Phi'_1 &= 2 V_0 \bigl[ - \sin(h) \Phi'_2 + \cos(h) \Phi^{\prime \prime}_2 \bigr] \\
\label{eq:48}
    \partial_\tau \Phi'_2 &= V_0 \bigl[ \cos(h) \sin(\pi \alpha) + \cos(\pi \alpha) \sin(h) ( 2 \Phi'_1 - 1) \bigr] \\
\label{eq:49}
  \partial_\tau \Phi^{\prime \prime}_2 &= V_0 \bigl[ \sin(\pi \alpha) \sin(h) - \cos(\pi \alpha) \cos(h) (2 \Phi'_1 - 1) \bigr]\,.
\end{align}
For a particular realization of the random height function $h(\tau)$, the time evolution described by these equations is \emph{not unitary}. The values of the different components $\{\Phi'_1, \Phi'_2, \Phi^{\prime \prime}_2\}$ are therefore not bounded. In contrast, we can restore the $\xi_j \rightarrow - \xi_j$ symmetry by performing two simulations with $h(\tau) = h_\epsilon(\tau) \pm h_s(\tau)$, which yield the two (unbounded) solutions $\Phi^{\pm}(t)$. We then take the average $\Phi(\tau) = \frac12 [ \Phi^+(\tau) + \Phi^-(\tau) ]$, and find that both $\Phi'_1(\tau)$ and $\Phi^{\prime \prime}_2(\tau)$ are bounded if and only if $\epsilon = 0$. In this case, the components are constrained to $\Phi'_1(\tau) \in [0,1]$ and $\Phi^{\prime \prime}_2(\tau) \in [-\frac12, \frac12]$. This significantly improves the numerical convergence properties, also for the remaining component $\Phi'_2(\tau)$ even though it is not bounded.





\section{Analogy to classical spin in random field and relation to NIBA}
\label{sec:anal-class-spin}
In this section, we show that we can interpret the stochastic equation of motion for $\av{\sigma^z(t)} = P(t)$ in the scaling limit $\Delta/\omega_c \ll 1$ and at zero bias $\epsilon = 0$ as that of a classical spin $\bfss(t)$ that rotates in a random magnetic field $\boldsymbol{H}(t)$. Quantum effects and dissipation follow from the non-commutativity of rotations around different axes, which are induced by the magnetic field, and due to the average over different random magnetic field configurations. 

Importantly, however, the quantum-classical analogy only applies to the $y$ and $z$-component of the spin. These components are related as $\av{\sigma^y(t)} = - \Delta^{-1} \frac{d}{dt} \av{\sigma^z(t)}$~\cite{weissdissipation}. The analogy cannot be applied to the component $\av{\sigma^x(t)}$, because the derivation below relies on an assumption that is not valid for $\av{\sigma^x(t)}$. In fact, the ``classical'' Bloch-type equation, which we derive in the following, predicts $\av{\sigma^x(t)} \equiv 0$ which is incorrect. 

We then exploit the analogy to the classical Bloch equations to explain the relation of the SSE method to the well-known non-interacting blip approximation (NIBA). We show that the NIBA follows in our approach from neglecting correlations between the classical trajectory of the spin and the random height function $h_s(t)$. 

\subsection{Analogy to classical Bloch equations}
\label{sec:anal-class-bloch}
We have previously shown how to obtain the spin expectation values $\av{\sigma^{x,y,z}(t)}$ by solving the stochastic equation~\eqref{eq:39} for different height functions $h(\tau)$ and averaging over the results. Different vector components $\av{\Phi_\alpha(t)}_{\mathcal{S}}$ determine different spin components [see Eqs.~\eqref{eq:41}-\eqref{eq:43}]. For a particular realization of the height function, however, the components $\Phi_\alpha(t)$ are unbounded and can therefore not be interpreted as components of a spin. The situation is different if we explicitly perform the sum over the sojourn variables in the expression of $\mathcal{Q}_1$ in Eq.~\eqref{eq:31}, which gives
\begin{align}
  \label{eq:50}
  \mathcal{Q}_1 &= e^{i \pi \alpha \xi_1 } (2 \cos \pi \alpha)^{n-1} \,.
\end{align}
For zero bias $\epsilon =0$ and if we are interested in calculating $\av{\sigma^z(t)}$, where the system ends in a sojourn state $\eta_n$, we may use that the real-time functional integral expression for $p(t)$ in Eq.~\eqref{eq:18} is invariant under the reversal of the sign of all blip variables $\xi_j \rightarrow - \xi_j$. The imaginary part in Eq.~\eqref{eq:50} does therefore not contribute to $\av{\sigma^z(t)}$, since it cancels after summation over $\xi_1$. This assumption does not hold for $\av{\sigma^x(t)}$, where the system ends in a blip state $\xi$. Neglecting the imaginary part in Eq.~\eqref{eq:50} is therefore not justified for $\av{\sigma^x(t)}$. In fact, $\av{\sigma^x(t)}$ is solely determined by the imaginary contribution in Eq.~\eqref{eq:50}, which is antisymmetric in $\xi_1$. Since we only keep the real part in the following, the resulting equations yield $\av{\sigma^x(t)} \equiv 0$.

Keeping only the real part in Eq.~\eqref{eq:50}, we find from Eq.~\eqref{eq:32} that $\av{\sigma^z(t)} = P(t) = 2 p(t) - 1$
is given by
\begin{align}
  \label{eq:51}
  \av{\sigma^z(\tau)} &= 1 + \int {\cal D}\mathcal{S} \sum_{n=1}^\infty \bigl(i \sqrt{2 \cos \pi \alpha} \, V_0 \bigr)^{2n} \nonumber \\ & \quad \times \int_0^\tau d\tau_{2n} \cdots \int_0^{\tau_2} d\tau_1 \sum_{\{\Xi_j\}} \prod_{j=1}^{2n} \exp{[i \Xi_j h_s(\tau_j)] }\,,
\end{align}
with $V_0$ given in Eq.~\eqref{eq:37}. Note that Eq.~\eqref{eq:51} only contains the random part of the height function $h_s(\tau)$, since we have assumed that $\epsilon = 0$. It is worth noting that the right-hand side of Eq.~\eqref{eq:51}vanishes at the Toulouse point $\alpha = 1/2$. This point marks the boundary between coherent and incoherent dynamics of $\av{\sigma^z(t)}$. For $\alpha=1/2$ the spin-boson model may be solved exactly via mapping to a non-interacting resonant level model, and one finds $\av{\sigma^z(t)}_{\alpha = 1/2} = \exp(- \frac{\pi \Delta^2}{2 \omega_c} t)$~\cite{RevModPhys.59.1,weissdissipation}.

Since the sojourn variables $\{\eta_j\}$ are now absent from the expression, we do not need to distinguish between the two diagonal states $\ketuu$ and $\ketdd$ anymore. It is thus sufficient to work with a three-dimensional basis $\{\ketuu, \ketud, \ketdu\}$. To perform the summation over the blip variables $\{\Xi_j\}$ via a matrix product, we introduce the three-dimensional matrix $V_3(\tau)$. Specifically, we can express $\av{\sigma^z(t)}$ as a time-ordered product, averaged over the noise variables, as
\begin{align}
  \label{eq:52}
  \av{\sigma^z(\tau)} = \int {\cal D} \mathcal{S} \braopket{\Phi_f}{T e^{- i \int_0^\tau d\tau' V_3(\tau')}}{\Phi_i}
\end{align}
with the matrix
\begin{align}
  \label{eq:53}
  V_3(\tau) = V_0 \sqrt{2 \cos \pi \alpha} \begin{pmatrix} 0 & e^{- i h_s(\tau)} & -e^{i h_s(\tau)} \\ e^{i h_s(\tau)} & 0 & 0 \\ -e^{-i h_s(\tau)} & 0 & 0 \\ \end{pmatrix}\,.
\end{align}
The random height function $h_s(t)$ is defined in Eq.~\eqref{eq:28}. To calculate $\av{\sigma^z(t)}$, we solve the differential equations
\begin{align}
  \label{eq:54}
  i \frac{\partial}{\partial \tau} \ket{\Phi(\tau)} = V_3(\tau) \ket{\Phi(\tau)}
\end{align}
with initial and final conditions $\ket{\Phi_{i,f}} = (1,0,0)^T$, and take the average over $N$ different noise configurations: $\av{\sigma^z(t)} = \av{\Phi_1(t_{\text{tot}} \tau + t_I)}_{\mathcal{S}}$. 

Similarly to Eq.~\eqref{eq:39}, the differential equations~\eqref{eq:54} with initial condition $\ket{\Phi_i} = (1,0,0)^T$ obey additional symmetries
\begin{align}
  \label{eq:55}
  \text{Im} \, \Phi_1(\tau) &= 0 \\
\label{eq:56}
   \Phi_3^*(\tau) &= \Phi_2(\tau) \,.
\end{align}
Only three out of the six variables $\text{Re}(\Phi_{1,2,3})$, $\text{Im}(\Phi_{1,2,3})$ are thus independent. As before, we choose $\{ \Phi'_1, \Phi'_2, \Phi^{\prime \prime}_2\} = \{ \text{Re} \Phi_1, \text{Re} \Phi_2, \text{Im} \Phi_2\}$, and the differential equations~\eqref{eq:54} explicitly read
\begin{align}
  \label{eq:57}
  \partial_\tau \Phi'_1 &= 2 V_0 \sqrt{ 2 \cos \pi \alpha} \bigl[ \cos(h_s) \Phi^{\prime \prime}_2 - \sin(h_s) \Phi'_2 \bigr] \\
\label{eq:58}
   \partial_\tau \Phi'_2 &= V_0 \sqrt{2 \cos \pi \alpha} \sin(h_s) \Phi'_1 \\
\label{eq:59}
  \partial_\tau \Phi^{\prime \prime}_2 &= - V_0 \sqrt{2 \cos \pi \alpha} \cos(h_s) \Phi'_1 \,.
\end{align}
Denoting a solution for a particular realization of the height function $h(\tau)$ by $\Phi_\alpha^{(+)}(\tau)$, then the solutions for $[-h(\tau)]$ are given by $\Phi_1^{\prime -}(\tau)  = \Phi_1^{\prime +}(\tau)$, $\Phi_2^{\prime -}(\tau) = - \Phi_2^{\prime +}(\tau)$ and $\Phi_2^{\prime \prime -}(\tau) = \Phi_2^{\prime \prime +}(\tau)$. Since we draw the random variables $\{s_m\}$ from a Gaussian distribution with mean $\av{s_m}_{\mathcal{S}} = 0$, both $h(\tau)$ and $[-h(\tau)]$ are equally probable and $\av{h(\tau)}_{\mathcal{S}} = 0$. As discussed earlier, Eq.~\eqref{eq:54} therefore (incorrectly) predicts that $\av{\sigma^x(\tau)} = 2  \av{\Phi'_2(\tau)}_{\mathcal{S}}= 0$. 

A crucial observation is that the system of differential equations in Eq.~\eqref{eq:54} has an integral of motion 
\begin{align}
  \label{eq:60}
  1 = \Phi_1(\tau)^2 + 2 |\Phi_2(\tau)|^2 = \Phi'_1(\tau)^2 + 2 \bigl[\Phi'_2(\tau)^2 + \Phi^{\prime \prime}_2(\tau)^2\bigr]\,.
\end{align}
This allows us to introduce an effective classical spin variable 
\begin{align}
  \label{eq:61}
  \boldsymbol{S} = (S^x, S^y, S^z) = \bigl(\sqrt{2} \Phi^{\prime}_2, \sqrt{2} \Phi^{\prime \prime}_2, \Phi^{\prime}_1 \bigr)\,,
\end{align}
which is normalized to unit length $| \boldsymbol{S} | = 1$. The equations of motion for the spin components in Eqs.~\eqref{eq:57}-\eqref{eq:59} can be compactly expressed as the \emph{classical Bloch equation}
\begin{align}
  \label{eq:62}
  \frac{d}{d \tau} \boldsymbol{S}(\tau) = \boldsymbol{H}(\tau) \times \boldsymbol{S}(\tau)\,.
\end{align}
The effective noisy magnetic field $\boldsymbol{H}(\tau)$ depends on the random height function $h_s(\tau)$ and lies in the $x$-$y$-plane
\begin{align}
  \label{eq:63}
  \boldsymbol{H} = H_0 \bigl(\cos h_s(\tau), \sin h_s(\tau), 0\bigr) \,.
\end{align}
The amplitude of the magnetic fields reads $H_0 = \sqrt{2} V_0 \sqrt{2 \cos \pi \alpha}$. The dissipative dynamics of the \emph{quantum} spin follows from averaging over different random field configurations as
\begin{align}
  \label{eq:64}
    \av{\sigma^z(t)} = \av{S^z(t)}_{\mathcal{S}}\,.
\end{align}
The quantum problem of the dissipative time evolution of the quantum spin component $\av{\sigma^z(t)}$ at zero bias can therefore be formulated as an evolution of a classical spin $\boldsymbol{S}(t)$ in a random magnetic field $\boldsymbol{H}(t)$. The quantum nature of the problem is hidden in the fact that spin rotations about different axes do not commute and through the averaging over different random field configurations. It is important to keep in mind, however, that the quantum-classical correspondence relies on an assumption that is not valid for $\sigma^x$. It is therefore restricted to the $y$ and $z$-components of the spin, and predicts that $\av{S^x(t)}_{\mathcal{S}} = 0 \neq \av{\sigma^x(t)}$. 

\subsection{Relation between SSE method and NIBA}
\label{sec:relation-sse-method}
We can employ the classical spin description of the previous section to make the relation of the SSE method to the NIBA transparent. Starting from the ``classical'' Bloch equations of motion in Eq.~\eqref{eq:62}, which yield the exact result for $\av{\sigma^z(t)} = \av{S^z(t)}_\mathcal{S}$ in the limit of $m_{\text{max}}, N \rightarrow \infty$, we derive an equation that describes $\av{\sigma^z(t)}$ within the NIBA.

We start from the ``classical'' Bloch equation~\eqref{eq:62} in the random magnetic field $\boldsymbol{H}=H_0 \bigl(\cos[h_s(\tau)], \sin[h_s(\tau)], 0\bigr)$. The different spin components obey $\dot{S}^x = H_y S^z$, $\dot{S}^y = - H_x S^z$ and $\dot{S}^z = H_x S^y - H_y S^x$. For the $z$-component we thus obtain
\begin{align}
  \label{eq:65}
  \dot{S}^z(t) &= - \Delta^2\cos(\pi \alpha) \int_0^t ds   \cos \bigl[ h_s(t) - h_s(s) \bigr] S^z(s)\,,
\end{align}
 since $H_0 = \Delta \sqrt{\cos \pi \alpha}$ for $m_{\text{max}} \rightarrow \infty$. For a particular realization of the noise $\{s_m\}$, which defines the height function $h_s(t)$, Eq.~\eqref{eq:65} is not in convoluted form. In general, one cannot write $[h_s(t) - h_s(s)]$ as a function of the time difference $(t-s)$ only. It is thus not possible to solve Eq.~\eqref{eq:65} via Laplace transformation. 

In order to find the time evolution of the quantum spin $\av{\sigma^z(t)}$, we have to average $S^z(t)$ over different magnetic field configurations
\begin{align}
  \label{eq:66}
  \frac{d}{dt} \av{\sigma^z(t)} &= \av{\dot{S}^z(t)}_{\mathcal{S}} =  - \Delta^2 \cos(\pi \alpha) \nonumber \\ & \qquad \times \int_0^t ds  \; \Bigl\langle \cos \bigl[ h_s(t) - h_s(s) \bigr] S^z(s)\Bigr\rangle_{\mathcal{S}} \,.
\end{align}
It is important to note that there exist correlations between the random height function $h_s(t)$ and the classical spin trajectory $S^z(s)$ such that in general
\begin{align}
  \label{eq:67}
    \Bigl\langle \cos \bigl[ &h_s(t) - h_s(s) \bigr] S^z(s)\Bigr\rangle_{\mathcal{S}} \nonumber \\ & \qquad  \neq \Bigl\langle \cos \bigl[ h_s(t) - h_s(s) \bigr] \Bigr\rangle_{\mathcal{S}} \Bigl\langle S^z(s)\Bigr\rangle_{\mathcal{S}} \,.
\end{align}
These correlations are absent in the initial state at $t=0$, but are generated over the course of time, as follows from the differential equation~\eqref{eq:65}. The correlations are thus small at short times $t$. Note also that since $h_s(t) \sim \sqrt{\alpha}$, the factor $\cos \bigl[ h_s(t) - h_s(s) \bigr] \approx 1$ for small $\alpha \ll 1$. At $\alpha = 0$, both classical and quantum spin undergo undamped Rabi oscillations with frequency $\Delta$. The correlations between $h_s(t)$ and $S^z(t)$ thus become more pronounced for larger values of $\alpha$. The mean-field decoupling anticipated in Eq.~\eqref{eq:67} can thus be justified at short times $t$ and/or small dissipation $\alpha$. Indeed, we now show that one recovers the NIBA from this mean-field decoupling.

Using the mean-field approximation of Eq.~\eqref{eq:67}, we obtain for the equation of motion of the quantum spin 
\begin{align}
  \label{eq:68}
  \frac{d}{dt} \av{\sigma^z(t)} &= \av{\dot{S}^z(t)}_{\mathcal{S}} \approx  - \Delta^2 \cos(\pi \alpha) \nonumber \\ & \quad \times \int_0^t ds  \; \Bigl\langle \cos \bigl[ h_s(t) - h_s(s) \Bigr] \Bigr\rangle_{\mathcal{S}} \Bigl\langle S^z(s)\Bigr\rangle_{\mathcal{S}} \,.
\end{align}
Using the statistical properties of the height function $\av{h_s(\tau)}_{\mathcal{S}} = 0$ and $\av{h_s(\tau) h_s(s)}_{\mathcal{S}} = \alpha G_0 - Q_2(\tau - s)/\pi$, one easily computes the expectation value
\begin{align}
  \label{eq:69}
  \Bigl\langle \cos \bigl[ h_s(t) - h_s(s) \bigr] \Bigr\rangle_{\mathcal{S}} = \exp \bigl[ - Q_2(t - s)/\pi \bigr]\,.
\end{align}
The equation of motion thus takes the form 
\begin{align}
  \label{eq:70}
  \frac{d}{dt} \av{\sigma^z(t)} &= - \Delta^2 \cos(\pi \alpha) \\ & \qquad \times \int_0^t ds \exp \bigl[ - Q_2(t - s)/\pi \bigr] \av{\sigma^z(s)} \nonumber \,,
\end{align}
which we recognize as the NIBA equation of motion as we show in Appendix~\ref{sec:deriv-using-heis} following Refs.~\onlinecite{PhysRevA.35.1436, aslangul_spin-boson_1986}.

We note that one also recovers the NIBA equations for finite bias $\epsilon_0 \neq 0$ within the same mean-field decoupling scheme of Eq.~\eqref{eq:67}. For non-zero bias, however, one must use the equations of motion in Eqs.~\eqref{eq:47}--\eqref{eq:49} to arrive at~\cite{RevModPhys.59.1}
\begin{align}
  \label{eq:143}
  \frac{d}{dt} \av{\sigma^z(t)} + \int_0^t d\tau \bigl[ f(t-\tau) + g(t - \tau) \av{\sigma^z(\tau)} \bigr]  = 0\,,
\end{align}
where $f(t-\tau) = \Delta^2 e^{-Q_2(t-\tau)/\pi} \sin(\pi \alpha) \sin[\epsilon_0(t-\tau)]$ and $g(t-\tau) = \Delta^2 e^{-Q_2(t-\tau)/\pi} \cos(\pi \alpha) \cos[\epsilon_0(t-\tau)]$. Since the norm of the vector $\bfss$ defined in Eq.~\eqref{eq:61} is not conserved for finite $\epsilon_0$ however, we cannot interpret each trajectory $\Phi_j(t)$ for a fixed realization of the noise $\mathcal{S}$ as the path of a classical spin in a random field. 

To summarize, within our approach we recover the NIBA when neglecting the statistical correlations between the classical spin trajectories in a random magnetic field and the random magnetic field itself (see Eq.~\eqref{eq:67}). These correlations develop over the course of time and thus become more pronounced at longer times. Since the magnetic field fluctuations grow with larger dissipation, the correlations also increase with $\alpha$. The derivation of the NIBA using our SSE approach thus makes the validity of the NIBA at short times and/or weak dissipation very transparent. In the following sections, we study the differences between these two methods. 


\section{Spin-bath preparation effects}
\label{sec:spin-bath-prep}
In this section, we investigate the effect of the initial spin-bath preparation on the dynamics of the spin. We show that the initial preparation can influence the dynamics even at long times. The SSE method is ideally suited to consider such situations, because it allows us to take the spin-bath preparation exactly into account. Specifically, we give two examples: the dynamics of the coherence $\sigma^x(t)$ and the behavior of $\sigma^z(t)$ under a linear Landau-Zener sweep of the bias.

We distinguish two different spin-bath preparation schemes: in the first one, spin and bath are brought into contact at time $t_0 \rightarrow - \infty$, but the spin is held fixed in state $\ketu$ by applying a large bias field until a much later time $t_I$, where $|t_I - t_0| \rightarrow \infty$. By the time $t_I$, the bath has relaxed to the shifted canonical equilibrium state 
\begin{align}
  \label{eq:71}
  \rho_B(\sigma_i) = \frac{\exp\bigl\{ - \beta [H_B + \frac{\sigma_i}{2} \sum_k \lambda_k (b^\dag_k + b_k)] \bigr\}}{\text{Tr} \bigl[ \exp\bigl( - \beta [H_B + \frac{\sigma_i}{2} \sum_k \lambda_k (b^\dag_k + b_k)] \bigr) \bigr] }\,,
\end{align}
with $\sigma_i = 1$ corresponding to initial spin state $\ketu$. In the second preparation scheme, spin and bath are brought into contact only at time $t_0 = t_I$. The initial bath state then equals the canonical thermal state, as given by Eq.~\eqref{eq:71} with $\sigma_i = 0$. 

We first explain how the exact consideration of the spin-bath preparation is technically implemented within the SSE method, and then investigate two physical situations.

\subsection{Exact consideration of spin-bath preparation within SSE method} 
\label{sec:exact-cons-spin}
The spin-bath preparation is encoded in the contribution of the initial sojourn between time $t_0$ and time $t_1$, which denotes the beginning of the first blip. To exactly take the spin-bath preparation into account, we must therefore use the full form of the bath kernel function $Q_1(t) = 2 \pi \alpha \tan^{-1}(\omega_c t)$ in the contribution of the first sojourn.

The contribution of the first sojourn corresponds to the $k=0,1$ terms in Eq.~\eqref{eq:11}, and explicitly reads 
\begin{align}
  \label{eq:72}
  \mathcal{Q}_1^{(0)} &=   \exp \biggl[ \frac{i}{\pi} \sum_{j=1}^{2n} \Xi_j \Bigl\{ Q_1(t_j - t_0) - Q_1(t_j - t_1) \Bigr\} \biggr] \,.
\end{align}
We can incorporate these terms in an exact manner by adding them to the height function $h(\tau)$ in Eq.~\eqref{eq:33}, which then explicitly depends on $\tau_1$, the beginning of the first blip, and reads
\begin{align}
  \label{eq:73}
  h(\tau, \tau_1) & = h_s(\tau) + h_\epsilon(\tau) - 2 \alpha \Bigl\{ \tan^{-1}\bigl[ \omega_c t_{\text{tot}} (\tau - \tau_1) \bigr] \nonumber \\ & \quad - \delta_{t_0,t_I} \tan^{-1}\bigl[ \omega_c t_{\text{tot}} \tau \bigr] \Bigr\} \,.
\end{align}
The terms $h_s(\tau)$ and $h_\epsilon(\tau)$ are defined in Eqs.~\eqref{eq:28} and~\eqref{eq:34}. The last term is present only for the preparation scheme $t_0 = t_I$ and absent for $t_0 \rightarrow - \infty$. 

The fact that the height function $h(\tau, \tau_1)$ in Eq.~\eqref{eq:73} now depends on $\tau_1$ forces us to explicitly perform the integration over $\tau_1 \in [0,1]$ in Eq.~\eqref{eq:32}. We thus randomly pick a uniformly distributed $\tau_1 \in [0,1]$, which determines the height function $h(\tau, \tau_1)$ in Eq.~\eqref{eq:73}. It also determines the initial state of the simulation via a single application of $V(\tau_1)$ on $\Phi_i = (1,0,0,0)^T$ as
\begin{align}
  \label{eq:74}
  \ket{\Phi_{\tau_1}}  & = - i \bigl( 0, e^{i h(\tau_1, \tau_1)}, - e^{-i h(\tau_1, \tau_1)}, 0 \bigr)^T\,.
\end{align}
Note that the factors $\exp(\pm i \pi \alpha)$ that we would expect from naively computing $-i V(\tau_1)\Phi_i$ do not occur in Eq.~\eqref{eq:74}, because they are taken into account by the last two terms in the height function in Eq.~\eqref{eq:73}. 

We then propagate this initial state $\ket{\Phi_{\tau_1}}$ in the interval $[\tau_1, 1]$ according to Eq.~\eqref{eq:39}, and calculate the probability to find the system in state $\ketu$ at time $t = \tau t_{\text{tot}} + t_I$ as
\begin{align}
  \label{eq:75}
  p(\tau) = 1 + \av{\Phi_1(\tau)}_{\mathcal{S}}\,.
\end{align}
From this we find $\av{\sigma^z(\tau)} = 2 p(\tau) - 1$. The average is over $N$ different choices of $\tau_1$ and sets of random variables $\{s_m\}$. In each individual run, we set $\ket{\Phi(\tau < \tau_1)} = 0$, since $\Phi(\tau)$ only accounts for the contribution of paths with at least one spin flip. This also explains the difference to Eq.~\eqref{eq:40}. The off-diagonal element $\brau \rho_S(t)\ketd = \av{\sigma^+}$ is still determined by $\av{\Phi_2}_{\mathcal{S}}$ and yields the spin expectation value $\av{\sigma^x(t)} = 2 (\Delta_r t_{\text{tot}}\Delta/\omega_c)^\alpha \text{Re} \, \av{\Phi_2}_{\mathcal{S}}$ (see Eq.~\eqref{eq:41}). 

\subsection{Dynamics of coherence $\av{\sigma^x(t)}$}
\label{sec:coherence-avsigmaxt}
In this section, we investigate the dynamics of the spin expectation value $\av{\sigma^x(t)}$, which describes phase coherence between states $\ketu$ and $\ketd$~\cite{lehur_entanglement_spinboson}. The quantum (de)coherence can be measured in a persistent current experiment or in a SQUID geometry as suggested by Refs.~\onlinecite{PhysRevLett.84.346,kopp:220401}. We show that the initial spin-bath preparation has a profound influence on its dynamics, not just at short times. Of course, in the long-time limit the system relaxes to its equilibrium state and $\av{\sigma^x(t)}$ becomes independent on the preparation scheme. 

\begin{figure}[h!]
  \centering
    \includegraphics[width=.95\linewidth]{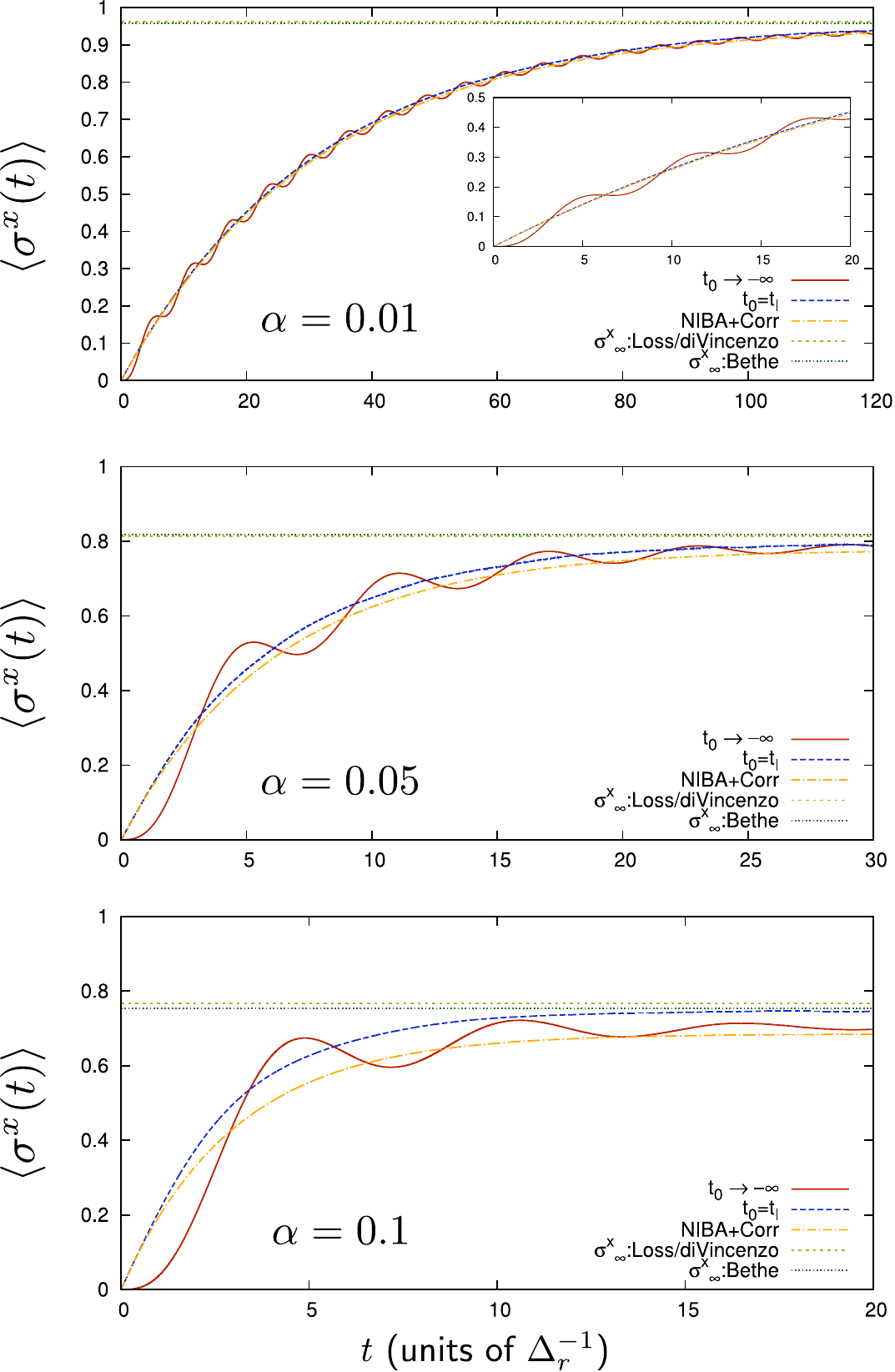}
  \caption{Coherence $\av{\sigma^x(t)}$ for different values of $\alpha = \{0.01, 0.05, 0.1\}$ with $\omega_c = 200 \Delta$ for $\alpha=\{0.01,0.05\}$ and $\omega_c = 50 \Delta$ for $\alpha=0.1$. Other parameters are $\Delta = 1$, $\epsilon = 0$, $m_{\text{max}} = 3000$ and $N = 9 \times 10^6$. We present SSE results for two different spin-bath preparations: $t_0 \rightarrow - \infty$ and $t_0 = t_I$. While $\av{\sigma^x(t)}$ oscillates for $t_0 \rightarrow - \infty$, it increases monotonously for $t_0 = t_I$. We include results of a weak-coupling theory beyond NIBA (``NIBA+Corr'')~\cite{weiss_dynamics_1989,goerlich_low-temperature_1989,weissdissipation}, that we discuss in Appendix~\ref{sec:weak-coupl-extens}. SSE curves approach the correct thermodynamic expectation value $\av{\sigma^x}_\infty$ at long times as calculated from Bethe ansatz~\cite{lehur_entanglement_spinboson}. We also show $\av{\sigma^x}_{\infty, \text{Loss/DiVincenzo}}$ calculated from a rigorous Born approximation to order $\alpha$ by Loss and DiVincenzo~\cite{PhysRevB.71.035318}. }
  \label{fig:1}
\end{figure}

In Fig.~\ref{fig:1}, we present $\av{\sigma^x(t)}$ for the two preparation schemes that we have introduced above. The spin-bath interaction is either turned on at $t_0 \rightarrow - \infty$ (red line) or at $t_0 = t_I$ (blue dashed line). While $\av{\sigma^x(t)}$ clearly exhibits oscillations if $t_0 \rightarrow -\infty$, it increases monotonously for $t_0 = t_I$. The oscillation frequency is of the order of the renormalized tunneling element $\Delta_r$ and independent of the bath cutoff frequency $\omega_c$. For both preparation protocols $\av{\sigma^x(t)}$ approaches the correct equilibrium value, which we have computed using the thermodynamic Bethe ansatz for the interacting resonant level model, which can be mapped to the spin-boson model.~\cite{lehur_entanglement_spinboson} The spin relaxation to the ground state occurs due to thermalization with the bath. 

We can intuitively understand the appearance of the oscillations for $t_0 \rightarrow -\infty$ in the following way. For this spin-bath preparation the initial bath state at $t=t_I$ is polarized, because the bath has relaxed to the state $\rho_B(1)$ in Eq.~\eqref{eq:71} due to the interaction with the fixed spin. At $t=t_I$, all harmonic oscillators are in the ground state of a shifted quadratic potential. This shifted bath state acts as a bias field $\epsilon_B(t)$ in the $z$-direction for the spin. At $t=t_I$, it reads 
\begin{align}
  \label{eq:76}
  \epsilon_{B}(t_I) = \av{\sum_k \lambda_k (b^\dag_k + b_k)}_{\rho_B(1)} = - 2 \alpha \omega_c \,.
\end{align}
Since the spin is released for $t>t_I$, it relaxes towards $\av{\sigma^z}_\infty = 0$ for $\epsilon =0$. The polarization of the harmonic oscillator bath therefore slowly disappears for $t > t_I$, and as a result $\epsilon_B(t) \rightarrow 0$. For each oscillator the relaxation process occurs on a timescale given by its frequency $\omega_k$. Due to the presence of many slow modes in the Ohmic bath with $\omega_k < \Delta_r$, the spin thus experiences the bath induced bias field $\epsilon_B(t)$ until times much larger than $\Delta_r^{-1}$. 

We can understand the oscillations in $\av{\sigma^x(t)}$ by noticing that the total (``magnetic'') field for the spin reads 
\begin{align}
  \label{eq:77}
  \bfbb = \bigl(\Delta_{r}, 0, \epsilon_B(t) \bigr) \,,
\end{align}
where initially $\epsilon_B(t_I) = - 2 \alpha \omega_c$ and we have taken into account the renormalization of $\Delta$ to $\Delta_{r}$. The spin rotates around the magnetic field vector $\bfbb(t)$ with a frequency $\Delta_B(t) = \sqrt{\Delta_{r}^2 + \epsilon_B(t)^2}$ that is given by the total field strength $|\bfbb|$. Different components $\av{\sigma^\alpha(t)}$ are given by projections on the different axes. Oscillations in $\sigma^x(t)$ thus only occur if the field does not point along the $x$-direction, \ie only as long as $\epsilon_B(t) \neq 0$. 

It is worth pointing out that although $\epsilon_B(t_I) \gg \Delta_{r}$ for the parameters in Fig.~\ref{fig:1}, we observe an oscillation frequency of the order of $\Delta_{r}$, independently of $\epsilon_B(t_I)$, which depends on the bath cutoff $\omega_c$. This can be easily understood from the fact that the bath oscillators with frequencies $\omega_k > \Delta_{r}$ relax on a fast timescale much smaller than $\Delta_{r}^{-1}$. The relevant bath induced detuning for $t>\Delta_{r}^{-1}$ is thus rather given by $\epsilon_B \approx - 2 \alpha \Delta_{r}$. 

The amplitude of the oscillations in $\av{\sigma^x(t)}$ is proportional to the angle $\gamma$ between $\bfbb(t)$ and the $x$-axis that reads $\gamma = \tan^{-1} (\epsilon_B(t)/\Delta_{r})$. In fact, measuring the oscillation amplitude of $\av{\sigma^x(t)}$ is a new kind of bath spectroscopy. It yields the bath relaxation function $\epsilon_B(t)$ in Eq.~\eqref{eq:76}, which contains information about the distribution of oscillators and their coupling to the spin. 

This clearly non-Markovian effect of the bath initial state is captured exactly within SSE, as shown in Fig.~\ref{fig:1}. Since the NIBA yields erroneous results for calculating $\av{\sigma^x(t)}$, we compare SSE to predictions of a weak-coupling theory beyond the NIBA (``NIBA+Corr''). This approach perturbatively accounts for all interblip correlations up to first order in $\alpha$~\cite{weissdissipation,weiss_dynamics_1989,goerlich_low-temperature_1989}. Details about NIBA+Corr are provided in Appendix~\ref{sec:weak-coupl-extens}. 

In the case of $t_0 = t_I$, where the bath initial state is unpolarized, SSE and NIBA+Corr agree well at weak dissipation ($\alpha = 0.01$). For slightly stronger dissipation of $\alpha \in \{0.05, 0.1\}$, however, the agreement is limited to short times only. This reflects the fact that blip-blip interactions become more important at longer times.
\begin{figure}[t!]
  \centering
    \includegraphics[width=\linewidth]{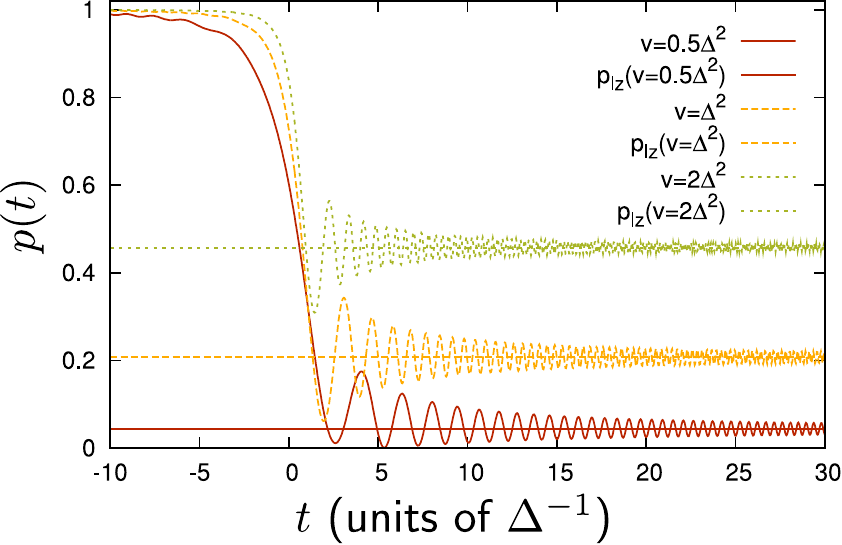}
  \caption{Survival probability $p(t)$ of a free spin for different sweep velocities $v/\Delta^2 = \{0.5, 1,2\}$ and $\Delta = 1$. After the jump at $\epsilon =0$, $p(t)$ rapidly converges to the classic Landau-Zener result $p_{lz}$ on a timescale $t \approx \Delta/v$.}
  \label{fig:2}
\end{figure}

Most importantly, in contrast to NIBA+Corr, the SSE results approach the correct thermodynamic stationary value $\av{\sigma^x}_\infty$ at long times, independently of the spin-bath preparation and for all values of $\alpha$. The relaxation to the ground state occurs since spin and bath thermalize. This clearly exemplifies the strength of the SSE approach. In Fig.~\ref{fig:1}, we include the result for $\av{\sigma^x}_\infty$ of two different calculations. First, using a thermodynamic Bethe ansatz for the interacting resonant level model~\cite{lehur_entanglement_spinboson} and, second, using a rigorous Born approximation to order $\alpha$~\cite{PhysRevB.71.035318} (see Appendices~\ref{sec:weak-coupl-extens} and~\ref{sec:exact-born-appr} for the analytical expressions).
\label{sec:land-zener-trans}

\subsection{Landau-Zener transition}
Another important situation where the spin-bath preparation affects the spin dynamics at long times is the famous Landau-Zener level crossing problem, where the bias field varies linearly in time like $\epsilon(t) = v t$ with $v>0$. Such a Landau-Zener sweep of the bias arises in a variety of physical areas such as molecular collisions~\cite{Child_molecular_collision_theory_book}, chemical reaction dynamics,~\cite{Nitzan_chem_dyn_in_condensed_phase_book} molecular nanomagnets,~\cite{W.Wernsdorfer04021999} quantum information and metrology,~\cite{Wallraff_Nature_2004,sillanpPRL,OrlandoLevitov_AmplitSpectr_Nature_2008,Manucharyan10022009,ZuecoHanggiKohler-LZInCavQED-NewJPhys-2008} and cold-atom systems.~\cite{zenesini:090403,ChenBlochAltman-LZ1DBose-NatPhys-2011,PhysRevLett.108.175303,UehlingerEsslinger-LZDirac-arXiv-2012}

In the absence of dissipation, the Landau-Zener problem can be solved exactly.~\cite{landau_lz,zener_lz,stueckelberg_lz,majorana_lz} As shown in Fig.~\ref{fig:2}, the (survival) probability $p(t)$ for the spin to remain in its initial state $\ketu$ shows a jump at the resonance $\epsilon = 0$ at $t=0$. After the resonance, $p(t)$ quickly converges to its asymptotic value
\begin{align}
  \label{eq:78}
  \lim_{t \rightarrow \infty} p(t) = \exp \Bigl[ - \frac{\pi \Delta^2}{2 v} \Bigr] \equiv p_{lz} \,.
\end{align}
The convergence occurs as soon as $\epsilon \sim \Delta$, \ie, on a timescale $t \approx \Delta/v$. 
\begin{figure}[b!]
  \centering
  \includegraphics[width=\linewidth]{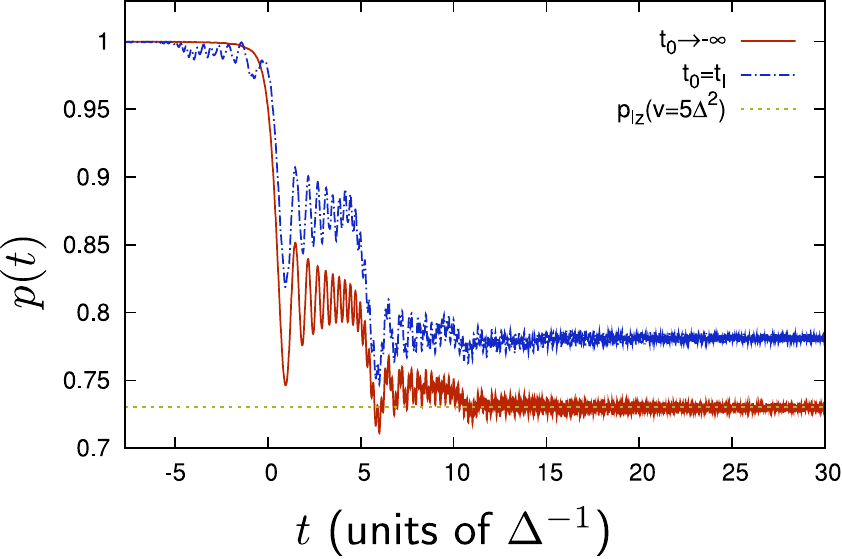}
  \caption{Toy model result for $p(t)$ under a Landau-Zener sweep for $v= 5\Delta^2$ and different bath initial states. It is only for $\rho_B(1)$, \emph{i.e.}, $t_0 \rightarrow - \infty$, that $p(t)$ converges towards $p_{lz}$ at long times of $\mathcal{O}(\omega_c/v)$. Other parameters are $\omega_c = 25 \Delta$, $\lambda = 8 \Delta$, $t_I = - 30 \Delta^{-1}$, and $\Delta=1$. }
  \label{fig:24}
\end{figure}

\begin{figure}[t!]
  \centering
  \includegraphics[width=\linewidth]{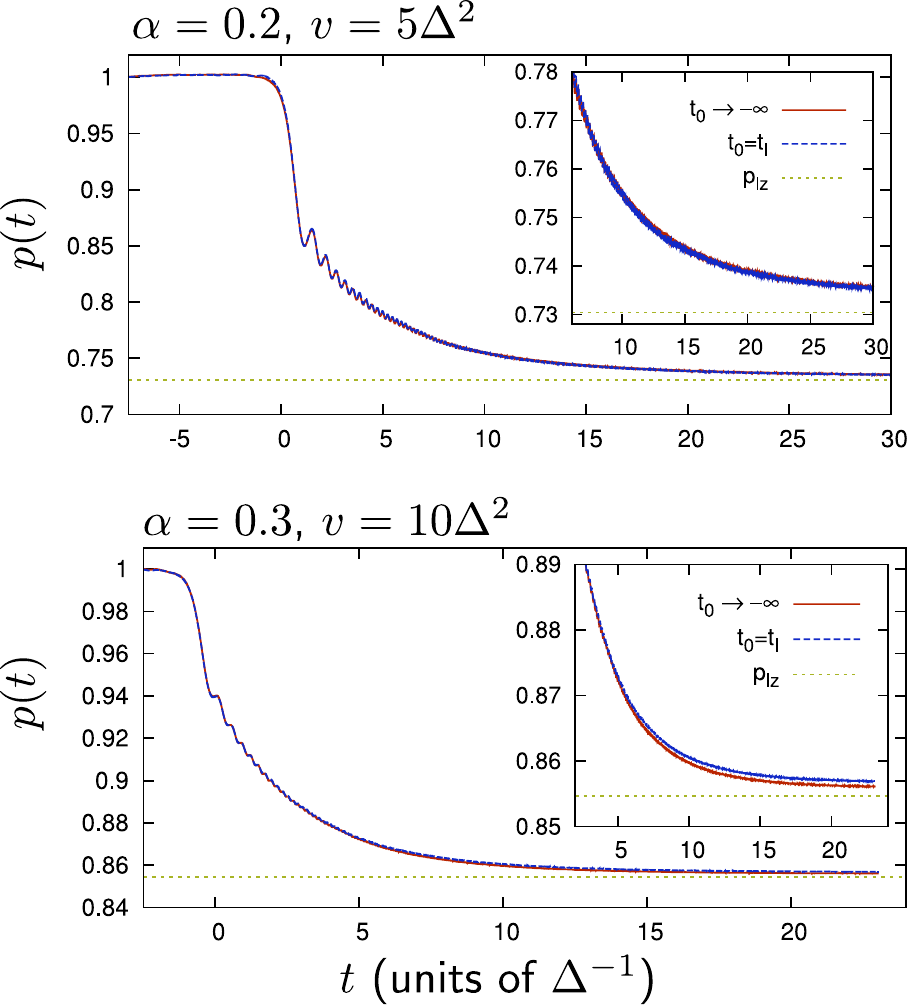}
  \caption{Survival probability $p(t)$ of spin coupled to an Ohmic bath under Landau-Zener sweep. In the upper panel, the detuning is swept with velocity $v = 5 \Delta^2$ and dissipation strength $\alpha = 0.2$. In the lower panel, velocity is $v = 10 \Delta^2$ and $\alpha = 0.3$. Other parameters read $\omega_c = 50 \Delta$, $\Delta = 1$, $m_{\text{max}} = 2000$, and $N \sim 10^6$.  }
  \label{fig:23}
\end{figure}

A fundamental question is how the coupling to an environment affects the dynamics and the asymptotic value of $p(t)$. Analytical results are only know in certain limits~\cite{PhysRevB.43.5397,PhysRevLett.62.3004,PhysRevB.57.13099,grifoni_driven_1998, PhysRevB.76.024310}. Quite surprisingly, it was proved rigorously in Refs.~\onlinecite{wubs:200404,saito:214308} that at zero temperature, the asymptotic transition probability in the presence of dissipation is still given by the classic Landau-Zener result $p_{lz}$, which is derived in the absence of dissipation, \emph{provided} the spin-bath coupling is purely longitudinal (\ie via $\sigma^z$) and the total system is initially prepared in its ground state. The proof is valid for any type of bath. 

The proof breaks down, however, for any other initial spin-bath state. In general one finds that the initial preparation affects the asymptotic long-time value of $p(t)$ in the Landau-Zener sweep. Furthermore, the timescale at which the spin reaches its asymptotic limit is governed by the bath cutoff frequency, which is typically orders of magnitude larger than $\Delta$. 

We exemplify this clearly by investigating a model of a spin coupled to a single bosonic oscillator mode as arises in cavity QED setups~\cite{Astafiev-SingleAtomLasing-Nature-2007,PhysRevLett.100.014101,PhysRevB.80.014519,EsslingerRMP-arXiv2012}. It is described by the Hamiltonian 
\begin{align}
  \label{eq:79}
  H_{\text{single-mode}} &= \frac{\Delta}{2} \sigma^x + \frac{v t}{2} \sigma^z + \frac{\sigma^z}{2} \lambda (b + b^\dag) + \omega_c b^\dag b\,.
\end{align}
In Fig.~\ref{fig:24}, we observe in this toy model that $p(t)$ undergoes a sequence of discrete steps separated in time by $\omega_c/v$. We change the detuning from an initial value of $\epsilon(t_I) = - 150 = - 6 \omega_c$ at $t_I = -30 \Delta^{-1}$ to a final value of $\epsilon(t_f) = 150 \Delta = 6 \omega_c$ at $t_f = 30 \Delta^{-1}$. This behavior occurs since the system is driven through a series of avoided crossings, which are separated in energy by $\omega_c$. Each spin state is dressed by a ladder of bosonic states $\ket{n_b}$ where $n_b \in \mathbb{N}$ denotes the occupation number of the bosonic mode. 

The probability $p(t)$ only converges towards $p_{lz}$ for the initial preparation $t_0 \rightarrow -\infty$, which corresponds to the initial bath state $\rho_B(1)$ at $t=t_I$ (see Eq.~\eqref{eq:71} with only one bosonic mode here). The timescale of the convergence is set by the oscillator frequency: $t \sim \omega_c/v$, and is much larger than $\Delta$ (compare to Fig.~\ref{fig:2}). If, on the other hand, the system does not start out from the ground state of the full Hamiltonian in Eq.~\eqref{eq:79}, as is the case for $t=t_I$ where the initial bath state reads $\rho_B(0) = \ket{0}\bra{0}$ with $b \ket{0} = 0$, we observe that $p(t)$ does \emph{not} approach $p_{lz}$ at long times. The long-time value of $p(t)$ thus depends on the initial spin-bath preparation. Of course, the difference in the final values depends on the coupling strength between spin and bath. 

We have also investigated the Landau-Zener sweep for a spin coupled to an Ohmic bath. This is shown in Fig.~\ref{fig:23} for two different velocities $v = \{5 \Delta^2, 10 \Delta^2\}$ and spin-bath coupling strengths $\alpha = \{0.2, 0.3\}$. Prominently, we find that the jump at resonance at $t=0$ is strongly suppressed due to the coupling to the bath. The size of the jump decreases with increasing $\alpha$. The series of steps, which occurred for the single-mode bath, is replaced with a smooth decay of $p(t)$ for the continuous bath. The decay occurs over a timescale governed by the bath cutoff frequency $\omega_c$. In this decay region, which occurs for intermediate times $\Delta/v \lesssim t \lesssim \omega_c/v$, the spin dynamics is universal~\cite{PhysRevA.82.032118}. 

At long times, the system converges to the classic Landau-Zener result $p_{lz}$. Interestingly, this is true for both initial spin-bath preparations, at least for sufficiently weak interaction $\alpha = 0.2$. While the proof of Refs.~\onlinecite{saito:214308,wubs:200404} is only valid for the preparation $t_0 \rightarrow -\infty$, we conclude that significant differences in the asymptotic limit of $p(t)$ require large spin-bath couplings, at least $\alpha \gtrsim 0.2$. This is in agreement with the result for $\alpha= 0.3$ in the lower panel of Fig.~\ref{fig:23}, where we notice first small differences in the long-time value of $p(t)$. 

In conclusion, the asymptotic long-time value of $p(t)$ depends on the spin-bath preparation scheme. Convergence to the classic Landau-Zener result $p_{lz}$ occurs if the full system starts out from the ground state, \emph{e.g.}, for an infinitely long sweep. Significant differences, however, require sufficiently strong spin-bath coupling.



\section{Correlation function}
\label{sec:correlation-function}
With the SSE method we can also access the spin-spin autocorrelation function
\begin{align}
  \label{eq:80}
  C_z(t) &= \av{\sigma^z(t) \sigma^z(0)}_T - \av{\sigma^z}_T^2 \,.
\end{align}
where $\sigma^z(t)$ is taken in the Heisenberg picture and $\av{\sigma^z}_T = \text{Tr} \bigl( \sigma^z e^{-\beta H} \bigr)/\text{Tr} \bigl(e^{-\beta H} \bigr)$ denotes the equilibrium expectation value at temperature $T$ with respect to the full spin-boson Hamiltonian $H$ in Eq.~\eqref{eq:1}. To obtain $\lim_{t \rightarrow \infty} C_z(t) = 0$, we subtracted the equilibrium value $\av{\sigma^z}_T = \frac{\epsilon}{\Delta_b} \tanh \frac{\Delta_b}{2 T}$ with $\Delta_b = (\Delta_{\text{eff}}^2 + \epsilon^2)^{1/2}$ and $\Delta_{\text{eff}} = [ \Gamma(1 - 2 \alpha) \cos ( \pi \alpha) ]^{1/ 2 (1-\alpha)} \Delta_r$. 

In the following, we focus on the symmetric autocorrelation function, which is the real part of $C_z(t)$. The antisymmetric part $\chi_z(t) = - 2 \theta(t) \text{Im} C_z(t)$ can be computed within the SSE approach in a similar way. 
The symmetric part of the autocorrelation function $S_z(t) = \text{Re} \, C_z(t)$ can be expressed as~\cite{PhysRevA.41.5383}
\begin{align}
  \label{eq:81}
  S_z(t) &= P_s(t) + \lim_{t_0 \rightarrow -\infty} Q_s(t, t_0) \,,
\end{align}
where $P_s(t)$ is the bias-symmetric part of $P(t) \equiv \av{\sigma^z(t)}$, \ie, $P_s(t) = \frac12 [ P(\epsilon,t) + P(-\epsilon,t)]$. 
The remaining part $Q_s(t, t_0)$ in Eq.~\eqref{eq:81} describes the difference between the equilibrium autocorrelation function $S_z(t)$ and the single-spin expectation value $P_s(t)$ (for zero bias) due to the different bath preparation protocols. We always use $t_I = 0$ in this section. 

It is worth pointing out that the initial condition for $P(t=0) = 1$ does not correspond to a small perturbation. Therefore, $P(t)$ can not be expressed in terms of equilibrium correlation functions.~\cite{RevModPhys.59.1} The initial spin-bath state is a product state $\rho(0) = \ketu \otimes \rho_B(1)$. In contrast, in the case of $S_z(t)$ the system is in its equilibrium state at $t=0$, where spin-bath correlations are present. These equilibrium spin-bath correlations are described by the term $Q_s(t,t_0)$ in Eq.~\eqref{eq:81}~\cite{PhysRevA.41.5383, PhysRevA.36.3509}.

The initial spin-bath correlations lead to significant differences between $S_z(t)$ and $P(t)$, especially at low temperatures. At $T=0$, for example, $P(t)$ decays exponentially~\cite{PhysRevLett.80.4370} (or exponentially $\times$ power-law~\cite{egger_crossover_1997}) at long times (see more later in Sec.~\ref{sec:spin-dynam-avsigm}). In contrast, the long-time decay of the autocorrelation is algebraically $S_z(t) \sim t^{-2}$ for all $\alpha < 1$. This includes the exactly solvable Toulouse point $\alpha = 1/2$, where $P(t) = \exp \bigl( - \gamma t \bigr)$ with $\gamma = \pi \Delta^2/2 \omega_c$ while $S_z(t) \approx - \bigl(4 /\pi \gamma t)^2$ for $t \rightarrow \infty$~\citep{PhysRevA.41.5383,weissdissipation}. We note that the fact that $S_z(t) \sim t^{-2}$ for $\alpha < 1$ follows very generally from the Shiba relation~\cite{PTP.54.967, sassetti_universality_1990,PhysRevLett.76.1683,PhysRevLett.80.1038}, which yields $\lim_{\omega \rightarrow 0} S_z(\omega) = 2 \pi \alpha (\bar{\chi}_z/2)^2 |\omega| \sim \alpha |\omega|$, where $\bar{\chi}_z = \text{Re} \chi_z(\omega=0)$ is the static susceptibility. 

\subsection{Computation of $S_z(t)$ with SSE method}
\label{sec:impl-with-sse}
We now show how to calculate $S_z(t)$ using the SSE approach. The additional term $Q_s(t,t_0)$ in Eq.~\eqref{eq:81}, that describes the spin-bath correlations in the equilibrium state reads explicitly~\citep{PhysRevA.41.5383}
\begin{align}
  \label{eq:82}
  &Q_s(t, t_0) = \bigl[ - \tan^2(\pi\alpha) \bigr] \sum_{n, m=1}^\infty \Bigl( \frac{i \Delta}{2} \Bigr)^{2n + 2m} \Bigl[ 2 \cos(\pi \alpha) \Bigr]^{n+m} \nonumber \\ 
  & \quad \times \int_{0}^t dt_{2n+2m} \cdots \int_{0}^{t_{2n + 2}} dt_{2n+1} \int_{t_0}^0 dt_{2n} \cdots \int_{t_0}^{t_2} dt_1 \nonumber \\ & \quad \times \sum_{\{\xi_j\}} \xi_1 \xi_{n+1} {\cal Q}_2 \cos \Bigl[ \epsilon \sum_{j=1}^{2n+2m} \Xi_j t_j \Bigr]\,.
\end{align}
Here, we have assumed a constant bias value $\epsilon$. For simplicity, we will focus on $\epsilon =0$ in the following, but it is straightforward to include a finite bias into our formalism. 

If there was no explicit dependence on the first blips at negative and positive times $\{\xi_1, \xi_{n+1} \}$, we could directly use the SSE formalism developed for $P(t)$ in Sec.~\ref{sec:anal-class-spin} to calculate $Q_s(t, t_0)$. There we learned that for $\epsilon=0$ it is possible to first perform the sum over sojourn states, and define the stochastic Schr\"odinger equation via a three-dimensional matrix $V_3(\tau)$ (see Eq.~\eqref{eq:53}). In order to keep track of the sign of the initial blips at negative and positive time, $\{\xi_1, \xi_{n+1}\}$, we add a fourth state to the SSE matrix formalism. For non-zero bias, we would simply introduce a fifth state in the version of our formalism with $V_4(\tau)$. This additional state serves as the initial state of the stochastic Schr\"odinger equation at times $t=t_0$ and $t=0$
\begin{align}
  \label{eq:83}
  \ket{\Phi(t=t_0)} = \ket{\Phi(t=0)} = (0,0,0,1)^T\,.
\end{align}
The enlarged matrix is obtained from $V_3(\tau)$ in Eq.~\eqref{eq:53} and now reads
\begin{align}
  \label{eq:84}
  V_Q(\tau) &= V_0 \sqrt{2 \cos(\pi \alpha)} \nonumber \\ & \times 
  \begin{pmatrix}
    0 & e^{- i h_s(\tau)} & - e^{i h_s(\tau)} & 0\\
    e^{i h_s(\tau)} & 0 & 0 & e^{i h_s(\tau)} \\
    -e^{- i h_s(\tau)} & 0 & 0 & e^{- i h_s(\tau)} \\
    0 & 0 & 0 & 0 
    \end{pmatrix}
\,,
\end{align}
with $h_s(\tau)$ and $V_0$ given in Eqs.~\eqref{eq:28} and~\eqref{eq:37}. 

We start our simulation at the early time $t_0$, which should in principle be sent to $t_0 \rightarrow -\infty$. In practice, we have to take this limit numerically. We use a large negative time $t_0 < 0$, where $|t_0| \gg \Delta^{-1}$, and check that the final result does not depend on $t_0$. This is to ensure that spin and bath have come to equilibrium by the time $t=0$. The final time of our simulation is given by $t_{\text{max}} > 0$. As before, we denote the total length by $t_{\text{tot}} =  t_{\text{max}} - t_0$, and the rescaled time by $\tau = (t-t_0)/t_{\text{tot}} \in [0,1]$. 

The function $Q_s(t,t_0)$ is then calculated from
\begin{align}
  \label{eq:85}
  Q_s(t, t_0) &= - \tan^2(\pi \alpha) \av{\Phi_1\bigl(|t_0|/t_{\text{tot}}\bigr)}_{{\cal S}}\nonumber \\ & \qquad \quad \times  \av{\Phi_1 \bigl[ \bigl( t + |t_0|\bigr)/t_{\text{tot}} \bigr]}_{{\cal S}}\,,
\end{align}
where $\av{\Phi_1\bigl(|t_0|/t_{\text{tot}}\bigr)}_{\cal S}$ is the first component of the solution at time $\tau = |t_0|/t_{\text{tot}}$ of the system of equations
\begin{align}
  \label{eq:86}
  i \frac{\partial}{\partial \tau} \ket{\Phi(\tau)} &= V_Q \ket{\Phi(\tau)} \\
  \label{eq:87}
  \tau &\in \bigl[ 0, |t_0|/t_{\text{tot}} \bigr] \\
\label{eq:88}
\ket{\Phi(\tau=0)} &= (0,0,0,1)^T\,,
\end{align}
averaged over the noise variables $\mathcal{S}$. 
The function $\av{\Phi_1 [ ( t + |t_0|)/t_{\text{tot}} ]}_{{\cal S}}$, on the other hand, is the first component of the solution at time $\tau = (t + |t_0|)/t_{\text{tot}}$ of the system of equations
\begin{align}
  \label{eq:89}
  i \frac{\partial}{\partial \tau} \ket{\Phi(\tau)} &= V_Q \ket{\Phi(\tau)} \\
\label{eq:90}
\tau &\in \bigl[ |t_0|/t_{\text{tot}}, t/t_{\text{tot}} \bigr] \\
\label{eq:91}
\ket{\Phi(\tau=|t_0|/t_{\text{tot}})} &= (0,0,0,1)^T\,,
\end{align}
again averaged over the noise $\mathcal{S}$. 

Initially, the system is in the sojourn state $\ket{\Phi(\tau = 0)} = (0,0,0,1)^T$. Applying $V_Q$ once moves the system to one of the blip states $\{\ketud, \ketdu\}$. The sign of the first blip $\xi_1$ is taken care of by the different signs in the fourth column compared to the first column in Eq.~\eqref{eq:84}. After an even number of spin transitions, the system has returned to a sojourn state at time $t=0$ corresponding to $\tau = |t_0|/t_{\text{tot}}$. Projecting onto the first component $\Phi_1(\tau = |t_0|/t_{\text{tot}})$ assures that we only consider paths in Eq.~\eqref{eq:85} that visit a sojourn state at $t=0$. This is required for the symmetric autocorrelation function. The sign of the first blip $\xi_{n+1}$ at $t>0$ is again taken care of by starting from state $(0,0,0,1)^T$ at $t =0$ and definition of $V_Q$. 

If we were interested in the antisymmetric autocorrelation function $\chi_z(t) = i \theta(t) \av{[ \sigma^z(t), \sigma^z(0)]}_T$ we would have to consider paths that visit a blip state at $t=0$. This quantity has been analyzed within the NRG via a mapping to the anisotropic Kondo model in Ref.~\onlinecite{PhysRevLett.76.1683, PhysRevLett.80.1038}. As in the case of the calculation of the coherence $\av{\sigma^x(t})$, if we want to compute $\chi_z(t)$ we are required to keep all four two-spin basis states in the SSE calculation and use the four-dimensional matrix $V_4(\tau)$. Apart from this difference, the calculation of $\chi_z(t)$ is similar to the one for $S_z(t)$. 
\subsection{Results for the symmetric autocorrelation function $S_z$}
\label{sec:results-symm-autoc}
\begin{figure}[h!]
  \centering
    \includegraphics[width=\linewidth]{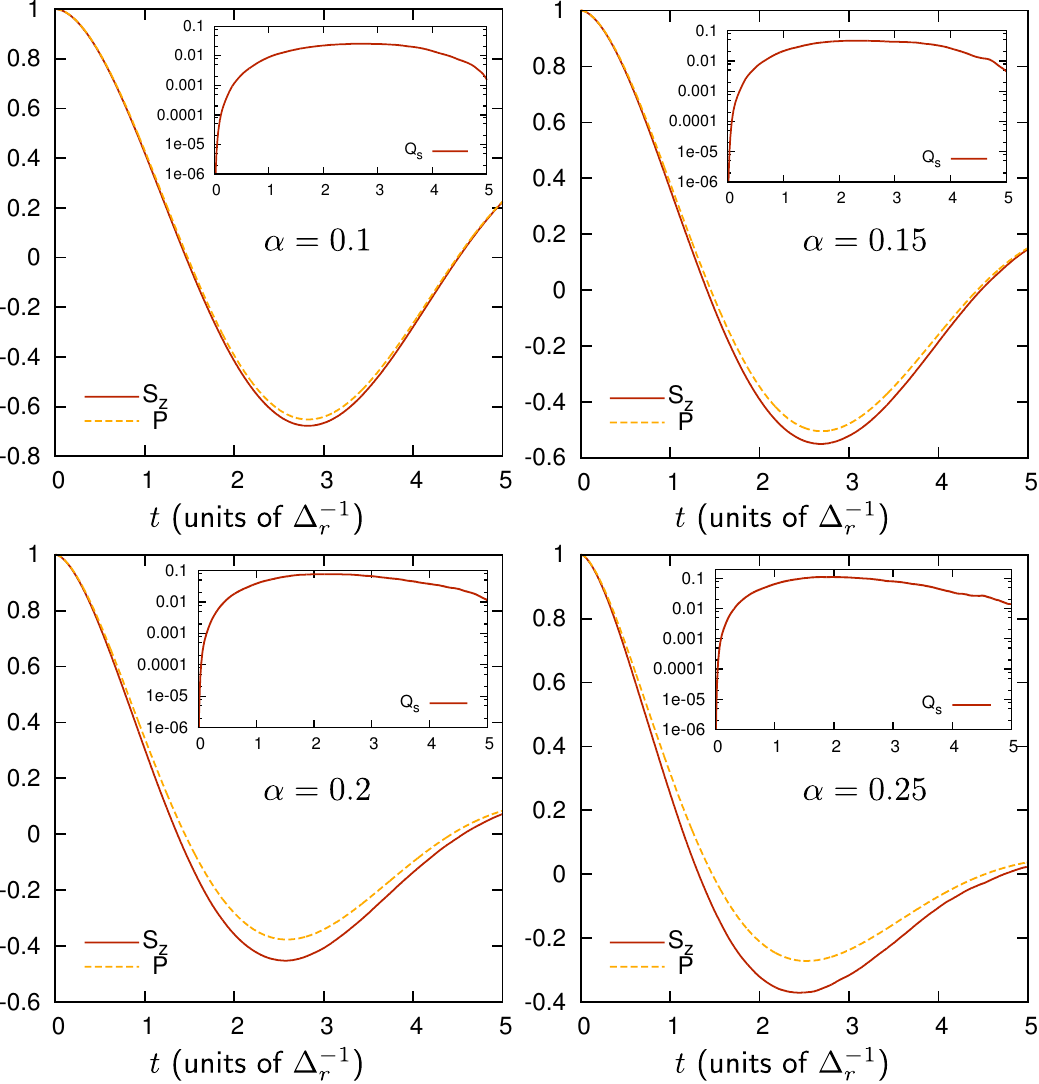}
  \caption{Symmetric autocorrelation function $S_z(t)$ and spin expectation value $P(t) \equiv \av{\sigma^z(t)}$ for different values of $\alpha = \{0.1, 0.15, 0.2, 0.25\}$. Inset shows $Q_s(t, t_0)$ defined in Eq.~\eqref{eq:82}. Other parameters read $\Delta = 1$, $\omega_c = 100 \Delta$, $\epsilon = 0$, and $t_0 = \{-30, -30, -20, -20\} \Delta^{-1}$. we observe a significant quantitative difference between $S_z(t)$ and $P(t)$ already over the first oscillation period. Within the NIBA both functions are predicted to be identical.} 
  \label{fig:3}
\end{figure}

In Fig.~\ref{fig:3}, we present results for $S_z(t)$ for different values of $\alpha$. Comparing $S_z(t)$ to $P(t)$ we find quantitative differences already within the first oscillation for $\alpha \gtrsim 0.1$. This is a result of the spin-bath correlations present in the initial state, and will become more pronounced at longer times. Since we must simulate the dynamics over a sufficiently long time interval at negative times $[t_0, 0]$ such that the system has reached equilibrium, the computation of the autocorrelation function $S_z(t)$ using SSE is limited to shorter times compared to the computation of $P(t)$. 

\begin{figure}[h!]
  \centering
  \includegraphics[width=\linewidth]{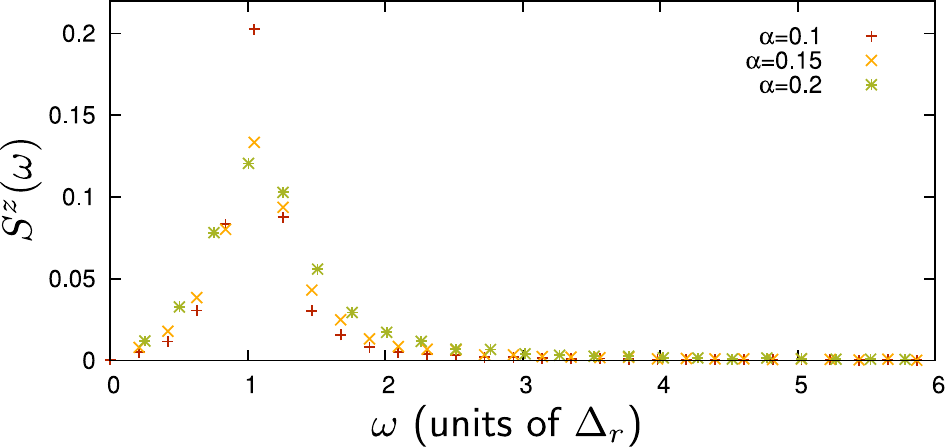}
  \caption{Fourier transform of symmetric autocorrelation function $S_z(\omega)$ for different values of $\alpha$. Other parameters are as in Fig.~\ref{fig:3}. We observe a peak close to the renormalized tunneling frequency $\Delta_r$ (independent of $\omega_c$. With increasing dissipation, the peak width increases while the peak height decreases. The low-frequency behavior is $S_z(\omega) \sim \alpha |\omega|$ as expected from the Shiba relation (Refs.~\onlinecite{PTP.54.967, sassetti_universality_1990,PhysRevLett.76.1683,PhysRevLett.76.1683,PhysRevLett.80.1038}). }
  \label{fig:10}
\end{figure}

Still, in Fig.~\ref{fig:10} we show the Fourier transform of the symmetric autocorrelation function
\begin{align}
  \label{eq:144}
  S_z(\omega) = \frac{1}{T} \int_{-T}^T dt \; S_z(t) e^{i \omega t} \,.
\end{align}
for different values of $\alpha$, where we note that $S_z(-t) = S_z(t)$. It exhibits a peak close to the renormalized tunneling frequency $\Delta_r$. The peak is universal, \emph{i.e.}, independent of the value of $\omega_c$. As expected, the peak width (height) increases (decreases) with increasing dissipation strength $\alpha$. Furthermore, in agreement with the Korringa-Shiba relation~\cite{PTP.54.967, sassetti_universality_1990,PhysRevLett.76.1683,PhysRevLett.76.1683,PhysRevLett.80.1038}, $S_z(\omega)$ shows linear behavior $S_z(\omega) \sim \alpha |\omega|$ at low frequencies. 

\section{Various applications}
\label{sec:spin-dynam-avsigm}
In this section, we discuss the dynamics of $\av{\sigma^z(t)}$ in a number of different physically relevant situations. It also serves to illustrate the capability of the SSE method. 

We first consider the dynamics of $\av{\sigma^z(t)}$ at zero bias $\epsilon =0$. We confirm the validity of SSE by comparison to the NIBA at short to intermediate times and not too strong coupling, where the NIBA is valid. We refer to Appendix~\ref{sec:non-interacting-blip} for details about the NIBA. 

We also closely investigate the long-time behavior of $\av{\sigma^z(t)}$, where the NIBA and corrections to it fail. We exploit the fact that we can compute $\av{\sigma^z(t)}$ with great numerical precision of $5 \times 10^{-4}$. Here, we find exponential decay, possibly with a power-law in the denominator, in agreement with a non-perturbative prediction from conformal field theory~\cite{PhysRevLett.80.4370} and an expansion around the Toulouse point~\cite{egger_crossover_1997,arXiv:1211.0293}.  

We then study the dynamics of $\av{\sigma^z(t)}$ for finite static bias fields $\epsilon \neq 0$. The SSE method is essentially (numerically) exact for $\alpha < 1/2$ in the Ohmic scaling regime of $\Delta/\omega_c \ll 1$ for any given bias field $\epsilon(t)$. This makes SSE particularly useful in parameter regions where no approximation scheme is known, for instance, at low temperatures $T$, small bias fields $\epsilon$ and intermediate coupling strength $\alpha \gtrsim 0.05$. In particular, we show that a correction to the NIBA for non-zero bias fails already for $\alpha \gtrsim 0.05$. In Appendix~\ref{sec:weak-coupl-extens} we include relevant predictions of this NIBA correction which we refer to as ``NIBA+Corr''.

\subsection{Zero bias dynamics of $\av{\sigma^z(t)}$}
\label{sec:zero-bias-dynamics-1}
In this section, we study the dynamics of $\av{\sigma^z(t)}$ for zero bias $\epsilon=0$. We consider both zero and finite temperature $T$. We first discuss the dynamics on short-to-intermediate timescales, where the NIBA works well. We then consider the long-time dynamics of $\av{\sigma^z(t)}$, where NIBA fails. Here, SSE predicts exponential decay, possibly with a power-law denominator, which is in agreement with non-perturbative analytical predictions~\cite{PhysRevLett.80.4370,egger_crossover_1997,arXiv:1211.0293}. The oscillation frequency is found to be in excellent agreement with predictions from conformal field theory~\cite{PhysRevLett.80.4370}.

\subsubsection{Dynamics at short-to-intermediate times }
\label{sec:dynam-at-interm}
At zero bias the NIBA is valid for short-to-intermediate times and not too strong coupling. We refer to Appendix~\ref{sec:non-interacting-blip} for details. 
\begin{figure}[t!]
  \centering
  \includegraphics[width=\linewidth]{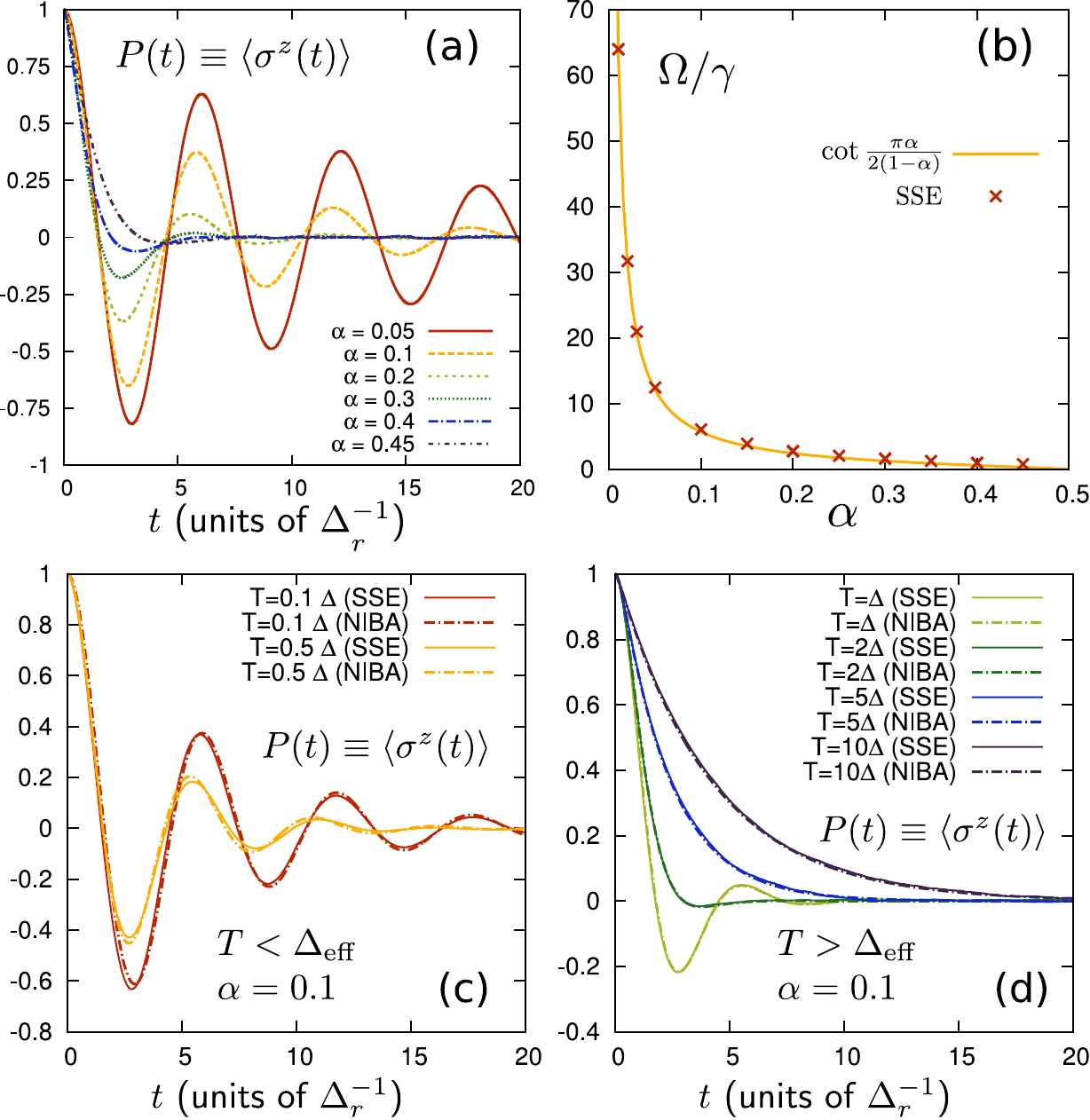}
  \caption{(a) $P(t) \equiv \av{\sigma^z(t)}$ from SSE for various values of $\alpha$, $\Delta = 1$, $\omega_c = 100$, $\epsilon = 0$, and $T =0$. For a given value of $\alpha$, curves corresponding to different values of $\omega_c/\Delta \gg 1$ scale on top of each other in units of the renormalized tunneling frequency $\Delta_r$. Numerical parameters are $m_{\text{max}} = 3000$, $ N = 5 \times 10^4$. 
(b) Quality factor $\Omega/\gamma$ of damped oscillations at $T=0$. Solid line shows prediction from CFT and NIBA.
(c-d) Finite temperature comparison of $\av{\sigma^z(t)}$ between SSE and NIBA at $\alpha = 0.1$. Other parameters read $\Delta = 1$, $\omega_c = 100 \Delta$, $\epsilon = 0$, $m_{\text{max}} = 2000$, $N=5 \times 10^4$. We find good agreement between SSE and NIBA, quantitative agreement improves for higher temperatures as expected. SSE also agrees with NIBA temperature $T^*(\alpha = 0.1) = 2.6 \Delta$, where the coherent-to-incoherent crossover occurs.}
  \label{fig:4}
\end{figure}

At $T=0$, the NIBA predicts that $\av{\sigma^z(t)}$ is a sum of a coherent part $P_{\text{coh}}$ and an incoherent part $P_{\text{inc}}$. The coherent part describes damped coherent oscillations with frequency $\Omega = \Delta_{\text{eff}} \sin \frac{\pi}{2 ( 1 - \alpha)}$ and quality factor $\Omega/\gamma = \cot \frac{\pi \alpha}{2 (1 - \alpha)}$. Surprisingly, the same result for the quality factor is obtained from a non-perturbative conformal field theory (CFT) calculation~\cite{PhysRevLett.80.4370}. Although the NIBA is a weak-coupling approximation, it yields the correct quality factor for the full range of $0 < \alpha < 1/2$. The predicted oscillation frequency, however, is slightly different from the NIBA and CFT. in Fig.~\ref{fig:4}(a), we present SSE results of $\av{\sigma^z(t)}$ for various values of $\alpha$. In Fig.~\ref{fig:4}(b), we show that the SSE quality factor precisely matches with this formula.

At finite temperature, the NIBA yields coherent behavior only below a temperature scale $T^*(\alpha)$. The coherent regime is further divided into low temperatures $T < \Delta_{\text{eff}}$ and $T > \Delta_{\text{eff}}$. Above $T^*$, the dynamics is fully incoherent. In Appendix~\ref{sec:finite-temp-dynam} we provide all relevant NIBA formulas in those parameter regions. In Figs.~\ref{fig:4}(c) and~\ref{fig:4}(d), we show that SSE agrees well with the NIBA over the full temperature range. The quantitative agreement improves for higher temperatures and for smaller values of $\alpha$ (weak-coupling). 

\subsubsection{Long-time behavior}
\label{sec:long-time-behavior}
Let us now investigate the asymptotic long-time limit of $\av{\sigma^z(t)}$, where $|\av{\sigma^z(t)}| \ll 1$. Within NIBA, the algebraically decaying incoherent part $P_{\text{inc}}(t)$ becomes larger than the exponentially decaying coherent part $P_{\text{coh}}(t)$ after a time $t \gg \Delta_{\text{eff}}^{-1}$ that depends on $\alpha$. For $\alpha = 0.3$, for instance, one finds that $|P_{\text{inc}}| > |P_{\text{coh}}|$ already after one half of an oscillation. 

Corrections of the NIBA that take further neighbor blip-blip correlations systematically into account modify the form of the algebraic power law, but all finite-order corrections to the NIBA predict the occurrence of an algebraically decaying incoherent part. The prediction of an algebraic decay of $\av{\sigma^z(t)}$ at long times is known to be an incorrect prediction of the NIBA (and its finite-order corrections)~\cite{weissdissipation}. 
 \begin{figure}[h!]
  \centering
  \includegraphics[width=\linewidth]{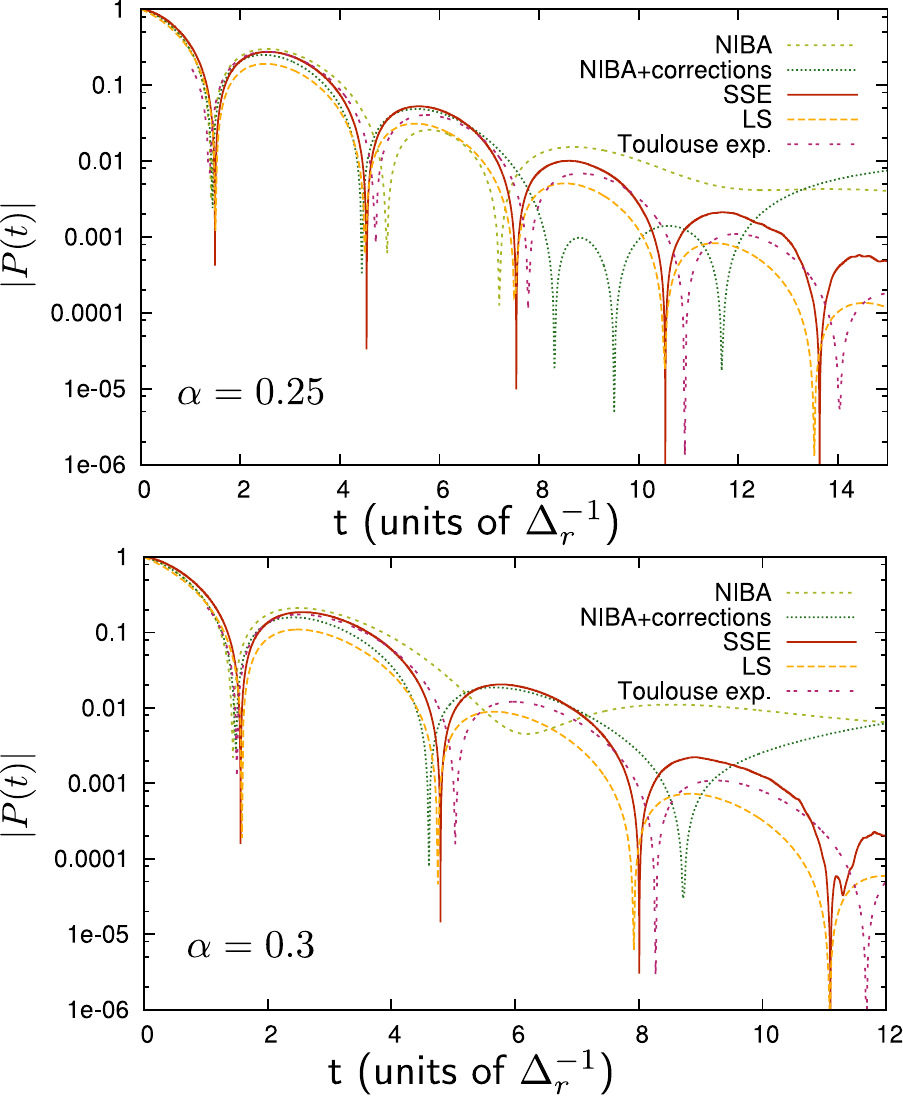}
  \caption{Long-time behavior of $|\av{\sigma^z(t)}|$ computed with different methods. Upper panel shows $\alpha = 0.25$, lower one $\alpha = 0.3$. Other parameters are $\Delta = 1$, $\omega_c = 50$, $\epsilon = 0$, $T=0$, $m_{\text{max}} = 1000$, and $N = 2\times 10^7$. ``NIBA+corrections'' refers to $P^{(1)}(\Delta_{\text{eff}} t)$ of Ref.~\onlinecite{RevModPhys.59.1}, which is a correction to the NIBA that takes nearest-neighbor blip-blip correlations into account. It improves the NIBA at shorter times, but also fails at longer times. ``LS'' refers to the CFT prediction of Lesage and Saleur~\cite{PhysRevLett.80.4370}, and ``Toulouse exp.'' result from an expansion around the Toulouse point~\cite{egger_crossover_1997}. Within numerical accuracy, our results are in agreement with exponential decay, possibly with a power-law in the denominator~\cite{egger_crossover_1997, arXiv:1211.0293}. We can certainly exclude purely algebraic contributions. SSE precisely confirms the CFT frequency scale $\Delta_{LS}$.}
  \label{fig:5}
\end{figure}

In contrast to the algebraic decay predicted by the NIBA and its corrections, the conformal field theory calculation in Ref.~\onlinecite{PhysRevLett.80.4370} predicts a purely exponential decay of $\av{\sigma^z(t)}$ at long times. The CFT oscillation frequency $\Delta_{LS}$ (and decay rate $\gamma_{LS}$) is also slightly different from the NIBA frequency $\Omega$, and reads
\begin{align}
  \label{eq:92}
  \Delta_{LS} = \sin \biggl[ \frac{\pi \alpha}{2 ( 1 - \alpha)} \biggr] a(\alpha) \Delta_{\text{eff}}
\end{align}
where
\begin{align}
  \label{eq:93}
  a(\alpha) &= \frac{\Gamma \bigl( \frac{\alpha}{2 (1-\alpha)} \bigr)}{\sqrt{\pi} \Gamma \bigl( \frac{1}{2(1-\alpha)} \bigr) } \biggl[ \frac{\Gamma(\frac12 + \alpha) \Gamma(1-\alpha)}{\sqrt{\pi}} \biggr]^{1/2(1-\alpha)}\,.
\end{align}
In addition, a systematic expansion about the exactly solvable Toulouse point $\alpha=1/2 - \kappa$ with $\kappa \ll 1$ yields an exponentially decaying $\av{\sigma^z(t)}$, since it yields an incoherent part of the form $P_{\text{inc}}(t) = -2\kappa \exp[-\Delta_{\text{eff}} t/2]/(\Delta_{\text{eff}}t)^{1 + 2 \kappa}$~\cite{egger_crossover_1997}. As shown in Ref.~\onlinecite{egger_crossover_1997}, a systematic expansion in $\kappa$ shows that interblip correlations shift the endpoint of the branch cut in $\av{\sigma^z(\lambda)}$, which is responsible for the algebraic decay of in real-time within the NIBA, from $\lambda = 0$ to the non-zero value $\lambda = - \Delta_{\text{eff}}/2$. This behavior is also found in a recent study using real-time renormalization grou (RG) and functional RG,~\cite{arXiv:1211.0293} where an analytical result for intermediate times, which is valid to $\mathcal{O}(|1-2 \alpha|)$, is reported as well.

In Fig.~\ref{fig:5}, we present SSE results of $|\av{\sigma^z(t)}|$ for $\alpha = \{0.25, 0.3\}$ in the long-time limit. We clearly observe exponential decay up to a numerical accuracy of about $5 \times 10^{-4}$, and no sign of a purely algebraic contribution. This agrees with predictions from CFT (``LS'') (Ref.~\onlinecite{PhysRevLett.80.4370}) and the expansion around the Toulouse point (``Toulouse exp.'').~\cite{egger_crossover_1997} It is worth pointing out that SSE oscillations precisely match with the CFT frequency scale in Eq.~\eqref{eq:92}. In Fig.~\ref{fig:5}, we clearly see that the erroneous algebraic term dominates the solution of the NIBA and its first-order correction (``NIBA+corrections'') already after a few oscillations. 


\subsection{Dynamics of $\av{\sigma^z(t)}$ at non-zero bias}
\label{sec:spin-dynamics-at}
In this section, we discuss the spin dynamics for non-zero bias $|\epsilon| > 0$. We focus on the case where $|\epsilon| \sim \Delta$. It is well-known that in this case the NIBA breaks down for temperatures below $\Delta_b = \sqrt{\Delta_{\text{eff}}^2 + \epsilon^2}$~\cite{weissdissipation}. One possibility to go beyond the NIBA is to consider interblip correlations up to first order in the spin-bath interaction strength $\alpha$. It is thus limited to weak spin-bath coupling. This approach was introduced in Refs.~\onlinecite{weiss_dynamics_1989,goerlich_low-temperature_1989} and we provide all relevant results in Appendix~\ref{sec:finite-temp-dynam} (see e.g. Eq.~\eqref{eq:133}). 

\begin{figure}[h!]
  \centering
  \includegraphics[width=.9\linewidth]{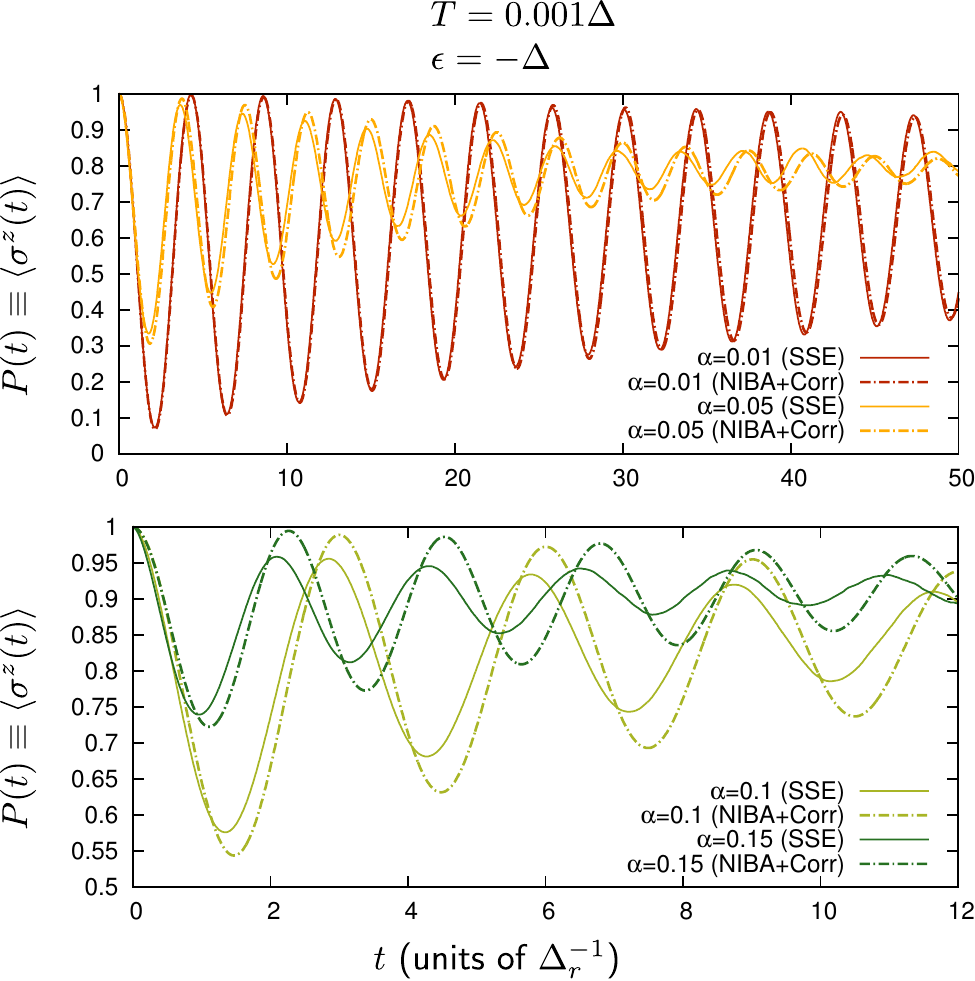}
  \caption[]{Comparison of $\av{\sigma^z(t)}$ for non-zero bias $\epsilon = - \Delta$ and low temperature $T = 10^{-3} \Delta$ between a weak-coupling extension of the NIBA (``NIBA+Corr'' dashed) and SSE (solid). Although both solutions agree for $\alpha = 0.01$, we observe significant deviations already for $\alpha = 0.05$. This is expected since the correction to the NIBA takes the interblip correlations only up to first-order in $\alpha$ into account. Other parameters are $\omega_c = 200 \Delta$, $m_{\text{max}} = 3000$ and $N = 4.5 \times 10^6$.}
  \label{fig:6}
\end{figure}

\begin{figure}[h!]
  \centering
  \includegraphics[width=.9\linewidth]{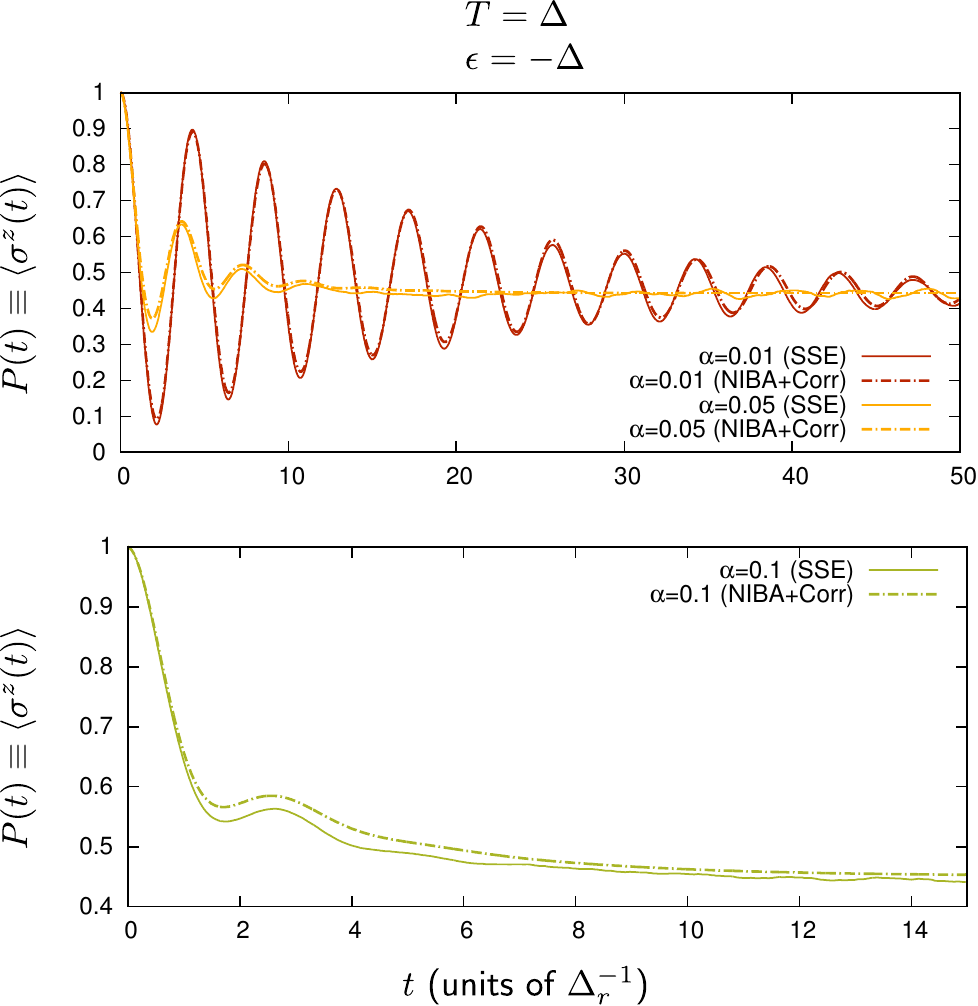}
  \caption[]{Comparison of $\av{\sigma^z(t)}$ for non-zero bias $\epsilon = - \Delta$ and temperature $T = \Delta$ between ``NIBA+Corr'' (dashed) and SSE (solid). For this larger value of temperature, the qualitative agreement between the two approaches has improved compared to Fig.~\ref{fig:6}. This is expected as the average blip length reduces with temperature~\cite{RevModPhys.59.1}. Other parameters are $\omega_c = 200 \Delta$, $m_{\text{max}} = 3000$ and $N = 4.5 \times 10^6$. }
  \label{fig:7}
\end{figure}

In Figs.~\ref{fig:6} and~\ref{fig:7}, we compare SSE to this weak-coupling extension of the NIBA (``NIBA+Corr'') for $\epsilon = - \Delta$ and two different temperatures $T = \{10^{-3} \Delta, \Delta\}$. We find that at low temperatures $T = 10^{-3} \Delta$, ``NIBA+Corr'' is limited to quite small values of $\alpha \lesssim 0.01$. Even for $\alpha= 0.05$, one observes large differences to SSE. At larger temperatures $T = \Delta$, the damping is much stronger and the qualitative agreement improves. This is similar to the zero bias case. The overall agreement between SSE and ``NIBA+Corr'' at non-zero bias is worse than the agreement between SSE and NIBA at zero bias. This makes the SSE approach a valuable tool to obtain the dynamics in the presence of non-zero bias, especially at smaller temperatures. 


\section{Summary and open questions}
\label{sec:open-quest-curr}
We want to end with a summary and a discussion of a number of open questions related to the SSE method and its application to problems beyond the spin-boson model. The spin-boson model finds abundant applications in physics from quantum computing to the study of dissipation induced quantum phase transitions. It is realized in a variety of experimental settings most notably tunable mesoscopic or cold-atom setups. 

In this article, we have exposed in detail a non-perturbative numerical method that allows us to exactly solve for the spin dynamics $\av{\sigma^\alpha(t)}$ in the Ohmic spin-boson model for $\alpha < 1/2$. The method can be applied provided the bath cutoff frequency is the largest frequency scale in the problem $\omega_c \gg \Delta$. The underlying idea of the SSE approach is very general and consists of employing Hubbard-Stratonovich identity to transform the quadratic and time non-local action of the spin into a linear and time-local action. The crucial advantage is that the functional integral over the spin path amplitudes can now be exactly calculated by solving a linear Schr\"odinger-type equation. The price to pay is the integration over the Gaussian distributed Hubbard-Stratonovich variables $\{s_m\}$. Since the Schr\"odinger equation contains the variables $\{s_m\}$, this integration corresponds to a numerical average over different Schr\"odinger equation solutions. 

The SSE method exhibits very nice convergence properties for the Ohmic model with $\omega_c \gg \Delta$ and $\alpha < 1/2$, since the random height function $h_s(t)$ is purely real in this case. As a result, each solution of the stochastic equation is bounded for zero bias. Even for non-zero bias $\epsilon \neq 0$, the individual solutions are well behaved and, for example, do not grow exponentially. The situation is different, however, for other bath spectral functions such as a sub-Ohmic bath. Here, the random height function $h_s(t)$ acquires an imaginary part, which leads to such bad convergence that the approach becomes impracticable. 

We obtain $\av{\sigma^\alpha(t)}$ as a statistical average over solutions of a time-dependent Schr\"odinger equation, that is easily solved numerically by a standard Runge-Kutta solver. Therefore, we can easily consider a time-dependent external bias field $\epsilon(t)$. Any non-pathological time-dependence can be implemented. As an example, in addition to constant external bias, we have investigated the case of a linear Landau-Zener sweep of the detuning. 

Finally, in contrast to earlier stochastic approaches, our method allows to take the initial spin-bath preparation exactly into account. In particular, a polarized bath initial state can have substantial effects on the spin dynamics, as we have shown for $\av{\sigma^x(t)}$ for example.

An interesting further direction is to apply this general idea~\cite{imambekov:063606,ImambekovGritsevDemler-FundNoiseFermiProceed2006} of using Hubbard-Stratonovich transformation to obtain a time-local linear action for the impurity degree of freedom to other impurity problems such as the Kondo model, the resonant level model, or the Holstein model.~\cite{PhysRevLett.107.256804} The method could also be generalized to the case of quantum transport through a quantum dot in the Coulomb blockade regime,~\cite{DuttLeHur-AnnPhys-2011} where quantum Monte-Carlo techniques on the real-time Keldysh contour have been implemented as well.~\cite{RevModPhys.83.349,PhysRevB.78.235110,PhysRevB.79.153302, PhysRevB.81.035108} Another possible extension of the SSE formalism is to consider a spin with $S> 1/2$. This increases the number of two-spin basis states, that are necessary and this approach is thus limited to $S \sim \mathcal{O}(1)$ in practice.


\acknowledgments
After we had obtained our results, Adilet Imambekov tragically died while mountaineering in Kazakhstan. We will always keep the memory of our dear friend as a great scientist, supportive mentor and collaborator and wonderful person in our hearts. The method presented in this article is based on his original ideas~\cite{imambek_jetp_02,ImambekovGritsevDemler-FundNoiseFermiProceed2006}. \\
The authors acknowledge useful discussions with A. Shnirman, V. Gritsev and particularly with D. Roosen and W. Hofstetter on a comparison between the stochastic method and the numerical renormalization group (unpublished). This work was supported from DOE under the Grant No. DE-FG02-08ER46541 (K.L.H.), from the NSF through the Yale Center for Quantum Information Physics (P.P.O. and K.L.H.), and from Ecole Polytechnique (K.L.H.). K.L.H. acknowledges KITP for hospitality and support from Grant No. NSF PHY11-25915. The Young Investigator Group of P.P.O. received financial support from the ``Concept for the Future'' of the Karlsruhe Institute of Technology within the framework of the German Excellence Initiative. 

\appendix

\section{Non-Interacting Blip Approximation (NIBA)}
\label{sec:non-interacting-blip}
In this section, we derive and discuss the well-known Non-Interacting Blip Approximation~\cite{RevModPhys.59.1, weissdissipation}. It is essentially a short-time and weak-coupling approximation. It becomes exact in the Markovian limit of an Ohmic bath at high temperatures. Neglecting blip-blip interactions simplifies the functional integral expression in such a way that it can be solved analytically via Laplace transformation. Even if the inverse transformation to real-time cannot be performed exactly, much can be learned from an investigation of the analytic structure (branch points, branch cuts) in Laplace space. 

The NIBA has many shortcomings. Since it is a short-time approximation, it always fails at long times (see Sec.~\ref{sec:relation-sse-method}). In addition, the NIBA can only be used for the spin component $\av{\sigma^z(t)}$. It cannot be used for calculating the coherence $\av{\sigma^x(t)}$ and the spin autocorrelation function $S_z(t)$ (except at large temperatures), where it fails even at weak spin-bath coupling. For $\av{\sigma^z(t)}$ it gives incorrect results at low temperatures and non-zero bias, except at very large bias, where is can be justified again.  

In all cases where the NIBA fails blip-blip interactions are important. A weak-coupling extension to the NIBA that takes blip-blip interactions to first order into account  in $\alpha$ is derived in Refs.~\onlinecite{weissdissipation, weiss_dynamics_1989,goerlich_low-temperature_1989} and is discussed in Appendix~\ref{sec:weak-coupl-extens}.
 
\subsection{Derivation from functional integral expression}
\label{sec:deriv-from-funct}
The starting point is the exact expression for the influence functional $F_n[\{\Xi_j\}, \{\Upsilon_j\}, \{t_j\}] = \mathcal{Q}_1 \mathcal{Q}_2$ with $\mathcal{Q}_{1,2}$ given in Eqs.~\eqref{eq:11} and~\eqref{eq:12}
\begin{align}
  \label{eq:94}
  {\cal Q}_1 &= \exp \biggl[ \frac{i}{\pi} \sum_{j > k \geq 0}^{2n} \Xi_j \Upsilon_k Q_1 (t_j - t_k) \biggr] \\
  \label{eq:95}
  {\cal Q}_2 &= \exp \biggl[ \frac{1}{\pi} \sum_{j> k \geq 1}^{2n} \Xi_j \Xi_k Q_2(t_j - t_k) \biggr]\,.
\end{align}
 The ${\cal Q}_1$-part greatly simplifies in the scaling limit $\Delta/\omega_c \rightarrow 0$ for $\alpha < 1/2$, since one may use $Q_1(t) = \pi^2 \alpha \theta(t)$. Summing over the sojourn variables $\{\eta_1, \ldots, \eta_{n-1} \}$ results in Eq.~\eqref{eq:30}
\begin{align}
  \label{eq:96}
  {\cal Q}_1 &= \exp \Bigl[ i \pi \alpha \sum_{k=0}^{n-1} \xi_{k+1} \eta_k \Bigr] = \bigl[ 2 \cos(\pi \alpha) \bigr]^{n-1} e^{i \pi \alpha \xi_1} \,.
\end{align}
For zero bias $\epsilon = 0$ and if we are interested in calculating $\av{\sigma^z(t)}$, where the system ends in a sojourn state $\eta_n$, the real-time functional integral expression for $p(t)$ in Eq.~\eqref{eq:18} is invariant under the simultaneous reversal of the sign of all blip variables $\{\xi_1, \ldots, \xi_{n} \} \rightarrow \{- \xi_1, \ldots,  - \xi_n\}$. Therefore, only the symmetric part $\cos( \pi \alpha)$ of the exponential $e^{i \pi \alpha \xi_1}$ contributes to $\av{\sigma^z(t)}$, and we find ${\cal Q}_1 = 2^{n-1} \bigl[ \cos(\pi \alpha) \bigr]^{n}$. 

The ${\cal Q}_2$-part of the influence functional in Eq.~\eqref{eq:75} contains the interactions between all blips. The NIBA consists of neglecting all blip-blip interactions apart from the blip self-interactions. This relies on the assumption that the average time that the system spends in a sojourn state is much longer than the average time it spends in a blip state~\citep{RevModPhys.59.1}. The expression of ${\cal Q}_2$ then becomes
\begin{align}
  \label{eq:97}
  {\cal Q}_2^{\text{NIBA}} = \exp \biggl[ - \frac{1}{\pi} \sum_{j=1}^{n} Q_2(t_{2j} - t_{2j-1} ) \biggr] \,.
\end{align}
A a result, the influence functional does not depend on the blip and sojourn variables $\{\xi_j, \eta_j \}$ anymore
\begin{align}
  \label{eq:98}
  F_n\bigl[\{t_j\}\bigr] &= {\cal Q}_1 {\cal Q}_2^{\text{NIBA}} \nonumber \\ & = 2^{n-1} \bigl[ \cos( \pi \alpha) \bigr]^{n} \prod_{j=1}^n \exp \biggl[ - \frac{1}{\pi} Q_2(t_{2j} - t_{2j-1} ) \biggr]\,.
\end{align}
The spin dynamics $\av{\sigma^z(t)}  \equiv P(t) = 2 p(t) - 1$ follows with $H_n = 1$ for zero bias from Eq.~\eqref{eq:18} as
\begin{align}
  \label{eq:99}
  \av{\sigma^z(t)} &= \sum_{n=0}^{\infty} \Bigl[- \Delta^2 \cos ( \pi \alpha) \Bigr]^n \int_{0}^t dt_{2n} \cdots \int_0^{t_2} dt_1 \nonumber \\ & \quad \times \prod_{j=1}^n \exp \biggl[ - \frac{1}{\pi} Q_2(t_{2j} - t_{2j-1} ) \biggr]\,,
\end{align}
where we have used that $\sum_{\{\xi_j\}} = 2^n$. One can solve Eq.~\eqref{eq:99} by Laplace transformation~\cite{RevModPhys.59.1}
\begin{align}
  \label{eq:100}
  \av{\sigma^z(\lambda)} &= \int_0^\infty dt \, e^{- \lambda t} \av{\sigma^z(t)}\,.
\end{align}
If we define the function 
\begin{align}
  \label{eq:101}
  f(t) = \Delta^2 \cos (\pi \alpha) \exp \biggl[ - \frac{1}{\pi} Q_2(t) \biggr]\,,
\end{align}
one finds after rearranging the order of integration
\begin{align}
  \label{eq:102}
  \av{\sigma^z(\lambda)} &= \sum_{n=0}^\infty (-1)^n \int_0^\infty dt \int_0^\infty dt_1 \cdots \int_0^\infty dt_{2n} \nonumber \\ & 
\times e^{- \lambda (t + t_1 + t_2 + \ldots + t_{2n})} \prod_{j=1}^n f(t_{2j}) \nonumber \\ & = \sum_{n=0}^\infty (-1)^n \frac{[f(\lambda)]^n}{\lambda^{n+1}} = \frac{1}{\lambda + f(\lambda)}\,,
\end{align}
where $f(\lambda)$ is the Laplace transform of $f(t)$. The solution in the time domain is obtained from an inverse Laplace transformation via the standard integral along the Bromwich contour $C$~\citep{GoldbartStone-MathMethodsBook}
\begin{align}
  \label{eq:103}
  \av{\sigma^z(t)} &= \frac{1}{ 2 \pi i} \int_C d\lambda \, e^{\lambda t} \av{\sigma^z(\lambda)} \,.
\end{align}
Even if the inverse transformation cannot be performed explicitly, much can be inferred from a study of the analytical properties of $\av{\sigma^z(\lambda)}$, \emph{i.e.},~its singularities, branch cuts and residua. 

\subsection{Derivation using Heisenberg equations of motion}
\label{sec:deriv-using-heis}
In this section, we present an alternative and physically more transparent derivation of the NIBA, which was derived in Ref.~\onlinecite{PhysRevA.35.1436}. It starts from the polaron transformed spin-boson Hamiltonian
\begin{align}
  \label{eq:104}
  \tilde{H} = U^\dag H U = \frac{\Delta}{2} \bigl( \sigma^+ e^{i \Omega}  + \text{h.c.} \bigr) + \sum_k \omega_k b^\dag_k b_k\,,
\end{align}
where $H$ is defined in Eq.~\eqref{eq:1} and the unitary transformation reads $U = \exp ( - \frac12 \sigma^z \Omega)$ with $\Omega = - i \sum_k \frac{\lambda_k}{\omega_k} (b^\dag_k - b_k)$. The Heisenberg equation of motion for $\sigma^z(t)$ then reads
\begin{align}
  \label{eq:105}
  \dot{\sigma}^z(t) = - i \Delta \sigma^+(t) e^{i \Omega(t)} + \text{h.c.}\,.
\end{align}
It contains $\sigma^{\pm}(t)$ which is calculated to
\begin{align}
  \label{eq:106}
  \sigma^{+}_j(t) = - \frac{i \Delta_j}{2} \int_{0}^t ds \, \sigma^z_j(s) e^{- i \Omega(s)}\,,
\end{align}
and $\sigma^-_j = (\sigma^+_j)^*$. Inserting Eq.~\eqref{eq:106} into Eq.~\eqref{eq:105} yields 
\begin{align}
  \label{eq:107}
  \dot{\sigma}^z(t) = - \frac{\Delta^2}{2} \int_{0}^t ds [\sigma^z(s) e^{i \Omega(t)} e^{- i \Omega(s)} + \text{h.c.} ]\,.
\end{align}
We now employ two approximations to recover the NIBA. First, we assume that the time evolution of the bath operators is governed by the free bath Hamiltonian $H_B = \sum_k \omega_k b^\dag_k b_k$. The reduced density matrix of the bath remains unperturbed by the spins. Second, we trace out the bath degrees of freedom in a weak-coupling sense by writing 
\begin{align}
  \label{eq:108}
  \text{Tr}_B [e^{i \Omega(t)} e^{- i \Omega(s) }] = \exp\Bigl\{ \frac{1}{\pi} \bigl[ i Q_1(t-s) - Q_2(t-s) \bigr] \Bigr\}\,,
\end{align}
which includes the bath correlation functions defined in Eqs.~\eqref{eq:13} and~\eqref{eq:14}. The equation of motion for the spin, averaged over the bath, thus becomes 
\begin{align}
  \label{eq:109}
  \av{\dot{\sigma}^z_j(t)} = - \Delta_j^2 \int_{0}^t ds \biggl\{ \av{\sigma_j^z(s)} \cos\biggl[ \frac{Q_1(t-s)}{\pi} \biggr] e^{- Q_2(t-s)/\pi} \biggl\}\,.
\end{align}
Using the definition of $f(t)$ in Eq.~\eqref{eq:101}, this can be written as
\begin{align}
  \label{eq:110}
  \av{\dot{\sigma}^z_j(t)} + \int_{0}^t ds f(t-s) \av{\sigma_j^z(s)} = 0 \,.
\end{align}
If we apply a Laplace transformation, we thus recover the result for $\av{\sigma^z(\lambda)}$ within the NIBA, that we have derived in the previous Sec.~\ref{sec:deriv-from-funct} in Eq.~\eqref{eq:102}
\begin{align}
  \label{eq:111}
  \av{\sigma^z(\lambda)} &= \frac{1}{\lambda + f(\lambda)} \,.
\end{align}

\subsection{Zero temperature dynamics}
\label{sec:zero-temp-dynam}
In this section, we discuss the predictions of the NIBA at zero temperature. At $T=0$, the Ohmic bath correlation function reads $Q_2(t) = \pi \alpha \ln \bigl[ 1 + \omega_c^2 t^2 \bigr]$. The Laplace transform of $f(t)$ in Eq.~\eqref{eq:101} is calculated to  
\begin{align}
  \label{eq:112}
  f(\lambda) = \Delta_{\text{eff}} (\Delta_{\text{eff}}/\lambda)^{1-2\alpha}\,.
\end{align}
It contains the effective tunneling element
\begin{align}
  \label{eq:113}
  \Delta_{\text{eff}} &= \Bigl[ \Gamma(1 - 2 \alpha) \cos ( \pi \alpha) \Bigr]^{1/ 2 (1-\alpha)} \Delta_r\,,
\end{align}
where the renormalized tunneling element is defined as
\begin{align}
  \label{eq:114}
  \Delta_r = \Delta (\Delta/\omega_c)^{\alpha/(1-\alpha)}\,.
\end{align}
Both $\Delta_{\text{eff}}$ and $\Delta_r$ are smaller than the bare value $\Delta$, because spin transitions are suppressed in the presence of a polaronic cloud of bath modes. From Eq.~\eqref{eq:102}, we find that the function $\av{\sigma^z(\lambda)}$ has a complex conjugate pair of simple poles at 
\begin{align}
  \label{eq:115}
  \lambda_{1,2} = - \gamma \pm i \Omega = \Delta_{\text{eff}} \exp \biggl[ \pm i \frac{\pi}{2 ( 1- \alpha)} \biggr]\,,
\end{align}
at which point $\lambda_{1,2} + f(\lambda_{1,2}) = 0$. It also has a branch cut along the negative real axis, which ends at the branch point $\lambda = 0$. We note that the branch cut is absent for $\alpha=0$ and $\alpha = 1/2$. In the time-domain, the complete solution within the NIBA thus reads
\begin{align}
  \label{eq:116}
  \av{\sigma^z(t)} \equiv P(t) = P_{\text{coh}}(t) + P_{\text{inc}}(t)\,.
\end{align}
The poles give rise to damped coherent oscillations of the form 
\begin{align}
  \label{eq:117}
  P_{\text{coh}}(t) = \frac{1}{1 - \alpha} e^{- \gamma t} \cos \Omega t
\end{align}
with frequency $\Omega = \Delta_{\text{eff}} \sin \frac{\pi}{2 (1-\alpha)}$ and decay rate $\gamma = \Delta_{\text{eff}} \cos \frac{\pi}{2 (1-\alpha)}$. The quality factor of the oscillations is thus independent of $\Delta_{\text{eff}}$ and reads
\begin{align}
  \label{eq:118}
  \frac{\Omega}{\gamma} = \cot \frac{\pi \alpha}{2 ( 1 - \alpha)}\,.
\end{align}
The branch cut, on the other hand, yields a (negative) incoherent contribution~\cite{RevModPhys.59.1} 
\begin{align}
  \label{eq:119}
  P_{\text{inc}}(t) &= - \frac{\sin 2 \pi \alpha}{\pi} \int_0^\infty dz \frac{z^{2 \alpha - 1} e^{- z \Delta_{\text{eff}} t}}{ z^2 + 2 z^{2 \alpha} \cos 2 \pi \alpha + z^{4 \alpha - 2}} \,.
\end{align}
The incoherent part dominates the dynamics at long times $\Delta_{\text{eff}} t \gg 1$, where it behaves like
\begin{align}
  \label{eq:120}
  P_{\text{inc}}(t) &\sim \frac{1}{(\Delta_{\text{eff}} t)^{2 - 2 \alpha}} \,.
\end{align}
This is known to be an incorrect prediction of the NIBA~\cite{weissdissipation}. Nevertheless, for short to intermediate times, the NIBA makes two correct predictions. First that the dynamics is universal, \ie $P(t)$ is a function of the dimensionless scaling variable $y = \Delta_{\text{eff}} t$ only. Results of $P(t)$ for different values of $\omega_c$ collapse on top of each other, if they are plotted as a function of $y = \Delta_{\text{eff}} t$.  Second, the quality factor of the oscillations in Eq.~\eqref{eq:118} exactly agrees with results from conformal field theory in the full range of $0 < \alpha < 1/2$~\cite{PhysRevLett.80.4370}.

\subsection{Finite temperature dynamics}
\label{sec:finite-temp-dynam}
In this section, we discuss the predictions of the NIBA at finite temperature. We include this section for completeness, it mostly follows Ref.~\onlinecite{weissdissipation}. In general, the quality of the NIBA improves with increasing temperature, since the average blip length decreases for larger temperatures. This follows directly from inserting the finite temperature bath correlation function
\begin{align}
  \label{eq:121}
    Q_2(t) &= \pi \alpha \ln (1 + \omega_c^2 t^2) + 2 \pi \alpha \ln \Bigl(\frac{\beta}{\pi t} \sinh \frac{\pi t}{\beta} \Bigr)
\end{align}
into $\mathcal{Q}_2^{\text{NIBA}}$ in Eq.~\eqref{eq:97}. 

The Laplace transform of $f(t)$ at $T>0$ is calculated to~\cite{weissdissipation,RevModPhys.59.1}
\begin{align}
  \label{eq:122}
  f(\lambda) &= \Delta_{\text{eff}} \Bigl( \frac{\beta \Delta_{\text{eff}}}{2 \pi} \Bigr)^{1-2\alpha} \frac{h(\lambda)}{\alpha + (\beta \lambda)/(2 \pi)}
\end{align}
with 
\begin{align}
  \label{eq:123}
  h(\lambda) &= \frac{\Gamma(1 + \alpha + \beta \lambda/2 \pi)}{\Gamma(1 - \alpha + \beta \lambda/2 \pi)}\,.
\end{align}
The branch point of $f(\lambda)|_{T=0}$ at $\lambda=0$ with the corresponding branch cut along the negative real axis, that occurred at $T=0$, turns into an infinite number of simple poles at finite temperatures. The poles $\lambda_n$ are located on the negative real axis in intervals $- \frac{2 \pi}{\beta} (n + \alpha) < \lambda_n < \frac{2 \pi}{\beta} (-n+\alpha)$ with $n=1,2,\ldots$. The spacing of the poles grows linearly with temperature. One can show~\cite{RevModPhys.59.1,weissdissipation}, that the resulting contribution $P_{\text{inc}}(t) = \sum_n A_n \exp( \lambda_n t)$ is still negative, like the contribution of the branch cut at $T=0$, and that this part may be neglected for weak dissipation. 

The dynamics is dominated by the behavior of the two simple poles of $\av{\sigma^z(\lambda)}$ as a function of temperature. The poles $\lambda_{1,2}(T) = - \gamma(T) \pm i \Omega(T)$ move toward the negative real axis for increasing temperature. At a temperature of $T=T^*(\alpha)$ they hit the real axis, where
\begin{align}
  \label{eq:124}
  T^* &= \frac{\Delta_{\text{eff}}}{2 \pi} \biggl\{ \frac{\Gamma(\alpha)}{ \alpha \Gamma(1-\alpha)}  \nonumber \\ & \qquad \times \Bigl[ 1+ \pi \alpha \cot(\pi \alpha) + 2 \sqrt{ W(\alpha)}  \Bigr] \biggr\}^{1/2(1-\alpha)}  \,.
\end{align}
For weak dissipation $\alpha \ll 1$, one finds $T^* \approx \frac{\Delta_{r}}{\pi \alpha}$. We have used the definitions
\begin{align}
  \label{eq:125}
  W(\alpha) &= \pi \alpha \cot(\pi \alpha) - \alpha^2 g_2(\alpha) \\
\label{eq:126}
  g_2(\alpha) &= \frac12 \bigl[ \psi'(1-\alpha) - \psi'(1 + \alpha) - g_1^2 \bigr] \\
\label{eq:127}
  g_1(\alpha) &= \alpha^{-1} - \pi \cot (\pi \alpha) \,.
\end{align}
Here, $\psi'(z)$ is the first derivative of the digamma function $\psi(z)$~\cite{ArfkenWeber-MathMethodsBook}. The temperature $T^*$ separates the regime, where $P(t)$ exhibits coherent oscillations $(T < T^*)$ from the regime, where it exhibits incoherent decay $(T > T^*)$. 

Specifically, as long as $T \lesssim \Delta_{\text{eff}}$, one finds that $\lambda_{1,2}(T) = - \gamma(T) \pm i \Omega(T)$ with~\cite{weissdissipation}
\begin{align}
  \label{eq:128}
  \Omega(T) &= \Delta_{\text{eff}} \Bigl\{ 1 + \alpha \bigl[ \text{Re} \,\psi(i \beta \Delta_{\text{eff}}/2 \pi) - \ln (\beta \Delta_{\text{eff}}/2 \pi) \bigr] \Bigr\} \\
  \label{eq:129}
    \gamma(T) &= \frac{\pi}{2} \alpha \Delta_{\text{eff}} \coth ( \beta \Delta_{\text{eff}}/2 )\,.
\end{align}

For larger temperatures $T \gtrsim \Delta_{\text{eff}}$, but potentially still $T < T^*$, one can expand $h(\lambda)$ in Eq.~\eqref{eq:122} up to second order in $\lambda$. From $\lambda + f(\lambda) = 0$, we find the poles $\lambda_{1,2}(T)$ by solving~\cite{weissdissipation}
\begin{align}
  \label{eq:130}
  (1 - g_2 u) x^2 + ( \alpha + g_1 u) x + u = 0\,.
\end{align}
Here, we have introduced $x = \beta \lambda/2 \pi$, $u = (\beta \Delta_{\text{eff}}/2 \pi)^{2-2\alpha} h(0)$. The two poles are complex conjugates $\lambda_{1,2} = - \gamma \pm i \Omega$ for $T < T^*$. On the other hand, for $T > T^*$ they are real and negative $\lambda_{1,2} = - \gamma_{1,2}$ and lie in the interval $(-2 \pi T \alpha, 0)$. As temperature is increased, both poles move in opposite directions. While one of them converges for large temperatures toward $\lambda_1(T \gg \Delta_{\text{eff}}) \rightarrow 0$, the other one converges toward $\lambda_2(T \gg \Delta_{\text{eff}}) \rightarrow - 2 \pi T \alpha$.

The real-time dynamics of the spin is again obtained from Laplace inversion. In the coherent regime $\Delta_{\text{eff}} \lesssim T < T^*$, we find that $\av{\sigma^z(t)}$ exhibits damped oscillations with $\{\Omega, \gamma\}$ defined by the location of the pole, and $\phi = -|\tan^{-1}(\gamma/\Omega)|$ being an initial phase shift. In the incoherent regime $T > T^*$, we obtain the dynamics
\begin{align}
  \label{eq:131}
  \av{\sigma^z(t)} &= \frac{\gamma_1}{\gamma_1 - \gamma_2} e^{- \gamma_2 t} - \frac{\gamma_2}{\gamma_1 - \gamma_2} e^{- \gamma_1 t}\,,
\end{align}
where $\gamma_1 > \gamma_2$. Fairly above $T^*$, the prefactor of the larger decay rate term, which reads $\gamma_2/(\gamma_1 - \gamma_2)$ becomes very small. The decay is thus dominated by the smaller decay rate $\gamma_2$, which shows an asymptotic temperature dependence of~\cite{weissdissipation}
\begin{align}
  \label{eq:132}
  \gamma_2(T) &= \frac{\sqrt{\pi} \Gamma(\alpha)}{2 \Gamma(\alpha + 1/2)} \frac{\Delta_r^2}{T} \Bigl( \frac{\pi T}{\Delta_r} \Bigr)^{2 \alpha} \sim T^{2 \alpha -1}\,.
\end{align}
In the main text, we compare SSE to these NIBA predictions and find good agreement at short to intermediate times. As expected, the quantitative agreement enhances for increasing temperatures. The NIBA fails at longer times, non-zero bias fields and off-diagonal elements of the reduced density matrix, \ie $\av{\sigma^x(t)}$. 

\section{Weak-coupling extension to the NIBA (NIBA+Corr)}
\label{sec:weak-coupl-extens}
The NIBA breaks down for finite bias fields $\epsilon \neq 0$ at temperatures below $\Delta_b = \sqrt{\Delta_{\text{eff}}^2 + \epsilon^2}$~\cite{weissdissipation}. Even for zero bias the NIBA cannot be used for calculating the coherence $\av{\sigma^x(t)}$. First-order blip-blip interactions are crucial in those situations. 

In this section, we state for completeness results of an approach that goes beyond the NIBA which was introduced in Refs.~\onlinecite{weiss_dynamics_1989,goerlich_low-temperature_1989}. This weak-coupling extension of the NIBA (``NIBA+Corr'') considers the interblip correlations up to first order in the spin-bath interaction strength $\alpha$. It is therefore limited to weak spin-bath coupling, \ie small values of $\alpha \ll 1$. 

The first-order interblip contribution to the influence functional can be exactly calculated and yields 
\begin{align}
  \label{eq:133}
  \av{\sigma^z(t)} &= \av{\sigma^z}_\infty + \biggl[ \frac{\epsilon^2}{\Delta_b^2} - \av{\sigma^z}_\infty \biggr] e^{- \gamma_r t} + \biggl\{ \frac{\Delta_{\text{eff}}^2}{\Delta_b^2} \cos \Omega t \nonumber \\ & \quad + \biggl[ \frac{\gamma_r \epsilon^2 + \gamma \Delta_{\text{eff}}^2}{\Omega^3} - \frac{\gamma_r P_\infty}{\Omega} \biggr] \sin \Omega t \biggr\} e^{-\gamma t}  \,,
\end{align}
with non-zero long time value
\begin{align}
  \label{eq:134}
  \av{\sigma^z}_\infty &= \frac{\epsilon}{\Delta_b} \tanh \frac{\Delta_b}{2 T} \,,
\end{align}
and
\begin{align}
  \label{eq:135}
\Delta_b &= \sqrt{\Delta_{\text{eff}}^2 + \epsilon^2} \\
\label{eq:136}
  \Omega &= \sqrt{ \Delta_b^2 + 2 \alpha \Delta_{\text{eff}}^2 \bigl[ \text{Re}\, \psi(i \Delta_b/2 \pi T) - \ln (\Delta_b/2 \pi T) \bigr] } \\
\label{eq:137}
\gamma &= \frac{\gamma_r}{2} + \frac{2 \pi \alpha \epsilon^2 T}{\Delta_b^2} \\
  \label{eq:138}
\gamma_r &= \pi \alpha \frac{\Delta^2_{\text{eff}}}{\Delta_b} \coth \frac{\Delta_b}{2 T} \,.
 \end{align}
Here, $\psi(z)$ is the digamma function. For the coherence $\av{\sigma^x(t)}$, one finds 
\begin{align}
  \label{eq:139}
     \av{\sigma^x(t)} &=  \Bigl[ - \frac{\epsilon \Delta_{\text{eff}}}{\Delta \Omega^2} \cos \Omega t + b_2 \sin \Omega t \Bigr] e^{- \gamma t}  \nonumber \\ & \quad + \Bigl[ \frac{\epsilon \Delta_{\text{eff}}}{\Delta \Omega^2} - \av{\sigma^x}_{\infty,wc} \Bigr] e^{- \gamma_r t}  + \av{\sigma^x}_{\infty,wc} \,,
\end{align}
with long-time value
\begin{align}
  \label{eq:140}
  \av{\sigma^x}_{\infty,wc} &= \frac{\Delta_{\text{eff}}^2}{\Delta \Omega} \tanh \frac{\Omega}{2 T}\,,
\end{align}
and
\begin{align}
  \label{eq:141}
  b_2 &= \frac{\Delta_{\text{eff}}^2}{\Delta \Omega} \Bigl[ \pi \alpha + \epsilon \frac{\gamma_r - \gamma}{\Omega^2} \Bigr] - \frac{\gamma_r \av{\sigma^x}_{\infty,wc}}{\Omega}
\end{align}


\section{Rigorous Born approximation results of Loss and DiVincenzo}
\label{sec:exact-born-appr}
Since we also compare the long-time limit of our results for the coherence $\av{\sigma^x(t)}$ with a formula derived in Ref.~\onlinecite{PhysRevB.71.035318} by Loss and DiVincenzo using an approximation scheme that does not apply any other approximation than the Born approximation. It is thus ``exact'' to first order in $\alpha$. They find the steady-state value of the coherence
\begin{align}
  \label{eq:142}
  \av{\sigma^x}_{\infty, \text{LDV}} = \frac{\Delta}{E} - \alpha \biggl[ - \frac{\Delta^3}{E^3} + \Bigl(C - \ln \frac{\omega_c}{E} \Bigr) \Bigl( \frac{\Delta^3}{E^3} - \frac{2 \Delta}{E} \Bigr) \biggr]\,,
\end{align}
with $E = \sqrt{\Delta^2 + \epsilon^2}$ and Euler-Mascheroni number $C$, which agrees perfectly with the Bethe ansatz prediction and our numerical SSE result for $\alpha \lesssim 0.1$. 


\begin{thebibliography}{163}%
\makeatletter
\providecommand \@ifxundefined [1]{%
 \@ifx{#1\undefined}
}%
\providecommand \@ifnum [1]{%
 \ifnum #1\expandafter \@firstoftwo
 \else \expandafter \@secondoftwo
 \fi
}%
\providecommand \@ifx [1]{%
 \ifx #1\expandafter \@firstoftwo
 \else \expandafter \@secondoftwo
 \fi
}%
\providecommand \natexlab [1]{#1}%
\providecommand \enquote  [1]{``#1''}%
\providecommand \bibnamefont  [1]{#1}%
\providecommand \bibfnamefont [1]{#1}%
\providecommand \citenamefont [1]{#1}%
\providecommand \href@noop [0]{\@secondoftwo}%
\providecommand \href [0]{\begingroup \@sanitize@url \@href}%
\providecommand \@href[1]{\@@startlink{#1}\@@href}%
\providecommand \@@href[1]{\endgroup#1\@@endlink}%
\providecommand \@sanitize@url [0]{\catcode `\\12\catcode `\$12\catcode
  `\&12\catcode `\#12\catcode `\^12\catcode `\_12\catcode `\%12\relax}%
\providecommand \@@startlink[1]{}%
\providecommand \@@endlink[0]{}%
\providecommand \url  [0]{\begingroup\@sanitize@url \@url }%
\providecommand \@url [1]{\endgroup\@href {#1}{\urlprefix }}%
\providecommand \urlprefix  [0]{URL }%
\providecommand \Eprint [0]{\href }%
\@ifxundefined \urlstyle {%
  \providecommand \doi  [0]{\begingroup \@sanitize@url \@doi}%
  \providecommand \@doi [1]{\endgroup \@@startlink {\doibase
  #1}doi:\discretionary {}{}{}#1\@@endlink }%
}{%
  \providecommand \doi  [0]{doi:\discretionary{}{}{}\begingroup
  \urlstyle{rm}\Url }%
}%
\providecommand \doibase [0]{http://dx.doi.org/}%
\providecommand \Doi [0]{\begingroup \@sanitize@url \@Doi }%
\providecommand \@Doi  [1]{\endgroup\@@startlink{\doibase#1}\@@Doi}%
\providecommand \@@Doi [1]{#1\@@endlink}%
\providecommand \selectlanguage [0]{\@gobble}%
\providecommand \bibinfo  [0]{\@secondoftwo}%
\providecommand \bibfield  [0]{\@secondoftwo}%
\providecommand \translation [1]{[#1]}%
\providecommand \BibitemOpen [0]{}%
\providecommand \bibitemStop [0]{}%
\providecommand \bibitemNoStop [0]{.\EOS\space}%
\providecommand \EOS [0]{\spacefactor3000\relax}%
\providecommand \BibitemShut  [1]{\csname bibitem#1\endcsname}%
\bibitem [{\citenamefont {{D}ittrich}\ \emph {et~al.}(1998)\citenamefont
  {{D}ittrich}, \citenamefont {{H}\"anggi}, \citenamefont {{I}ngold},
  \citenamefont {{K}ramer}, \citenamefont {{S}ch\"on},\ and\ \citenamefont
  {{Z}werger}}]{QuantumTransportAndDissipation-Dittrich-Book}%
  \BibitemOpen
  \bibfield  {author} {\bibinfo {author} {\bibfnamefont {T.}~\bibnamefont
  {{D}ittrich}}, \bibinfo {author} {\bibfnamefont {P.}~\bibnamefont
  {{H}\"anggi}}, \bibinfo {author} {\bibfnamefont {G.-L.}\ \bibnamefont
  {{I}ngold}}, \bibinfo {author} {\bibfnamefont {B.}~\bibnamefont {{K}ramer}},
  \bibinfo {author} {\bibfnamefont {G.}~\bibnamefont {{S}ch\"on}}, \ and\
  \bibinfo {author} {\bibfnamefont {W.}~\bibnamefont {{Z}werger}},\ }\href@noop
  {} {\emph {\bibinfo {title} {{Q}uantum transport and dissipation}}}\
  (\bibinfo  {publisher} {Wiley-VCH},\ \bibinfo {year} {1998})\BibitemShut
  {NoStop}%
\bibitem [{\citenamefont {{V}an
  {K}ampen}(2007)}]{VanKampen-StochasticProcesses-Book}%
  \BibitemOpen
  \bibfield  {author} {\bibinfo {author} {\bibfnamefont {N.~G.}\ \bibnamefont
  {{V}an {K}ampen}},\ }\href@noop {} {\emph {\bibinfo {title} {{S}tochastic
  {P}rocesses in {P}hysics and {C}hemistry}}}\ (\bibinfo  {publisher}
  {Elsevier},\ \bibinfo {year} {2007})\BibitemShut {NoStop}%
\bibitem [{\citenamefont {{W}eiss}(2008)}]{weissdissipation}%
  \BibitemOpen
  \bibfield  {author} {\bibinfo {author} {\bibfnamefont {U.}~\bibnamefont
  {{W}eiss}},\ }\href@noop {} {\emph {\bibinfo {title} {{Q}uantum {D}issipative
  {S}ystems}}},\ \bibinfo {edition} {3rd}\ ed.,\ \bibinfo {series} {Series in
  Modern Condensed Matter Physics}, Vol.~\bibinfo {volume} {13}\ (\bibinfo
  {publisher} {World Scientific},\ \bibinfo {address} {Singapore},\ \bibinfo
  {year} {2008})\BibitemShut {NoStop}%
\bibitem [{\citenamefont {{U}hlenbeck}\ and\ \citenamefont
  {{O}rnstein}(1930)}]{PhysRev.36.823}%
  \BibitemOpen
  \bibfield  {author} {\bibinfo {author} {\bibfnamefont {G.~E.}\ \bibnamefont
  {{U}hlenbeck}}\ and\ \bibinfo {author} {\bibfnamefont {L.~S.}\ \bibnamefont
  {{O}rnstein}},\ }\Doi {10.1103/PhysRev.36.823} {\bibfield  {journal}
  {\bibinfo  {journal} {Phys. Rev.},\ }\textbf {\bibinfo {volume} {36}},\
  \bibinfo {pages} {823} (\bibinfo {year} {1930})}\BibitemShut {NoStop}%
\bibitem [{\citenamefont {{M}ori}(1965)}]{PTP.33.423}%
  \BibitemOpen
  \bibfield  {author} {\bibinfo {author} {\bibfnamefont {H.}~\bibnamefont
  {{M}ori}},\ }\Doi {10.1143/PTP.33.423} {\bibfield  {journal} {\bibinfo
  {journal} {Prog. Theor. Phys.},\ }\textbf {\bibinfo {volume} {33}},\ \bibinfo
  {pages} {423} (\bibinfo {year} {1965})}\BibitemShut {NoStop}%
\bibitem [{\citenamefont {{G}rabert}\ \emph {et~al.}(1988)\citenamefont
  {{G}rabert}, \citenamefont {{S}chramm},\ and\ \citenamefont
  {{I}ngold}}]{Grabert1988115}%
  \BibitemOpen
  \bibfield  {author} {\bibinfo {author} {\bibfnamefont {H.}~\bibnamefont
  {{G}rabert}}, \bibinfo {author} {\bibfnamefont {P.}~\bibnamefont
  {{S}chramm}}, \ and\ \bibinfo {author} {\bibfnamefont {G.-L.}\ \bibnamefont
  {{I}ngold}},\ }\Doi {10.1016/0370-1573(88)90023-3} {\bibfield  {journal}
  {\bibinfo  {journal} {Phys. Rep.},\ }\textbf {\bibinfo {volume} {168}},\
  \bibinfo {pages} {115} (\bibinfo {year} {1988})},\ ISSN \bibinfo {issn}
  {0370-1573}\BibitemShut {NoStop}%
\bibitem [{\citenamefont {{C}aldeira}\ and\ \citenamefont
  {{L}eggett}(1983){\natexlab{a}}}]{Caldeira1983587}%
  \BibitemOpen
  \bibfield  {author} {\bibinfo {author} {\bibfnamefont {A.}~\bibnamefont
  {{C}aldeira}}\ and\ \bibinfo {author} {\bibfnamefont {A.}~\bibnamefont
  {{L}eggett}},\ }\Doi {10.1016/0378-4371(83)90013-4} {\bibfield  {journal}
  {\bibinfo  {journal} {Physica A},\ }\textbf {\bibinfo {volume} {121}},\
  \bibinfo {pages} {587 } (\bibinfo {year} {1983}{\natexlab{a}})}\BibitemShut
  {NoStop}%
\bibitem [{\citenamefont {{Z}urek}(2003)}]{RevModPhys.75.715}%
  \BibitemOpen
  \bibfield  {author} {\bibinfo {author} {\bibfnamefont {W.~H.}\ \bibnamefont
  {{Z}urek}},\ }\href@noop {} {\bibfield  {journal} {\bibinfo  {journal} {Rev.
  Mod. Phys.},\ }\textbf {\bibinfo {volume} {75}},\ \bibinfo {pages} {715}
  (\bibinfo {year} {2003})}\BibitemShut {NoStop}%
\bibitem [{\citenamefont {Lerner}\ \emph {et~al.}(2004)\citenamefont {Lerner},
  \citenamefont {Altshuler},\ and\ \citenamefont
  {Gefen}}]{DecoherenceNATOProceedings-2004}%
  \BibitemOpen
  \bibinfo {editor} {\bibfnamefont {I.~V.}\ \bibnamefont {Lerner}}, \bibinfo
  {editor} {\bibfnamefont {B.~L.}\ \bibnamefont {Altshuler}}, \ and\ \bibinfo
  {editor} {\bibfnamefont {Y.}~\bibnamefont {Gefen}},\ eds.,\ \href@noop {}
  {\emph {\bibinfo {title} {{F}undamental {P}roblems of {M}esoscopic {P}hysics:
  {I}nteractions and {D}ecoherence}}},\ \bibinfo {series} {Nato Science
  Series}, Vol.\ \bibinfo {volume} {154}\ (\bibinfo  {publisher} {Kluwer
  Academic Publishers},\ \bibinfo {year} {2004})\BibitemShut {NoStop}%
\bibitem [{\citenamefont {{N}ielsen}\ and\ \citenamefont
  {{C}huang}(2000)}]{nielsen_chuang_qc}%
  \BibitemOpen
  \bibfield  {author} {\bibinfo {author} {\bibfnamefont {M.~A.}\ \bibnamefont
  {{N}ielsen}}\ and\ \bibinfo {author} {\bibfnamefont {I.~L.}\ \bibnamefont
  {{C}huang}},\ }\href@noop {} {\emph {\bibinfo {title} {{Q}uantum computation
  and quantum information}}}\ (\bibinfo  {publisher} {Cambridge University
  Press},\ \bibinfo {address} {Cambridge, U.K.},\ \bibinfo {year}
  {2000})\BibitemShut {NoStop}%
\bibitem [{\citenamefont {{M}akhlin}\ \emph {et~al.}(2001)\citenamefont
  {{M}akhlin}, \citenamefont {{S}ch\"on},\ and\ \citenamefont
  {{S}hnirman}}]{RevModPhys.73.357}%
  \BibitemOpen
  \bibfield  {author} {\bibinfo {author} {\bibfnamefont {Y.}~\bibnamefont
  {{M}akhlin}}, \bibinfo {author} {\bibfnamefont {G.}~\bibnamefont
  {{S}ch\"on}}, \ and\ \bibinfo {author} {\bibfnamefont {A.}~\bibnamefont
  {{S}hnirman}},\ }\Doi {10.1103/RevModPhys.73.357} {\bibfield  {journal}
  {\bibinfo  {journal} {Rev. Mod. Phys.},\ }\textbf {\bibinfo {volume} {73}},\
  \bibinfo {pages} {357} (\bibinfo {year} {2001})}\BibitemShut {NoStop}%
\bibitem [{\citenamefont {{V}ion}\ \emph {et~al.}(2002)\citenamefont {{V}ion},
  \citenamefont {{A}assime}, \citenamefont {{C}ottet}, \citenamefont {{J}oyez},
  \citenamefont {{P}othier}, \citenamefont {{U}rbina}, \citenamefont
  {{E}steve},\ and\ \citenamefont {{D}evoret}}]{Vion03052002}%
  \BibitemOpen
  \bibfield  {author} {\bibinfo {author} {\bibfnamefont {D.}~\bibnamefont
  {{V}ion}}, \bibinfo {author} {\bibfnamefont {A.}~\bibnamefont {{A}assime}},
  \bibinfo {author} {\bibfnamefont {A.}~\bibnamefont {{C}ottet}}, \bibinfo
  {author} {\bibfnamefont {P.}~\bibnamefont {{J}oyez}}, \bibinfo {author}
  {\bibfnamefont {H.}~\bibnamefont {{P}othier}}, \bibinfo {author}
  {\bibfnamefont {C.}~\bibnamefont {{U}rbina}}, \bibinfo {author}
  {\bibfnamefont {D.}~\bibnamefont {{E}steve}}, \ and\ \bibinfo {author}
  {\bibfnamefont {M.~H.}\ \bibnamefont {{D}evoret}},\ }\Doi
  {10.1126/science.1069372} {\bibfield  {journal} {\bibinfo  {journal}
  {Science},\ }\textbf {\bibinfo {volume} {296}},\ \bibinfo {pages} {886}
  (\bibinfo {year} {2002})}\BibitemShut {NoStop}%
\bibitem [{\citenamefont {{K}och}\ \emph {et~al.}(2007)\citenamefont {{K}och},
  \citenamefont {{Y}u}, \citenamefont {{G}ambetta}, \citenamefont {{H}ouck},
  \citenamefont {{S}chuster}, \citenamefont {{M}ajer}, \citenamefont {{B}lais},
  \citenamefont {{D}evoret}, \citenamefont {{G}irvin},\ and\ \citenamefont
  {{S}choelkopf}}]{PhysRevA.76.042319}%
  \BibitemOpen
  \bibfield  {author} {\bibinfo {author} {\bibfnamefont {J.}~\bibnamefont
  {{K}och}}, \bibinfo {author} {\bibfnamefont {T.~M.}\ \bibnamefont {{Y}u}},
  \bibinfo {author} {\bibfnamefont {J.}~\bibnamefont {{G}ambetta}}, \bibinfo
  {author} {\bibfnamefont {A.~A.}\ \bibnamefont {{H}ouck}}, \bibinfo {author}
  {\bibfnamefont {D.~I.}\ \bibnamefont {{S}chuster}}, \bibinfo {author}
  {\bibfnamefont {J.}~\bibnamefont {{M}ajer}}, \bibinfo {author} {\bibfnamefont
  {A.}~\bibnamefont {{B}lais}}, \bibinfo {author} {\bibfnamefont {M.~H.}\
  \bibnamefont {{D}evoret}}, \bibinfo {author} {\bibfnamefont {S.~M.}\
  \bibnamefont {{G}irvin}}, \ and\ \bibinfo {author} {\bibfnamefont {R.~J.}\
  \bibnamefont {{S}choelkopf}},\ }\Doi {10.1103/PhysRevA.76.042319} {\bibfield
  {journal} {\bibinfo  {journal} {Phys. Rev. A},\ }\textbf {\bibinfo {volume}
  {76}},\ \bibinfo {pages} {042319} (\bibinfo {year} {2007})}\BibitemShut
  {NoStop}%
\bibitem [{\citenamefont {{S}choelkopf}\ and\ \citenamefont
  {{G}irvin}(2008)}]{Schoelkopf_Nature_2008}%
  \BibitemOpen
  \bibfield  {author} {\bibinfo {author} {\bibfnamefont {R.~J.}\ \bibnamefont
  {{S}choelkopf}}\ and\ \bibinfo {author} {\bibfnamefont {S.~M.}\ \bibnamefont
  {{G}irvin}},\ }\href@noop {} {\bibfield  {journal} {\bibinfo  {journal}
  {Nature (London)},\ }\textbf {\bibinfo {volume} {451}},\ \bibinfo {pages}
  {664} (\bibinfo {year} {2008})}\BibitemShut {NoStop}%
\bibitem [{\citenamefont
  {{N}itzan}(2006)}]{Nitzan_chem_dyn_in_condensed_phase_book}%
  \BibitemOpen
  \bibfield  {author} {\bibinfo {author} {\bibfnamefont {A.}~\bibnamefont
  {{N}itzan}},\ }\href@noop {} {\emph {\bibinfo {title} {{C}hemical {D}ynamics
  in {C}ondensed {P}hase}}}\ (\bibinfo  {publisher} {Oxford University Press},\
  \bibinfo {address} {Oxford, U.K.},\ \bibinfo {year} {2006})\BibitemShut
  {NoStop}%
\bibitem [{\citenamefont {{G}olding}\ \emph {et~al.}(1992)\citenamefont
  {{G}olding}, \citenamefont {{Z}immerman},\ and\ \citenamefont
  {{C}oppersmith}}]{PhysRevLett.68.998}%
  \BibitemOpen
  \bibfield  {author} {\bibinfo {author} {\bibfnamefont {B.}~\bibnamefont
  {{G}olding}}, \bibinfo {author} {\bibfnamefont {M.~N.}\ \bibnamefont
  {{Z}immerman}}, \ and\ \bibinfo {author} {\bibfnamefont {S.~N.}\ \bibnamefont
  {{C}oppersmith}},\ }\Doi {10.1103/PhysRevLett.68.998} {\bibfield  {journal}
  {\bibinfo  {journal} {Phys. Rev. Lett.},\ }\textbf {\bibinfo {volume} {68}},\
  \bibinfo {pages} {998} (\bibinfo {year} {1992})}\BibitemShut {NoStop}%
\bibitem [{\citenamefont {{X}u}\ and\ \citenamefont
  {{S}chulten}(1994)}]{XuSchulten-ChemPhys1994}%
  \BibitemOpen
  \bibfield  {author} {\bibinfo {author} {\bibfnamefont {D.}~\bibnamefont
  {{X}u}}\ and\ \bibinfo {author} {\bibfnamefont {K.}~\bibnamefont
  {{S}chulten}},\ }\href@noop {} {\bibfield  {journal} {\bibinfo  {journal}
  {Chem. Phys.},\ }\textbf {\bibinfo {volume} {182}},\ \bibinfo {pages} {91}
  (\bibinfo {year} {1994})}\BibitemShut {NoStop}%
\bibitem [{\citenamefont {{S}tockburger}\ and\ \citenamefont
  {{M}ak}(1996)}]{StockburgerMak-ChemPhys-1996}%
  \BibitemOpen
  \bibfield  {author} {\bibinfo {author} {\bibfnamefont {J.~T.}\ \bibnamefont
  {{S}tockburger}}\ and\ \bibinfo {author} {\bibfnamefont {C.~H.}\ \bibnamefont
  {{M}ak}},\ }\href@noop {} {\bibfield  {journal} {\bibinfo  {journal} {J.
  Chem. Phys.},\ }\textbf {\bibinfo {volume} {105}},\ \bibinfo {pages} {8126}
  (\bibinfo {year} {1996})}\BibitemShut {NoStop}%
\bibitem [{\citenamefont {{C}amalet}\ and\ \citenamefont
  {{C}hitra}(2007)}]{PhysRevLett.99.267202}%
  \BibitemOpen
  \bibfield  {author} {\bibinfo {author} {\bibfnamefont {S.}~\bibnamefont
  {{C}amalet}}\ and\ \bibinfo {author} {\bibfnamefont {R.}~\bibnamefont
  {{C}hitra}},\ }\Doi {10.1103/PhysRevLett.99.267202} {\bibfield  {journal}
  {\bibinfo  {journal} {Phys. Rev. Lett.},\ }\textbf {\bibinfo {volume} {99}},\
  \bibinfo {pages} {267202} (\bibinfo {year} {2007})}\BibitemShut {NoStop}%
\bibitem [{\citenamefont {{R}estrepo}\ \emph {et~al.}(2011)\citenamefont
  {{R}estrepo}, \citenamefont {{C}hitra}, \citenamefont {{C}amalet},\ and\
  \citenamefont {{D}upont}}]{PhysRevB.84.245109}%
  \BibitemOpen
  \bibfield  {author} {\bibinfo {author} {\bibfnamefont {J.}~\bibnamefont
  {{R}estrepo}}, \bibinfo {author} {\bibfnamefont {R.}~\bibnamefont
  {{C}hitra}}, \bibinfo {author} {\bibfnamefont {S.}~\bibnamefont {{C}amalet}},
  \ and\ \bibinfo {author} {\bibfnamefont {E.}~\bibnamefont {{D}upont}},\ }\Doi
  {10.1103/PhysRevB.84.245109} {\bibfield  {journal} {\bibinfo  {journal}
  {Phys. Rev. B},\ }\textbf {\bibinfo {volume} {84}},\ \bibinfo {pages}
  {245109} (\bibinfo {year} {2011})}\BibitemShut {NoStop}%
\bibitem [{\citenamefont {{C}hin}\ \emph {et~al.}(2012)\citenamefont {{C}hin},
  \citenamefont {{H}uelga},\ and\ \citenamefont {{P}lenio}}]{Chin13082012}%
  \BibitemOpen
  \bibfield  {author} {\bibinfo {author} {\bibfnamefont {A.~W.}\ \bibnamefont
  {{C}hin}}, \bibinfo {author} {\bibfnamefont {S.~F.}\ \bibnamefont
  {{H}uelga}}, \ and\ \bibinfo {author} {\bibfnamefont {M.~B.}\ \bibnamefont
  {{P}lenio}},\ }\Doi {10.1098/rsta.2011.0224} {\bibfield  {journal} {\bibinfo
  {journal} {Phil. Trans. R. Soc. A},\ }\textbf {\bibinfo {volume} {370}},\
  \bibinfo {pages} {3638} (\bibinfo {year} {2012})}\BibitemShut {NoStop}%
\bibitem [{\citenamefont {{G}eorges}\ \emph {et~al.}(1996)\citenamefont
  {{G}eorges}, \citenamefont {{K}otliar}, \citenamefont {{K}rauth},\ and\
  \citenamefont {{R}ozenberg}}]{RevModPhys.68.13}%
  \BibitemOpen
  \bibfield  {author} {\bibinfo {author} {\bibfnamefont {A.}~\bibnamefont
  {{G}eorges}}, \bibinfo {author} {\bibfnamefont {G.}~\bibnamefont
  {{K}otliar}}, \bibinfo {author} {\bibfnamefont {W.}~\bibnamefont {{K}rauth}},
  \ and\ \bibinfo {author} {\bibfnamefont {M.~J.}\ \bibnamefont
  {{R}ozenberg}},\ }\Doi {10.1103/RevModPhys.68.13} {\bibfield  {journal}
  {\bibinfo  {journal} {Rev. Mod. Phys.},\ }\textbf {\bibinfo {volume} {68}},\
  \bibinfo {pages} {13} (\bibinfo {year} {1996})}\BibitemShut {NoStop}%
\bibitem [{\citenamefont {{H}aule}\ \emph {et~al.}(2002)\citenamefont
  {{H}aule}, \citenamefont {{R}osch}, \citenamefont {{K}roha},\ and\
  \citenamefont {{W}\"olfle}}]{PhysRevLett.89.236402}%
  \BibitemOpen
  \bibfield  {author} {\bibinfo {author} {\bibfnamefont {K.}~\bibnamefont
  {{H}aule}}, \bibinfo {author} {\bibfnamefont {A.}~\bibnamefont {{R}osch}},
  \bibinfo {author} {\bibfnamefont {J.}~\bibnamefont {{K}roha}}, \ and\
  \bibinfo {author} {\bibfnamefont {P.}~\bibnamefont {{W}\"olfle}},\ }\Doi
  {10.1103/PhysRevLett.89.236402} {\bibfield  {journal} {\bibinfo  {journal}
  {Phys. Rev. Lett.},\ }\textbf {\bibinfo {volume} {89}},\ \bibinfo {pages}
  {236402} (\bibinfo {year} {2002})}\BibitemShut {NoStop}%
\bibitem [{\citenamefont {{S}mith}\ and\ \citenamefont
  {{S}i}(2000)}]{PhysRevB.61.5184}%
  \BibitemOpen
  \bibfield  {author} {\bibinfo {author} {\bibfnamefont {J.~L.}\ \bibnamefont
  {{S}mith}}\ and\ \bibinfo {author} {\bibfnamefont {Q.}~\bibnamefont {{S}i}},\
  }\Doi {10.1103/PhysRevB.61.5184} {\bibfield  {journal} {\bibinfo  {journal}
  {Phys. Rev. B},\ }\textbf {\bibinfo {volume} {61}},\ \bibinfo {pages} {5184}
  (\bibinfo {year} {2000})}\BibitemShut {NoStop}%
\bibitem [{\citenamefont {{S}un}\ and\ \citenamefont
  {{K}otliar}(2002)}]{PhysRevB.66.085120}%
  \BibitemOpen
  \bibfield  {author} {\bibinfo {author} {\bibfnamefont {P.}~\bibnamefont
  {{S}un}}\ and\ \bibinfo {author} {\bibfnamefont {G.}~\bibnamefont
  {{K}otliar}},\ }\Doi {10.1103/PhysRevB.66.085120} {\bibfield  {journal}
  {\bibinfo  {journal} {Phys. Rev. B},\ }\textbf {\bibinfo {volume} {66}},\
  \bibinfo {pages} {085120} (\bibinfo {year} {2002})}\BibitemShut {NoStop}%
\bibitem [{\citenamefont {{L}eggett}\ \emph {et~al.}(1987)\citenamefont
  {{L}eggett}, \citenamefont {{C}hakravarty}, \citenamefont {{D}orsey},
  \citenamefont {{F}isher}, \citenamefont {{G}arg},\ and\ \citenamefont
  {{Z}werger}}]{RevModPhys.59.1}%
  \BibitemOpen
  \bibfield  {author} {\bibinfo {author} {\bibfnamefont {A.~J.}\ \bibnamefont
  {{L}eggett}}, \bibinfo {author} {\bibfnamefont {S.}~\bibnamefont
  {{C}hakravarty}}, \bibinfo {author} {\bibfnamefont {A.~T.}\ \bibnamefont
  {{D}orsey}}, \bibinfo {author} {\bibfnamefont {M.~P.~A.}\ \bibnamefont
  {{F}isher}}, \bibinfo {author} {\bibfnamefont {A.}~\bibnamefont {{G}arg}}, \
  and\ \bibinfo {author} {\bibfnamefont {W.}~\bibnamefont {{Z}werger}},\ }\Doi
  {10.1103/RevModPhys.59.1} {\bibfield  {journal} {\bibinfo  {journal} {Rev.
  Mod. Phys.},\ }\textbf {\bibinfo {volume} {59}},\ \bibinfo {pages} {1}
  (\bibinfo {year} {1987})}\BibitemShut {NoStop}%
\bibitem [{\citenamefont {{C}aldeira}\ and\ \citenamefont
  {{L}eggett}(1981)}]{PhysRevLett.46.211}%
  \BibitemOpen
  \bibfield  {author} {\bibinfo {author} {\bibfnamefont {A.~O.}\ \bibnamefont
  {{C}aldeira}}\ and\ \bibinfo {author} {\bibfnamefont {A.~J.}\ \bibnamefont
  {{L}eggett}},\ }\Doi {10.1103/PhysRevLett.46.211} {\bibfield  {journal}
  {\bibinfo  {journal} {Phys. Rev. Lett.},\ }\textbf {\bibinfo {volume} {46}},\
  \bibinfo {pages} {211} (\bibinfo {year} {1981})}\BibitemShut {NoStop}%
\bibitem [{\citenamefont {{C}aldeira}\ and\ \citenamefont
  {{L}eggett}(1983){\natexlab{b}}}]{caldeira_quantum_1983}%
  \BibitemOpen
  \bibfield  {author} {\bibinfo {author} {\bibfnamefont {A.~O.}\ \bibnamefont
  {{C}aldeira}}\ and\ \bibinfo {author} {\bibfnamefont {A.~J.}\ \bibnamefont
  {{L}eggett}},\ }\Doi {10.1016/0003-4916(83)90202-6} {\bibfield  {journal}
  {\bibinfo  {journal} {Ann. Phys. (NY)},\ }\textbf {\bibinfo {volume} {149}},\
  \bibinfo {pages} {374} (\bibinfo {year} {1983}{\natexlab{b}})},\ ISSN
  \bibinfo {issn} {0003-4916}\BibitemShut {NoStop}%
\bibitem [{\citenamefont {{B}lume}\ \emph {et~al.}(1970)\citenamefont
  {{B}lume}, \citenamefont {{E}mery},\ and\ \citenamefont
  {{L}uther}}]{PhysRevLett.25.450}%
  \BibitemOpen
  \bibfield  {author} {\bibinfo {author} {\bibfnamefont {M.}~\bibnamefont
  {{B}lume}}, \bibinfo {author} {\bibfnamefont {V.~J.}\ \bibnamefont
  {{E}mery}}, \ and\ \bibinfo {author} {\bibfnamefont {A.}~\bibnamefont
  {{L}uther}},\ }\Doi {10.1103/PhysRevLett.25.450} {\bibfield  {journal}
  {\bibinfo  {journal} {Phys. Rev. Lett.},\ }\textbf {\bibinfo {volume} {25}},\
  \bibinfo {pages} {450} (\bibinfo {year} {1970})}\BibitemShut {NoStop}%
\bibitem [{\citenamefont {{G}uinea}\ \emph
  {et~al.}(1985){\natexlab{a}}\citenamefont {{G}uinea}, \citenamefont
  {{H}akim},\ and\ \citenamefont {{M}uramatsu}}]{PhysRevLett.54.263}%
  \BibitemOpen
  \bibfield  {author} {\bibinfo {author} {\bibfnamefont {F.}~\bibnamefont
  {{G}uinea}}, \bibinfo {author} {\bibfnamefont {V.}~\bibnamefont {{H}akim}}, \
  and\ \bibinfo {author} {\bibfnamefont {A.}~\bibnamefont {{M}uramatsu}},\
  }\Doi {10.1103/PhysRevLett.54.263} {\bibfield  {journal} {\bibinfo  {journal}
  {Phys. Rev. Lett.},\ }\textbf {\bibinfo {volume} {54}},\ \bibinfo {pages}
  {263} (\bibinfo {year} {1985}{\natexlab{a}})}\BibitemShut {NoStop}%
\bibitem [{\citenamefont {{V}ojta}(2006)}]{vojta_philmag_2006}%
  \BibitemOpen
  \bibfield  {author} {\bibinfo {author} {\bibfnamefont {M.}~\bibnamefont
  {{V}ojta}},\ }\href@noop {} {\bibfield  {journal} {\bibinfo  {journal} {Phil.
  Mag.},\ }\textbf {\bibinfo {volume} {86}},\ \bibinfo {pages} {1807} (\bibinfo
  {year} {2006})}\BibitemShut {NoStop}%
\bibitem [{\citenamefont {{L}e {H}ur}(2008)}]{lehur_entanglement_spinboson}%
  \BibitemOpen
  \bibfield  {author} {\bibinfo {author} {\bibfnamefont {K.}~\bibnamefont {{L}e
  {H}ur}},\ }\href@noop {} {\bibfield  {journal} {\bibinfo  {journal} {Ann.
  Phys. (NY)},\ }\textbf {\bibinfo {volume} {323}},\ \bibinfo {pages} {2208}
  (\bibinfo {year} {2008})}\BibitemShut {NoStop}%
\bibitem [{\citenamefont {{F}lorens}\ \emph {et~al.}(2010)\citenamefont
  {{F}lorens}, \citenamefont {{V}enturelli},\ and\ \citenamefont
  {{N}arayanan}}]{Florens-SBBookChapter-2010}%
  \BibitemOpen
  \bibfield  {author} {\bibinfo {author} {\bibfnamefont {S.}~\bibnamefont
  {{F}lorens}}, \bibinfo {author} {\bibfnamefont {D.}~\bibnamefont
  {{V}enturelli}}, \ and\ \bibinfo {author} {\bibfnamefont {R.}~\bibnamefont
  {{N}arayanan}},\ }\enquote {\bibinfo {title} {{Q}uantum quenching, annealing
  and computation},}\ \ (\bibinfo  {publisher} {Springer},\ \bibinfo {year}
  {2010})\ Chap.\ \bibinfo {chapter} {Quantum phase transition in the
  spin-boson model}, pp.\ \bibinfo {pages} {145--162}\BibitemShut {NoStop}%
\bibitem [{\citenamefont {{L}e
  {H}ur}(2010)}]{KarynLeHur-UnderstandingQPT-Article}%
  \BibitemOpen
  \bibfield  {author} {\bibinfo {author} {\bibfnamefont {K.}~\bibnamefont {{L}e
  {H}ur}},\ }\enquote {\bibinfo {title} {{U}nderstanding {Q}uantum {P}hase
  {T}ransitions},}\ \ (\bibinfo  {publisher} {CRC Press, Cleveland/Taylor \&
  Francis},\ \bibinfo {address} {Cleveland, OH, USA},\ \bibinfo {year} {2010})\
  Chap.\ \bibinfo {chapter} {Quantum Phase Transitions in Spin-Boson Systems:
  Dissipation and Light Phenomena}, pp.\ \bibinfo {pages} {217--237},\ \bibinfo
  {edition} {1st}\ ed.\BibitemShut {Stop}%
\bibitem [{\citenamefont {{C}hakravarty}(1982)}]{PhysRevLett.49.681}%
  \BibitemOpen
  \bibfield  {author} {\bibinfo {author} {\bibfnamefont {S.}~\bibnamefont
  {{C}hakravarty}},\ }\Doi {10.1103/PhysRevLett.49.681} {\bibfield  {journal}
  {\bibinfo  {journal} {Phys. Rev. Lett.},\ }\textbf {\bibinfo {volume} {49}},\
  \bibinfo {pages} {681} (\bibinfo {year} {1982})}\BibitemShut {NoStop}%
\bibitem [{\citenamefont {{B}ray}\ and\ \citenamefont
  {{M}oore}(1982)}]{PhysRevLett.49.1545}%
  \BibitemOpen
  \bibfield  {author} {\bibinfo {author} {\bibfnamefont {A.~J.}\ \bibnamefont
  {{B}ray}}\ and\ \bibinfo {author} {\bibfnamefont {M.~A.}\ \bibnamefont
  {{M}oore}},\ }\href@noop {} {\bibfield  {journal} {\bibinfo  {journal} {Phys.
  Rev. Lett.},\ }\textbf {\bibinfo {volume} {49}},\ \bibinfo {pages} {1545}
  (\bibinfo {year} {1982})}\BibitemShut {NoStop}%
\bibitem [{\citenamefont {{G}uinea}\ \emph
  {et~al.}(1985){\natexlab{b}}\citenamefont {{G}uinea}, \citenamefont
  {{H}akim},\ and\ \citenamefont {{M}uramatsu}}]{PhysRevB.32.4410}%
  \BibitemOpen
  \bibfield  {author} {\bibinfo {author} {\bibfnamefont {F.}~\bibnamefont
  {{G}uinea}}, \bibinfo {author} {\bibfnamefont {V.}~\bibnamefont {{H}akim}}, \
  and\ \bibinfo {author} {\bibfnamefont {A.}~\bibnamefont {{M}uramatsu}},\
  }\Doi {10.1103/PhysRevB.32.4410} {\bibfield  {journal} {\bibinfo  {journal}
  {Phys. Rev. B},\ }\textbf {\bibinfo {volume} {32}},\ \bibinfo {pages} {4410}
  (\bibinfo {year} {1985}{\natexlab{b}})}\BibitemShut {NoStop}%
\bibitem [{\citenamefont {{A}nderson}\ \emph {et~al.}(1970)\citenamefont
  {{A}nderson}, \citenamefont {{Y}uval},\ and\ \citenamefont
  {{H}amann}}]{PhysRevB.1.4464}%
  \BibitemOpen
  \bibfield  {author} {\bibinfo {author} {\bibfnamefont {P.~W.}\ \bibnamefont
  {{A}nderson}}, \bibinfo {author} {\bibfnamefont {G.}~\bibnamefont {{Y}uval}},
  \ and\ \bibinfo {author} {\bibfnamefont {D.~R.}\ \bibnamefont {{H}amann}},\
  }\Doi {10.1103/PhysRevB.1.4464} {\bibfield  {journal} {\bibinfo  {journal}
  {Phys. Rev. B},\ }\textbf {\bibinfo {volume} {1}},\ \bibinfo {pages} {4464}
  (\bibinfo {year} {1970})}\BibitemShut {NoStop}%
\bibitem [{\citenamefont {{S}pohn}\ and\ \citenamefont
  {{D}{\"u}mcke}(1985)}]{SpohnDumcke1985}%
  \BibitemOpen
  \bibfield  {author} {\bibinfo {author} {\bibfnamefont {H.}~\bibnamefont
  {{S}pohn}}\ and\ \bibinfo {author} {\bibfnamefont {R.}~\bibnamefont
  {{D}{\"u}mcke}},\ }\href@noop {} {\bibfield  {journal} {\bibinfo  {journal}
  {J. Stat. Phys.},\ }\textbf {\bibinfo {volume} {41}},\ \bibinfo {pages} {389}
  (\bibinfo {year} {1985})}\BibitemShut {NoStop}%
\bibitem [{\citenamefont {{A}nderson}\ and\ \citenamefont
  {{Y}uval}(1971)}]{0022-3719-4-5-011}%
  \BibitemOpen
  \bibfield  {author} {\bibinfo {author} {\bibfnamefont {P.~W.}\ \bibnamefont
  {{A}nderson}}\ and\ \bibinfo {author} {\bibfnamefont {G.}~\bibnamefont
  {{Y}uval}},\ }\href@noop {} {\bibfield  {journal} {\bibinfo  {journal} {J.
  Phys. C},\ }\textbf {\bibinfo {volume} {4}},\ \bibinfo {pages} {607}
  (\bibinfo {year} {1971})}\BibitemShut {NoStop}%
\bibitem [{\citenamefont {{C}astellanos {B}eltra}\ and\ \citenamefont
  {{L}ehnert}(2007)}]{CastellanosBeltran-Lehnert-ApplPhysLett_2007}%
  \BibitemOpen
  \bibfield  {author} {\bibinfo {author} {\bibfnamefont {M.~A.}\ \bibnamefont
  {{C}astellanos {B}eltra}}\ and\ \bibinfo {author} {\bibfnamefont {K.~W.}\
  \bibnamefont {{L}ehnert}},\ }\href@noop {} {\bibfield  {journal} {\bibinfo
  {journal} {Appl. Phys. Lett.},\ }\textbf {\bibinfo {volume} {91}},\ \bibinfo
  {pages} {083509} (\bibinfo {year} {2007})}\BibitemShut {NoStop}%
\bibitem [{\citenamefont {{M}anucharyan}\ \emph {et~al.}(2009)\citenamefont
  {{M}anucharyan}, \citenamefont {{K}och}, \citenamefont {{G}lazman},\ and\
  \citenamefont {{D}evoret}}]{Manucharyan10022009}%
  \BibitemOpen
  \bibfield  {author} {\bibinfo {author} {\bibfnamefont {V.~E.}\ \bibnamefont
  {{M}anucharyan}}, \bibinfo {author} {\bibfnamefont {J.}~\bibnamefont
  {{K}och}}, \bibinfo {author} {\bibfnamefont {L.~I.}\ \bibnamefont
  {{G}lazman}}, \ and\ \bibinfo {author} {\bibfnamefont {M.~H.}\ \bibnamefont
  {{D}evoret}},\ }\Doi {10.1126/science.1175552} {\bibfield  {journal}
  {\bibinfo  {journal} {Science},\ }\textbf {\bibinfo {volume} {326}},\
  \bibinfo {pages} {113} (\bibinfo {year} {2009})}\BibitemShut {NoStop}%
\bibitem [{\citenamefont {{C}hung}\ \emph {et~al.}(2009)\citenamefont
  {{C}hung}, \citenamefont {{L}e {H}ur}, \citenamefont {{V}ojta},\ and\
  \citenamefont {{W}\"{o}lfle}}]{chung:216803}%
  \BibitemOpen
  \bibfield  {author} {\bibinfo {author} {\bibfnamefont {C.-H.}\ \bibnamefont
  {{C}hung}}, \bibinfo {author} {\bibfnamefont {K.}~\bibnamefont {{L}e {H}ur}},
  \bibinfo {author} {\bibfnamefont {M.}~\bibnamefont {{V}ojta}}, \ and\
  \bibinfo {author} {\bibfnamefont {P.}~\bibnamefont {{W}\"{o}lfle}},\ }\Doi
  {10.1103/PhysRevLett.102.216803} {\bibfield  {journal} {\bibinfo  {journal}
  {Phys. Rev. Lett.},\ }\textbf {\bibinfo {volume} {102}},\ \bibinfo {eid}
  {216803} (\bibinfo {year} {2009})}\BibitemShut {NoStop}%
\bibitem [{\citenamefont {{M}ebrahtu}\ \emph {et~al.}(2012)\citenamefont
  {{M}ebrahtu}, \citenamefont {{B}orzenets}, \citenamefont {{L}iu},
  \citenamefont {{Z}heng}, \citenamefont {{B}omze}, \citenamefont {{S}mirnov},
  \citenamefont {{B}aranger},\ and\ \citenamefont
  {{F}inkelstein}}]{Mebrahtu-QPTResLev-Nature-2012}%
  \BibitemOpen
  \bibfield  {author} {\bibinfo {author} {\bibfnamefont {H.~T.}\ \bibnamefont
  {{M}ebrahtu}}, \bibinfo {author} {\bibfnamefont {I.~V.}\ \bibnamefont
  {{B}orzenets}}, \bibinfo {author} {\bibfnamefont {D.~E.}\ \bibnamefont
  {{L}iu}}, \bibinfo {author} {\bibfnamefont {H.}~\bibnamefont {{Z}heng}},
  \bibinfo {author} {\bibfnamefont {Y.~V.}\ \bibnamefont {{B}omze}}, \bibinfo
  {author} {\bibfnamefont {A.~I.}\ \bibnamefont {{S}mirnov}}, \bibinfo {author}
  {\bibfnamefont {H.~U.}\ \bibnamefont {{B}aranger}}, \ and\ \bibinfo {author}
  {\bibfnamefont {G.}~\bibnamefont {{F}inkelstein}},\ }\href@noop {} {\bibfield
   {journal} {\bibinfo  {journal} {Nature},\ }\textbf {\bibinfo {volume}
  {488}},\ \bibinfo {pages} {61} (\bibinfo {year} {2012})}\BibitemShut
  {NoStop}%
\bibitem [{\citenamefont {{P}op}\ \emph {et~al.}(2012)\citenamefont {{P}op},
  \citenamefont {{P}rotopopov}, \citenamefont {{L}ecocq}, \citenamefont
  {{P}eng}, \citenamefont {{P}annetier}, \citenamefont {{B}uisson},\ and\
  \citenamefont {{G}uichard}}]{Pop-PhaseSlips-NatPhys-2010}%
  \BibitemOpen
  \bibfield  {author} {\bibinfo {author} {\bibfnamefont {I.~M.}\ \bibnamefont
  {{P}op}}, \bibinfo {author} {\bibfnamefont {I.}~\bibnamefont {{P}rotopopov}},
  \bibinfo {author} {\bibfnamefont {F.}~\bibnamefont {{L}ecocq}}, \bibinfo
  {author} {\bibfnamefont {Z.}~\bibnamefont {{P}eng}}, \bibinfo {author}
  {\bibfnamefont {B.}~\bibnamefont {{P}annetier}}, \bibinfo {author}
  {\bibfnamefont {O.}~\bibnamefont {{B}uisson}}, \ and\ \bibinfo {author}
  {\bibfnamefont {W.}~\bibnamefont {{G}uichard}},\ }\href@noop {} {\bibfield
  {journal} {\bibinfo  {journal} {Nat. Phys.},\ }\textbf {\bibinfo {volume}
  {6}},\ \bibinfo {pages} {589} (\bibinfo {year} {2012})}\BibitemShut {NoStop}%
\bibitem [{\citenamefont {{H}ofheinz}\ \emph {et~al.}(2011)\citenamefont
  {{H}ofheinz}, \citenamefont {{P}ortier}, \citenamefont {{B}audouin},
  \citenamefont {{J}oyez}, \citenamefont {{V}ion}, \citenamefont {{B}ertet},
  \citenamefont {{R}oche},\ and\ \citenamefont
  {{E}steve}}]{PhysRevLett.106.217005}%
  \BibitemOpen
  \bibfield  {author} {\bibinfo {author} {\bibfnamefont {M.}~\bibnamefont
  {{H}ofheinz}}, \bibinfo {author} {\bibfnamefont {F.}~\bibnamefont
  {{P}ortier}}, \bibinfo {author} {\bibfnamefont {Q.}~\bibnamefont
  {{B}audouin}}, \bibinfo {author} {\bibfnamefont {P.}~\bibnamefont {{J}oyez}},
  \bibinfo {author} {\bibfnamefont {D.}~\bibnamefont {{V}ion}}, \bibinfo
  {author} {\bibfnamefont {P.}~\bibnamefont {{B}ertet}}, \bibinfo {author}
  {\bibfnamefont {P.}~\bibnamefont {{R}oche}}, \ and\ \bibinfo {author}
  {\bibfnamefont {D.}~\bibnamefont {{E}steve}},\ }\Doi
  {10.1103/PhysRevLett.106.217005} {\bibfield  {journal} {\bibinfo  {journal}
  {Phys. Rev. Lett.},\ }\textbf {\bibinfo {volume} {106}},\ \bibinfo {pages}
  {217005} (\bibinfo {year} {2011})}\BibitemShut {NoStop}%
\bibitem [{\citenamefont {{M}asluk}\ \emph {et~al.}(2012)\citenamefont
  {{M}asluk}, \citenamefont {{P}op}, \citenamefont {{K}amal}, \citenamefont
  {{M}inev},\ and\ \citenamefont {{D}evoret}}]{PhysRevLett.109.137002}%
  \BibitemOpen
  \bibfield  {author} {\bibinfo {author} {\bibfnamefont {N.~A.}\ \bibnamefont
  {{M}asluk}}, \bibinfo {author} {\bibfnamefont {I.~M.}\ \bibnamefont {{P}op}},
  \bibinfo {author} {\bibfnamefont {A.}~\bibnamefont {{K}amal}}, \bibinfo
  {author} {\bibfnamefont {Z.~K.}\ \bibnamefont {{M}inev}}, \ and\ \bibinfo
  {author} {\bibfnamefont {M.~H.}\ \bibnamefont {{D}evoret}},\ }\Doi
  {10.1103/PhysRevLett.109.137002} {\bibfield  {journal} {\bibinfo  {journal}
  {Phys. Rev. Lett.},\ }\textbf {\bibinfo {volume} {109}},\ \bibinfo {pages}
  {137002} (\bibinfo {year} {2012})}\BibitemShut {NoStop}%
\bibitem [{\citenamefont {{C}edraschi}\ \emph {et~al.}(2000)\citenamefont
  {{C}edraschi}, \citenamefont {{P}onomarenko},\ and\ \citenamefont
  {{B}\"uttiker}}]{PhysRevLett.84.346}%
  \BibitemOpen
  \bibfield  {author} {\bibinfo {author} {\bibfnamefont {P.}~\bibnamefont
  {{C}edraschi}}, \bibinfo {author} {\bibfnamefont {V.~V.}\ \bibnamefont
  {{P}onomarenko}}, \ and\ \bibinfo {author} {\bibfnamefont {M.}~\bibnamefont
  {{B}\"uttiker}},\ }\Doi {10.1103/PhysRevLett.84.346} {\bibfield  {journal}
  {\bibinfo  {journal} {Phys. Rev. Lett.},\ }\textbf {\bibinfo {volume} {84}},\
  \bibinfo {pages} {346} (\bibinfo {year} {2000})}\BibitemShut {NoStop}%
\bibitem [{\citenamefont {{K}opp}\ and\ \citenamefont {{L}e
  {H}ur}(2007)}]{kopp:220401}%
  \BibitemOpen
  \bibfield  {author} {\bibinfo {author} {\bibfnamefont {A.}~\bibnamefont
  {{K}opp}}\ and\ \bibinfo {author} {\bibfnamefont {K.}~\bibnamefont {{L}e
  {H}ur}},\ }\Doi {10.1103/PhysRevLett.98.220401} {\bibfield  {journal}
  {\bibinfo  {journal} {Phys. Rev. Lett.},\ }\textbf {\bibinfo {volume} {98}},\
  \bibinfo {eid} {220401} (\bibinfo {year} {2007})}\BibitemShut {NoStop}%
\bibitem [{\citenamefont {{F}urusaki}\ and\ \citenamefont
  {{M}atveev}(2002)}]{PhysRevLett.88.226404}%
  \BibitemOpen
  \bibfield  {author} {\bibinfo {author} {\bibfnamefont {A.}~\bibnamefont
  {{F}urusaki}}\ and\ \bibinfo {author} {\bibfnamefont {K.~A.}\ \bibnamefont
  {{M}atveev}},\ }\Doi {10.1103/PhysRevLett.88.226404} {\bibfield  {journal}
  {\bibinfo  {journal} {Phys. Rev. Lett.},\ }\textbf {\bibinfo {volume} {88}},\
  \bibinfo {pages} {226404} (\bibinfo {year} {2002})}\BibitemShut {NoStop}%
\bibitem [{\citenamefont {{M}\"under}\ \emph {et~al.}(2012)\citenamefont
  {{M}\"under}, \citenamefont {{W}eichselbaum}, \citenamefont {{G}oldstein},
  \citenamefont {{G}efen},\ and\ \citenamefont {von
  {D}elft}}]{PhysRevB.85.235104}%
  \BibitemOpen
  \bibfield  {author} {\bibinfo {author} {\bibfnamefont {W.}~\bibnamefont
  {{M}\"under}}, \bibinfo {author} {\bibfnamefont {A.}~\bibnamefont
  {{W}eichselbaum}}, \bibinfo {author} {\bibfnamefont {M.}~\bibnamefont
  {{G}oldstein}}, \bibinfo {author} {\bibfnamefont {Y.}~\bibnamefont
  {{G}efen}}, \ and\ \bibinfo {author} {\bibfnamefont {J.}~\bibnamefont {von
  {D}elft}},\ }\Doi {10.1103/PhysRevB.85.235104} {\bibfield  {journal}
  {\bibinfo  {journal} {Phys. Rev. B},\ }\textbf {\bibinfo {volume} {85}},\
  \bibinfo {pages} {235104} (\bibinfo {year} {2012})}\BibitemShut {NoStop}%
\bibitem [{\citenamefont {{N}issen}\ \emph {et~al.}(2012)\citenamefont
  {{N}issen}, \citenamefont {{S}chmidt}, \citenamefont {{B}iondi},
  \citenamefont {{B}latter}, \citenamefont {{T}\"ureci},\ and\ \citenamefont
  {{K}eeling}}]{PhysRevLett.108.233603}%
  \BibitemOpen
  \bibfield  {author} {\bibinfo {author} {\bibfnamefont {F.}~\bibnamefont
  {{N}issen}}, \bibinfo {author} {\bibfnamefont {S.}~\bibnamefont {{S}chmidt}},
  \bibinfo {author} {\bibfnamefont {M.}~\bibnamefont {{B}iondi}}, \bibinfo
  {author} {\bibfnamefont {G.}~\bibnamefont {{B}latter}}, \bibinfo {author}
  {\bibfnamefont {H.~E.}\ \bibnamefont {{T}\"ureci}}, \ and\ \bibinfo {author}
  {\bibfnamefont {J.}~\bibnamefont {{K}eeling}},\ }\Doi
  {10.1103/PhysRevLett.108.233603} {\bibfield  {journal} {\bibinfo  {journal}
  {Phys. Rev. Lett.},\ }\textbf {\bibinfo {volume} {108}},\ \bibinfo {pages}
  {233603} (\bibinfo {year} {2012})}\BibitemShut {NoStop}%
\bibitem [{\citenamefont {{B}ishop}\ \emph {et~al.}(2010)\citenamefont
  {{B}ishop}, \citenamefont {{G}inossar},\ and\ \citenamefont
  {{G}irvin}}]{PhysRevLett.105.100505}%
  \BibitemOpen
  \bibfield  {author} {\bibinfo {author} {\bibfnamefont {L.~S.}\ \bibnamefont
  {{B}ishop}}, \bibinfo {author} {\bibfnamefont {E.}~\bibnamefont
  {{G}inossar}}, \ and\ \bibinfo {author} {\bibfnamefont {S.~M.}\ \bibnamefont
  {{G}irvin}},\ }\Doi {10.1103/PhysRevLett.105.100505} {\bibfield  {journal}
  {\bibinfo  {journal} {Phys. Rev. Lett.},\ }\textbf {\bibinfo {volume}
  {105}},\ \bibinfo {pages} {100505} (\bibinfo {year} {2010})}\BibitemShut
  {NoStop}%
\bibitem [{\citenamefont {{D}alla {T}orre}\ \emph {et~al.}(2012)\citenamefont
  {{D}alla {T}orre}, \citenamefont {{D}emler}, \citenamefont {{G}iamarchi},\
  and\ \citenamefont {{A}ltman}}]{PhysRevB.85.184302}%
  \BibitemOpen
  \bibfield  {author} {\bibinfo {author} {\bibfnamefont {E.~G.}\ \bibnamefont
  {{D}alla {T}orre}}, \bibinfo {author} {\bibfnamefont {E.}~\bibnamefont
  {{D}emler}}, \bibinfo {author} {\bibfnamefont {T.}~\bibnamefont
  {{G}iamarchi}}, \ and\ \bibinfo {author} {\bibfnamefont {E.}~\bibnamefont
  {{A}ltman}},\ }\Doi {10.1103/PhysRevB.85.184302} {\bibfield  {journal}
  {\bibinfo  {journal} {Phys. Rev. B},\ }\textbf {\bibinfo {volume} {85}},\
  \bibinfo {pages} {184302} (\bibinfo {year} {2012})}\BibitemShut {NoStop}%
\bibitem [{\citenamefont {{L}e{C}lair}\ \emph {et~al.}(1997)\citenamefont
  {{L}e{C}lair}, \citenamefont {{L}esage}, \citenamefont {{L}ukyanov},\ and\
  \citenamefont {{S}aleur}}]{LeClair-PhysLettA-1997}%
  \BibitemOpen
  \bibfield  {author} {\bibinfo {author} {\bibfnamefont {A.}~\bibnamefont
  {{L}e{C}lair}}, \bibinfo {author} {\bibfnamefont {F.}~\bibnamefont
  {{L}esage}}, \bibinfo {author} {\bibfnamefont {S.}~\bibnamefont
  {{L}ukyanov}}, \ and\ \bibinfo {author} {\bibfnamefont {H.}~\bibnamefont
  {{S}aleur}},\ }\href@noop {} {\bibfield  {journal} {\bibinfo  {journal}
  {Phys. Lett. A},\ }\textbf {\bibinfo {volume} {235}},\ \bibinfo {pages} {203}
  (\bibinfo {year} {1997})}\BibitemShut {NoStop}%
\bibitem [{\citenamefont {{K}onik}\ and\ \citenamefont
  {{L}e{C}lair}(1998)}]{PhysRevB.58.1872}%
  \BibitemOpen
  \bibfield  {author} {\bibinfo {author} {\bibfnamefont {R.}~\bibnamefont
  {{K}onik}}\ and\ \bibinfo {author} {\bibfnamefont {A.}~\bibnamefont
  {{L}e{C}lair}},\ }\Doi {10.1103/PhysRevB.58.1872} {\bibfield  {journal}
  {\bibinfo  {journal} {Phys. Rev. B},\ }\textbf {\bibinfo {volume} {58}},\
  \bibinfo {pages} {1872} (\bibinfo {year} {1998})}\BibitemShut {NoStop}%
\bibitem [{\citenamefont {{Z}heng}\ \emph {et~al.}(2010)\citenamefont
  {{Z}heng}, \citenamefont {{G}authier},\ and\ \citenamefont
  {{B}aranger}}]{PhysRevA.82.063816}%
  \BibitemOpen
  \bibfield  {author} {\bibinfo {author} {\bibfnamefont {H.}~\bibnamefont
  {{Z}heng}}, \bibinfo {author} {\bibfnamefont {D.~J.}\ \bibnamefont
  {{G}authier}}, \ and\ \bibinfo {author} {\bibfnamefont {H.~U.}\ \bibnamefont
  {{B}aranger}},\ }\Doi {10.1103/PhysRevA.82.063816} {\bibfield  {journal}
  {\bibinfo  {journal} {Phys. Rev. A},\ }\textbf {\bibinfo {volume} {82}},\
  \bibinfo {pages} {063816} (\bibinfo {year} {2010})}\BibitemShut {NoStop}%
\bibitem [{\citenamefont {{L}e {H}ur}(2012)}]{PhysRevB.85.140506}%
  \BibitemOpen
  \bibfield  {author} {\bibinfo {author} {\bibfnamefont {K.}~\bibnamefont {{L}e
  {H}ur}},\ }\Doi {10.1103/PhysRevB.85.140506} {\bibfield  {journal} {\bibinfo
  {journal} {Phys. Rev. B},\ }\textbf {\bibinfo {volume} {85}},\ \bibinfo
  {pages} {140506} (\bibinfo {year} {2012})}\BibitemShut {NoStop}%
\bibitem [{\citenamefont {{G}oldstein}\ \emph {et~al.}(2013)\citenamefont
  {{G}oldstein}, \citenamefont {{D}evoret}, \citenamefont {{H}ouzet},\ and\
  \citenamefont {{G}lazman}}]{PhysRevLett.110.017002}%
  \BibitemOpen
  \bibfield  {author} {\bibinfo {author} {\bibfnamefont {M.}~\bibnamefont
  {{G}oldstein}}, \bibinfo {author} {\bibfnamefont {M.~H.}\ \bibnamefont
  {{D}evoret}}, \bibinfo {author} {\bibfnamefont {M.}~\bibnamefont {{H}ouzet}},
  \ and\ \bibinfo {author} {\bibfnamefont {L.~I.}\ \bibnamefont {{G}lazman}},\
  }\Doi {10.1103/PhysRevLett.110.017002} {\bibfield  {journal} {\bibinfo
  {journal} {Phys. Rev. Lett.},\ }\textbf {\bibinfo {volume} {110}},\ \bibinfo
  {pages} {017002} (\bibinfo {year} {2013})}\BibitemShut {NoStop}%
\bibitem [{\citenamefont {{H}oi}\ \emph {et~al.}(2012)\citenamefont {{H}oi},
  \citenamefont {{W}ilson}, \citenamefont {{J}ohansson}, \citenamefont
  {{L}indkvist}, \citenamefont {{P}eropadre}, \citenamefont {{P}alomaki},\ and\
  \citenamefont {{D}elsing}}]{Delsing-arXiv2012}%
  \BibitemOpen
  \bibfield  {author} {\bibinfo {author} {\bibfnamefont {I.-C.}\ \bibnamefont
  {{H}oi}}, \bibinfo {author} {\bibfnamefont {C.~M.}\ \bibnamefont {{W}ilson}},
  \bibinfo {author} {\bibfnamefont {G.}~\bibnamefont {{J}ohansson}}, \bibinfo
  {author} {\bibfnamefont {J.}~\bibnamefont {{L}indkvist}}, \bibinfo {author}
  {\bibfnamefont {B.}~\bibnamefont {{P}eropadre}}, \bibinfo {author}
  {\bibfnamefont {T.}~\bibnamefont {{P}alomaki}}, \ and\ \bibinfo {author}
  {\bibfnamefont {P.}~\bibnamefont {{D}elsing}},\ }\href@noop {} {\bibfield
  {journal} {\bibinfo  {journal} {arXiv:1210.4303}} (\bibinfo {year}
  {2012})}\BibitemShut {NoStop}%
\bibitem [{\citenamefont {{P}orras}\ \emph {et~al.}(2008)\citenamefont
  {{P}orras}, \citenamefont {{M}arquardt}, \citenamefont {von {D}elft},\ and\
  \citenamefont {{C}irac}}]{porras:010101}%
  \BibitemOpen
  \bibfield  {author} {\bibinfo {author} {\bibfnamefont {D.}~\bibnamefont
  {{P}orras}}, \bibinfo {author} {\bibfnamefont {F.}~\bibnamefont
  {{M}arquardt}}, \bibinfo {author} {\bibfnamefont {J.}~\bibnamefont {von
  {D}elft}}, \ and\ \bibinfo {author} {\bibfnamefont {J.~I.}\ \bibnamefont
  {{C}irac}},\ }\Doi {10.1103/PhysRevA.78.010101} {\bibfield  {journal}
  {\bibinfo  {journal} {Phys. Rev. A},\ }\textbf {\bibinfo {volume} {78}},\
  \bibinfo {eid} {010101(R)} (\bibinfo {year} {2008})}\BibitemShut {NoStop}%
\bibitem [{\citenamefont {{R}ecati}\ \emph {et~al.}(2005)\citenamefont
  {{R}ecati}, \citenamefont {{F}edichev}, \citenamefont {{Z}werger},
  \citenamefont {von {D}elft},\ and\ \citenamefont {{Z}oller}}]{recati:040404}%
  \BibitemOpen
  \bibfield  {author} {\bibinfo {author} {\bibfnamefont {A.}~\bibnamefont
  {{R}ecati}}, \bibinfo {author} {\bibfnamefont {P.~O.}\ \bibnamefont
  {{F}edichev}}, \bibinfo {author} {\bibfnamefont {W.}~\bibnamefont
  {{Z}werger}}, \bibinfo {author} {\bibfnamefont {J.}~\bibnamefont {von
  {D}elft}}, \ and\ \bibinfo {author} {\bibfnamefont {P.}~\bibnamefont
  {{Z}oller}},\ }\Doi {10.1103/PhysRevLett.94.040404} {\bibfield  {journal}
  {\bibinfo  {journal} {Phys. Rev. Lett.},\ }\textbf {\bibinfo {volume} {94}},\
  \bibinfo {eid} {040404} (\bibinfo {year} {2005})}\BibitemShut {NoStop}%
\bibitem [{\citenamefont {{O}rth}\ \emph {et~al.}(2008)\citenamefont {{O}rth},
  \citenamefont {{S}tanic},\ and\ \citenamefont {{L}e {H}ur}}]{orth:051601}%
  \BibitemOpen
  \bibfield  {author} {\bibinfo {author} {\bibfnamefont {P.~P.}\ \bibnamefont
  {{O}rth}}, \bibinfo {author} {\bibfnamefont {I.}~\bibnamefont {{S}tanic}}, \
  and\ \bibinfo {author} {\bibfnamefont {K.}~\bibnamefont {{L}e {H}ur}},\ }\Doi
  {10.1103/PhysRevA.77.051601} {\bibfield  {journal} {\bibinfo  {journal}
  {Phys. Rev. A},\ }\textbf {\bibinfo {volume} {77}},\ \bibinfo {eid}
  {051601(R)} (\bibinfo {year} {2008})}\BibitemShut {NoStop}%
\bibitem [{\citenamefont {{O}rth}\ \emph
  {et~al.}(2010){\natexlab{a}}\citenamefont {{O}rth}, \citenamefont {{R}oosen},
  \citenamefont {{H}ofstetter},\ and\ \citenamefont {{L}e
  {H}ur}}]{PhysRevB.82.144423}%
  \BibitemOpen
  \bibfield  {author} {\bibinfo {author} {\bibfnamefont {P.~P.}\ \bibnamefont
  {{O}rth}}, \bibinfo {author} {\bibfnamefont {D.}~\bibnamefont {{R}oosen}},
  \bibinfo {author} {\bibfnamefont {W.}~\bibnamefont {{H}ofstetter}}, \ and\
  \bibinfo {author} {\bibfnamefont {K.}~\bibnamefont {{L}e {H}ur}},\ }\Doi
  {10.1103/PhysRevB.82.144423} {\bibfield  {journal} {\bibinfo  {journal}
  {Phys. Rev. B},\ }\textbf {\bibinfo {volume} {82}},\ \bibinfo {pages}
  {144423} (\bibinfo {year} {2010}{\natexlab{a}})}\BibitemShut {NoStop}%
\bibitem [{\citenamefont {{L}amacraft}(2009)}]{PhysRevB.79.241105}%
  \BibitemOpen
  \bibfield  {author} {\bibinfo {author} {\bibfnamefont {A.}~\bibnamefont
  {{L}amacraft}},\ }\Doi {10.1103/PhysRevB.79.241105} {\bibfield  {journal}
  {\bibinfo  {journal} {Phys. Rev. B},\ }\textbf {\bibinfo {volume} {79}},\
  \bibinfo {pages} {241105} (\bibinfo {year} {2009})}\BibitemShut {NoStop}%
\bibitem [{\citenamefont {{S}ortais}\ \emph {et~al.}(2007)\citenamefont
  {{S}ortais}, \citenamefont {{M}arion}, \citenamefont {{T}uchendler},
  \citenamefont {{L}ance}, \citenamefont {{L}amare}, \citenamefont {{F}ournet},
  \citenamefont {{A}rmellin}, \citenamefont {{M}ercier}, \citenamefont
  {{M}essin}, \citenamefont {{B}rowaeys},\ and\ \citenamefont
  {{G}rangier}}]{PhysRevA.75.013406}%
  \BibitemOpen
  \bibfield  {author} {\bibinfo {author} {\bibfnamefont {Y.~R.~P.}\
  \bibnamefont {{S}ortais}}, \bibinfo {author} {\bibfnamefont {H.}~\bibnamefont
  {{M}arion}}, \bibinfo {author} {\bibfnamefont {C.}~\bibnamefont
  {{T}uchendler}}, \bibinfo {author} {\bibfnamefont {A.~M.}\ \bibnamefont
  {{L}ance}}, \bibinfo {author} {\bibfnamefont {M.}~\bibnamefont {{L}amare}},
  \bibinfo {author} {\bibfnamefont {P.}~\bibnamefont {{F}ournet}}, \bibinfo
  {author} {\bibfnamefont {C.}~\bibnamefont {{A}rmellin}}, \bibinfo {author}
  {\bibfnamefont {R.}~\bibnamefont {{M}ercier}}, \bibinfo {author}
  {\bibfnamefont {G.}~\bibnamefont {{M}essin}}, \bibinfo {author}
  {\bibfnamefont {A.}~\bibnamefont {{B}rowaeys}}, \ and\ \bibinfo {author}
  {\bibfnamefont {P.}~\bibnamefont {{G}rangier}},\ }\Doi
  {10.1103/PhysRevA.75.013406} {\bibfield  {journal} {\bibinfo  {journal}
  {Phys. Rev. A},\ }\textbf {\bibinfo {volume} {75}},\ \bibinfo {pages}
  {013406} (\bibinfo {year} {2007})}\BibitemShut {NoStop}%
\bibitem [{\citenamefont {{P}ertot}\ \emph {et~al.}(2010)\citenamefont
  {{P}ertot}, \citenamefont {{G}adway},\ and\ \citenamefont
  {{S}chneble}}]{PhysRevLett.104.200402}%
  \BibitemOpen
  \bibfield  {author} {\bibinfo {author} {\bibfnamefont {D.}~\bibnamefont
  {{P}ertot}}, \bibinfo {author} {\bibfnamefont {B.}~\bibnamefont {{G}adway}},
  \ and\ \bibinfo {author} {\bibfnamefont {D.}~\bibnamefont {{S}chneble}},\
  }\href@noop {} {\bibfield  {journal} {\bibinfo  {journal} {Phys. Rev.
  Lett.},\ }\textbf {\bibinfo {volume} {104}},\ \bibinfo {pages} {200402}
  (\bibinfo {year} {2010})}\BibitemShut {NoStop}%
\bibitem [{\citenamefont {{G}adway}\ \emph {et~al.}(2010)\citenamefont
  {{G}adway}, \citenamefont {{P}ertot}, \citenamefont {{R}eimann},\ and\
  \citenamefont {{S}chneble}}]{PhysRevLett.105.045303}%
  \BibitemOpen
  \bibfield  {author} {\bibinfo {author} {\bibfnamefont {B.}~\bibnamefont
  {{G}adway}}, \bibinfo {author} {\bibfnamefont {D.}~\bibnamefont {{P}ertot}},
  \bibinfo {author} {\bibfnamefont {R.}~\bibnamefont {{R}eimann}}, \ and\
  \bibinfo {author} {\bibfnamefont {D.}~\bibnamefont {{S}chneble}},\ }\Doi
  {10.1103/PhysRevLett.105.045303} {\bibfield  {journal} {\bibinfo  {journal}
  {Phys. Rev. Lett.},\ }\textbf {\bibinfo {volume} {105}},\ \bibinfo {pages}
  {045303} (\bibinfo {year} {2010})}\BibitemShut {NoStop}%
\bibitem [{\citenamefont {{G}adway}\ \emph {et~al.}(2011)\citenamefont
  {{G}adway}, \citenamefont {{P}ertot}, \citenamefont {{R}eeves}, \citenamefont
  {{V}ogt},\ and\ \citenamefont {{S}chneble}}]{PhysRevLett.107.145306}%
  \BibitemOpen
  \bibfield  {author} {\bibinfo {author} {\bibfnamefont {B.}~\bibnamefont
  {{G}adway}}, \bibinfo {author} {\bibfnamefont {D.}~\bibnamefont {{P}ertot}},
  \bibinfo {author} {\bibfnamefont {J.}~\bibnamefont {{R}eeves}}, \bibinfo
  {author} {\bibfnamefont {M.}~\bibnamefont {{V}ogt}}, \ and\ \bibinfo {author}
  {\bibfnamefont {D.}~\bibnamefont {{S}chneble}},\ }\Doi
  {10.1103/PhysRevLett.107.145306} {\bibfield  {journal} {\bibinfo  {journal}
  {Phys. Rev. Lett.},\ }\textbf {\bibinfo {volume} {107}},\ \bibinfo {pages}
  {145306} (\bibinfo {year} {2011})}\BibitemShut {NoStop}%
\bibitem [{\citenamefont {{W}eitenberg}\ \emph {et~al.}(2011)\citenamefont
  {{W}eitenberg}, \citenamefont {{E}ndres}, \citenamefont {{S}herson},
  \citenamefont {{C}heneau}, \citenamefont {{S}chauss}, \citenamefont
  {{F}ukuhara}, \citenamefont {{B}loch},\ and\ \citenamefont
  {{K}uhr}}]{WeitenbergBlochKuhr-Nature-2011}%
  \BibitemOpen
  \bibfield  {author} {\bibinfo {author} {\bibfnamefont {C.}~\bibnamefont
  {{W}eitenberg}}, \bibinfo {author} {\bibfnamefont {M.}~\bibnamefont
  {{E}ndres}}, \bibinfo {author} {\bibfnamefont {J.~F.}\ \bibnamefont
  {{S}herson}}, \bibinfo {author} {\bibfnamefont {M.}~\bibnamefont
  {{C}heneau}}, \bibinfo {author} {\bibfnamefont {P.}~\bibnamefont
  {{S}chauss}}, \bibinfo {author} {\bibfnamefont {T.}~\bibnamefont
  {{F}ukuhara}}, \bibinfo {author} {\bibfnamefont {I.}~\bibnamefont {{B}loch}},
  \ and\ \bibinfo {author} {\bibfnamefont {S.}~\bibnamefont {{K}uhr}},\
  }\href@noop {} {\bibfield  {journal} {\bibinfo  {journal} {Nature},\ }\textbf
  {\bibinfo {volume} {471}},\ \bibinfo {pages} {319} (\bibinfo {year}
  {2011})}\BibitemShut {NoStop}%
\bibitem [{\citenamefont {{N}esi}\ \emph {et~al.}(2007)\citenamefont {{N}esi},
  \citenamefont {{P}aladino}, \citenamefont {{T}horwart},\ and\ \citenamefont
  {{G}rifoni}}]{0295-5075-80-4-40005}%
  \BibitemOpen
  \bibfield  {author} {\bibinfo {author} {\bibfnamefont {F.}~\bibnamefont
  {{N}esi}}, \bibinfo {author} {\bibfnamefont {E.}~\bibnamefont {{P}aladino}},
  \bibinfo {author} {\bibfnamefont {M.}~\bibnamefont {{T}horwart}}, \ and\
  \bibinfo {author} {\bibfnamefont {M.}~\bibnamefont {{G}rifoni}},\ }\href@noop
  {} {\bibfield  {journal} {\bibinfo  {journal} {Europhys. Lett.},\ }\textbf
  {\bibinfo {volume} {80}},\ \bibinfo {pages} {40005} (\bibinfo {year}
  {2007})}\BibitemShut {NoStop}%
\bibitem [{\citenamefont {{G}\"orlich}\ \emph {et~al.}(1989)\citenamefont
  {{G}\"orlich}, \citenamefont {{S}assetti},\ and\ \citenamefont
  {{W}eiss}}]{goerlich_low-temperature_1989}%
  \BibitemOpen
  \bibfield  {author} {\bibinfo {author} {\bibfnamefont {R.}~\bibnamefont
  {{G}\"orlich}}, \bibinfo {author} {\bibfnamefont {M.}~\bibnamefont
  {{S}assetti}}, \ and\ \bibinfo {author} {\bibfnamefont {U.}~\bibnamefont
  {{W}eiss}},\ }\Doi {10.1209/0295-5075/10/6/001} {\bibfield  {journal}
  {\bibinfo  {journal} {Europhys. Lett.},\ }\textbf {\bibinfo {volume} {10}},\
  \bibinfo {pages} {507} (\bibinfo {year} {1989})},\ ISSN \bibinfo {issn}
  {0295-5075}\BibitemShut {NoStop}%
\bibitem [{\citenamefont {{W}eiss}\ and\ \citenamefont
  {{W}ollensak}(1989)}]{weiss_dynamics_1989}%
  \BibitemOpen
  \bibfield  {author} {\bibinfo {author} {\bibfnamefont {U.}~\bibnamefont
  {{W}eiss}}\ and\ \bibinfo {author} {\bibfnamefont {M.}~\bibnamefont
  {{W}ollensak}},\ }\Doi {10.1103/PhysRevLett.62.1663} {\bibfield  {journal}
  {\bibinfo  {journal} {Phys. Rev. Lett.},\ }\textbf {\bibinfo {volume} {62}},\
  \bibinfo {pages} {1663} (\bibinfo {year} {1989})}\BibitemShut {NoStop}%
\bibitem [{\citenamefont {{K}och}\ \emph {et~al.}(2008)\citenamefont {{K}och},
  \citenamefont {{G}ro\ss mann}, \citenamefont {{S}tockburger},\ and\
  \citenamefont {{A}nkerhold}}]{koch:230402}%
  \BibitemOpen
  \bibfield  {author} {\bibinfo {author} {\bibfnamefont {W.}~\bibnamefont
  {{K}och}}, \bibinfo {author} {\bibfnamefont {F.}~\bibnamefont {{G}ro\ss
  mann}}, \bibinfo {author} {\bibfnamefont {J.~T.}\ \bibnamefont
  {{S}tockburger}}, \ and\ \bibinfo {author} {\bibfnamefont {J.}~\bibnamefont
  {{A}nkerhold}},\ }\href@noop {} {\bibfield  {journal} {\bibinfo  {journal}
  {Phys. Rev. Lett.},\ }\textbf {\bibinfo {volume} {100}},\ \bibinfo {eid}
  {230402} (\bibinfo {year} {2008})}\BibitemShut {NoStop}%
\bibitem [{\citenamefont {{H}artmann}\ \emph {et~al.}(2000)\citenamefont
  {{H}artmann}, \citenamefont {{G}oychuk}, \citenamefont {{G}rifoni},\ and\
  \citenamefont {{H}\"anggi}}]{PhysRevE.61.R4687}%
  \BibitemOpen
  \bibfield  {author} {\bibinfo {author} {\bibfnamefont {L.}~\bibnamefont
  {{H}artmann}}, \bibinfo {author} {\bibfnamefont {I.}~\bibnamefont
  {{G}oychuk}}, \bibinfo {author} {\bibfnamefont {M.}~\bibnamefont
  {{G}rifoni}}, \ and\ \bibinfo {author} {\bibfnamefont {P.}~\bibnamefont
  {{H}\"anggi}},\ }\href@noop {} {\bibfield  {journal} {\bibinfo  {journal}
  {Phys. Rev. E},\ }\textbf {\bibinfo {volume} {61}},\ \bibinfo {eid} {4687(R)}
  (\bibinfo {year} {2000})}\BibitemShut {NoStop}%
\bibitem [{\citenamefont {{D}i{V}incenzo}\ and\ \citenamefont
  {{L}oss}(2005)}]{PhysRevB.71.035318}%
  \BibitemOpen
  \bibfield  {author} {\bibinfo {author} {\bibfnamefont {D.~P.}\ \bibnamefont
  {{D}i{V}incenzo}}\ and\ \bibinfo {author} {\bibfnamefont {D.}~\bibnamefont
  {{L}oss}},\ }\Doi {10.1103/PhysRevB.71.035318} {\bibfield  {journal}
  {\bibinfo  {journal} {Phys. Rev. B},\ }\textbf {\bibinfo {volume} {71}},\
  \bibinfo {pages} {035318} (\bibinfo {year} {2005})}\BibitemShut {NoStop}%
\bibitem [{\citenamefont {{B}reuer}\ and\ \citenamefont
  {{P}etruccione}(2002)}]{BreuerPetruccione-Book}%
  \BibitemOpen
  \bibfield  {author} {\bibinfo {author} {\bibfnamefont {H.-P.}\ \bibnamefont
  {{B}reuer}}\ and\ \bibinfo {author} {\bibfnamefont {F.}~\bibnamefont
  {{P}etruccione}},\ }\href@noop {} {\emph {\bibinfo {title} {{T}he theory of
  open quantum systems}}}\ (\bibinfo  {publisher} {Oxford University Press},\
  \bibinfo {year} {2002})\BibitemShut {NoStop}%
\bibitem [{\citenamefont {{W}hitney}\ \emph {et~al.}(2011)\citenamefont
  {{W}hitney}, \citenamefont {{C}lusel},\ and\ \citenamefont
  {{Z}iman}}]{PhysRevLett.107.210402}%
  \BibitemOpen
  \bibfield  {author} {\bibinfo {author} {\bibfnamefont {R.~S.}\ \bibnamefont
  {{W}hitney}}, \bibinfo {author} {\bibfnamefont {M.}~\bibnamefont {{C}lusel}},
  \ and\ \bibinfo {author} {\bibfnamefont {T.}~\bibnamefont {{Z}iman}},\ }\Doi
  {10.1103/PhysRevLett.107.210402} {\bibfield  {journal} {\bibinfo  {journal}
  {Phys. Rev. Lett.},\ }\textbf {\bibinfo {volume} {107}},\ \bibinfo {pages}
  {210402} (\bibinfo {year} {2011})}\BibitemShut {NoStop}%
\bibitem [{\citenamefont {{K}amleitner}\ and\ \citenamefont
  {{S}hnirman}(2011)}]{PhysRevB.84.235140}%
  \BibitemOpen
  \bibfield  {author} {\bibinfo {author} {\bibfnamefont {I.}~\bibnamefont
  {{K}amleitner}}\ and\ \bibinfo {author} {\bibfnamefont {A.}~\bibnamefont
  {{S}hnirman}},\ }\Doi {10.1103/PhysRevB.84.235140} {\bibfield  {journal}
  {\bibinfo  {journal} {Phys. Rev. B},\ }\textbf {\bibinfo {volume} {84}},\
  \bibinfo {pages} {235140} (\bibinfo {year} {2011})}\BibitemShut {NoStop}%
\bibitem [{\citenamefont {{S}cala}\ \emph {et~al.}(2011)\citenamefont
  {{S}cala}, \citenamefont {{M}ilitello}, \citenamefont {{M}essina},\ and\
  \citenamefont {{V}itanov}}]{PhysRevA.84.023416}%
  \BibitemOpen
  \bibfield  {author} {\bibinfo {author} {\bibfnamefont {M.}~\bibnamefont
  {{S}cala}}, \bibinfo {author} {\bibfnamefont {B.}~\bibnamefont
  {{M}ilitello}}, \bibinfo {author} {\bibfnamefont {A.}~\bibnamefont
  {{M}essina}}, \ and\ \bibinfo {author} {\bibfnamefont {N.~V.}\ \bibnamefont
  {{V}itanov}},\ }\Doi {10.1103/PhysRevA.84.023416} {\bibfield  {journal}
  {\bibinfo  {journal} {Phys. Rev. A},\ }\textbf {\bibinfo {volume} {84}},\
  \bibinfo {pages} {023416} (\bibinfo {year} {2011})}\BibitemShut {NoStop}%
\bibitem [{\citenamefont {{N}azir}\ \emph {et~al.}(2012)\citenamefont
  {{N}azir}, \citenamefont {{M}c{C}utcheon},\ and\ \citenamefont
  {{C}hin}}]{PhysRevB.85.224301}%
  \BibitemOpen
  \bibfield  {author} {\bibinfo {author} {\bibfnamefont {A.}~\bibnamefont
  {{N}azir}}, \bibinfo {author} {\bibfnamefont {D.~P.~S.}\ \bibnamefont
  {{M}c{C}utcheon}}, \ and\ \bibinfo {author} {\bibfnamefont {A.~W.}\
  \bibnamefont {{C}hin}},\ }\Doi {10.1103/PhysRevB.85.224301} {\bibfield
  {journal} {\bibinfo  {journal} {Phys. Rev. B},\ }\textbf {\bibinfo {volume}
  {85}},\ \bibinfo {pages} {224301} (\bibinfo {year} {2012})}\BibitemShut
  {NoStop}%
\bibitem [{\citenamefont {{E}gger}\ and\ \citenamefont
  {{W}eiss}(1992)}]{springerlink:10.1007/BF01320834}%
  \BibitemOpen
  \bibfield  {author} {\bibinfo {author} {\bibfnamefont {R.}~\bibnamefont
  {{E}gger}}\ and\ \bibinfo {author} {\bibfnamefont {U.}~\bibnamefont
  {{W}eiss}},\ }\href {http://dx.doi.org/10.1007/BF01320834} {\bibfield
  {journal} {\bibinfo  {journal} {Z. Phys. B},\ }\textbf {\bibinfo {volume}
  {89}},\ \bibinfo {pages} {97} (\bibinfo {year} {1992})}\BibitemShut {NoStop}%
\bibitem [{\citenamefont {{E}gger}\ and\ \citenamefont
  {{M}ak}(1994)}]{PhysRevB.50.15210}%
  \BibitemOpen
  \bibfield  {author} {\bibinfo {author} {\bibfnamefont {R.}~\bibnamefont
  {{E}gger}}\ and\ \bibinfo {author} {\bibfnamefont {C.~H.}\ \bibnamefont
  {{M}ak}},\ }\Doi {10.1103/PhysRevB.50.15210} {\bibfield  {journal} {\bibinfo
  {journal} {Phys. Rev. B},\ }\textbf {\bibinfo {volume} {50}},\ \bibinfo
  {pages} {15210} (\bibinfo {year} {1994})}\BibitemShut {NoStop}%
\bibitem [{\citenamefont {{M}akarov}\ and\ \citenamefont
  {{M}akri}(1994)}]{Makri-QUAPI-ChemPhysLett-1994}%
  \BibitemOpen
  \bibfield  {author} {\bibinfo {author} {\bibfnamefont {D.~E.}\ \bibnamefont
  {{M}akarov}}\ and\ \bibinfo {author} {\bibfnamefont {N.}~\bibnamefont
  {{M}akri}},\ }\href@noop {} {\bibfield  {journal} {\bibinfo  {journal} {Chem.
  Phys. Lett.},\ }\textbf {\bibinfo {volume} {221}},\ \bibinfo {pages} {482}
  (\bibinfo {year} {1994})}\BibitemShut {NoStop}%
\bibitem [{\citenamefont {{M}akri}(1995)}]{makri_numerical_1995}%
  \BibitemOpen
  \bibfield  {author} {\bibinfo {author} {\bibfnamefont {N.}~\bibnamefont
  {{M}akri}},\ }\Doi {10.1063/1.531046} {\bibfield  {journal} {\bibinfo
  {journal} {J. Math. Phys.},\ }\textbf {\bibinfo {volume} {36}},\ \bibinfo
  {pages} {2430} (\bibinfo {year} {1995})}\BibitemShut {NoStop}%
\bibitem [{\citenamefont {{G}rifoni}\ and\ \citenamefont
  {{H}\"anggi}(1998)}]{grifoni_driven_1998}%
  \BibitemOpen
  \bibfield  {author} {\bibinfo {author} {\bibfnamefont {M.}~\bibnamefont
  {{G}rifoni}}\ and\ \bibinfo {author} {\bibfnamefont {P.}~\bibnamefont
  {{H}\"anggi}},\ }\Doi {16/S0370-1573(98)00022-2} {\bibfield  {journal}
  {\bibinfo  {journal} {Phys. Rep.},\ }\textbf {\bibinfo {volume} {304}},\
  \bibinfo {pages} {229} (\bibinfo {year} {1998})}\BibitemShut {NoStop}%
\bibitem [{\citenamefont {{W}ang}\ and\ \citenamefont
  {{T}hoss}(2008)}]{1367-2630-10-11-115005}%
  \BibitemOpen
  \bibfield  {author} {\bibinfo {author} {\bibfnamefont {H.}~\bibnamefont
  {{W}ang}}\ and\ \bibinfo {author} {\bibfnamefont {M.}~\bibnamefont
  {{T}hoss}},\ }\href@noop {} {\bibfield  {journal} {\bibinfo  {journal} {New
  J. Phys.},\ }\textbf {\bibinfo {volume} {10}},\ \bibinfo {pages} {115005}
  (\bibinfo {year} {2008})}\BibitemShut {NoStop}%
\bibitem [{\citenamefont {{N}albach}\ and\ \citenamefont
  {{T}horwart}(2009)}]{nalbach:220401}%
  \BibitemOpen
  \bibfield  {author} {\bibinfo {author} {\bibfnamefont {P.}~\bibnamefont
  {{N}albach}}\ and\ \bibinfo {author} {\bibfnamefont {M.}~\bibnamefont
  {{T}horwart}},\ }\href@noop {} {\bibfield  {journal} {\bibinfo  {journal}
  {Phys. Rev. Lett.},\ }\textbf {\bibinfo {volume} {103}},\ \bibinfo {eid}
  {220401} (\bibinfo {year} {2009})}\BibitemShut {NoStop}%
\bibitem [{\citenamefont {{W}eiss}\ \emph {et~al.}(2008)\citenamefont
  {{W}eiss}, \citenamefont {{E}ckel}, \citenamefont {{T}horwart},\ and\
  \citenamefont {{E}gger}}]{PhysRevB.77.195316}%
  \BibitemOpen
  \bibfield  {author} {\bibinfo {author} {\bibfnamefont {S.}~\bibnamefont
  {{W}eiss}}, \bibinfo {author} {\bibfnamefont {J.}~\bibnamefont {{E}ckel}},
  \bibinfo {author} {\bibfnamefont {M.}~\bibnamefont {{T}horwart}}, \ and\
  \bibinfo {author} {\bibfnamefont {R.}~\bibnamefont {{E}gger}},\ }\Doi
  {10.1103/PhysRevB.77.195316} {\bibfield  {journal} {\bibinfo  {journal}
  {Phys. Rev. B},\ }\textbf {\bibinfo {volume} {77}},\ \bibinfo {pages}
  {195316} (\bibinfo {year} {2008})}\BibitemShut {NoStop}%
\bibitem [{\citenamefont {{A}lvermann}\ and\ \citenamefont
  {{F}ehske}(2009)}]{PhysRevLett.102.150601}%
  \BibitemOpen
  \bibfield  {author} {\bibinfo {author} {\bibfnamefont {A.}~\bibnamefont
  {{A}lvermann}}\ and\ \bibinfo {author} {\bibfnamefont {H.}~\bibnamefont
  {{F}ehske}},\ }\Doi {10.1103/PhysRevLett.102.150601} {\bibfield  {journal}
  {\bibinfo  {journal} {Phys. Rev. Lett.},\ }\textbf {\bibinfo {volume}
  {102}},\ \bibinfo {pages} {150601} (\bibinfo {year} {2009})}\BibitemShut
  {NoStop}%
\bibitem [{\citenamefont {{W}u}\ \emph {et~al.}(2012)\citenamefont {{W}u},
  \citenamefont {{Y}u},\ and\ \citenamefont {{S}egal}}]{arXiv:1207.6995}%
  \BibitemOpen
  \bibfield  {author} {\bibinfo {author} {\bibfnamefont {L.-A.}\ \bibnamefont
  {{W}u}}, \bibinfo {author} {\bibfnamefont {C.~X.}\ \bibnamefont {{Y}u}}, \
  and\ \bibinfo {author} {\bibfnamefont {D.}~\bibnamefont {{S}egal}},\
  }\href@noop {} {\bibfield  {journal} {\bibinfo  {journal} {arXiv:1207.6995}}
  (\bibinfo {year} {2012})}\BibitemShut {NoStop}%
\bibitem [{\citenamefont {{K}ast}\ and\ \citenamefont
  {{A}nkerhold}(2013)}]{PhysRevLett.110.010402}%
  \BibitemOpen
  \bibfield  {author} {\bibinfo {author} {\bibfnamefont {D.}~\bibnamefont
  {{K}ast}}\ and\ \bibinfo {author} {\bibfnamefont {J.}~\bibnamefont
  {{A}nkerhold}},\ }\Doi {10.1103/PhysRevLett.110.010402} {\bibfield  {journal}
  {\bibinfo  {journal} {Phys. Rev. Lett.},\ }\textbf {\bibinfo {volume}
  {110}},\ \bibinfo {pages} {010402} (\bibinfo {year} {2013})}\BibitemShut
  {NoStop}%
\bibitem [{\citenamefont {{G}ull}\ \emph {et~al.}(2011)\citenamefont {{G}ull},
  \citenamefont {{M}illis}, \citenamefont {{L}ichtenstein}, \citenamefont
  {{R}ubtsov}, \citenamefont {{T}royer},\ and\ \citenamefont
  {{W}erner}}]{RevModPhys.83.349}%
  \BibitemOpen
  \bibfield  {author} {\bibinfo {author} {\bibfnamefont {E.}~\bibnamefont
  {{G}ull}}, \bibinfo {author} {\bibfnamefont {A.~J.}\ \bibnamefont
  {{M}illis}}, \bibinfo {author} {\bibfnamefont {A.~I.}\ \bibnamefont
  {{L}ichtenstein}}, \bibinfo {author} {\bibfnamefont {A.~N.}\ \bibnamefont
  {{R}ubtsov}}, \bibinfo {author} {\bibfnamefont {M.}~\bibnamefont {{T}royer}},
  \ and\ \bibinfo {author} {\bibfnamefont {P.}~\bibnamefont {{W}erner}},\ }\Doi
  {10.1103/RevModPhys.83.349} {\bibfield  {journal} {\bibinfo  {journal} {Rev.
  Mod. Phys.},\ }\textbf {\bibinfo {volume} {83}},\ \bibinfo {pages} {349}
  (\bibinfo {year} {2011})}\BibitemShut {NoStop}%
\bibitem [{\citenamefont {{S}chmidt}\ \emph {et~al.}(2008)\citenamefont
  {{S}chmidt}, \citenamefont {{W}erner}, \citenamefont {{M}\"uhlbacher},\ and\
  \citenamefont {{K}omnik}}]{PhysRevB.78.235110}%
  \BibitemOpen
  \bibfield  {author} {\bibinfo {author} {\bibfnamefont {T.~L.}\ \bibnamefont
  {{S}chmidt}}, \bibinfo {author} {\bibfnamefont {P.}~\bibnamefont {{W}erner}},
  \bibinfo {author} {\bibfnamefont {L.}~\bibnamefont {{M}\"uhlbacher}}, \ and\
  \bibinfo {author} {\bibfnamefont {A.}~\bibnamefont {{K}omnik}},\ }\href@noop
  {} {\bibfield  {journal} {\bibinfo  {journal} {Phys. Rev. B},\ }\textbf
  {\bibinfo {volume} {78}},\ \bibinfo {pages} {235110} (\bibinfo {year}
  {2008})}\BibitemShut {NoStop}%
\bibitem [{\citenamefont {{S}chir\'o}\ and\ \citenamefont
  {{F}abrizio}(2009)}]{PhysRevB.79.153302}%
  \BibitemOpen
  \bibfield  {author} {\bibinfo {author} {\bibfnamefont {M.}~\bibnamefont
  {{S}chir\'o}}\ and\ \bibinfo {author} {\bibfnamefont {M.}~\bibnamefont
  {{F}abrizio}},\ }\Doi {10.1103/PhysRevB.79.153302} {\bibfield  {journal}
  {\bibinfo  {journal} {Phys. Rev. B},\ }\textbf {\bibinfo {volume} {79}},\
  \bibinfo {pages} {153302} (\bibinfo {year} {2009})}\BibitemShut {NoStop}%
\bibitem [{\citenamefont {{W}erner}\ \emph {et~al.}(2010)\citenamefont
  {{W}erner}, \citenamefont {{O}ka}, \citenamefont {{E}ckstein},\ and\
  \citenamefont {{M}illis}}]{PhysRevB.81.035108}%
  \BibitemOpen
  \bibfield  {author} {\bibinfo {author} {\bibfnamefont {P.}~\bibnamefont
  {{W}erner}}, \bibinfo {author} {\bibfnamefont {T.}~\bibnamefont {{O}ka}},
  \bibinfo {author} {\bibfnamefont {M.}~\bibnamefont {{E}ckstein}}, \ and\
  \bibinfo {author} {\bibfnamefont {A.~J.}\ \bibnamefont {{M}illis}},\ }\Doi
  {10.1103/PhysRevB.81.035108} {\bibfield  {journal} {\bibinfo  {journal}
  {Phys. Rev. B},\ }\textbf {\bibinfo {volume} {81}},\ \bibinfo {pages}
  {035108} (\bibinfo {year} {2010})}\BibitemShut {NoStop}%
\bibitem [{\citenamefont {{A}nders}\ and\ \citenamefont
  {{S}chiller}(2006)}]{PhysRevB.74.245113}%
  \BibitemOpen
  \bibfield  {author} {\bibinfo {author} {\bibfnamefont {F.~B.}\ \bibnamefont
  {{A}nders}}\ and\ \bibinfo {author} {\bibfnamefont {A.}~\bibnamefont
  {{S}chiller}},\ }\href@noop {} {\bibfield  {journal} {\bibinfo  {journal}
  {Phys. Rev. B},\ }\textbf {\bibinfo {volume} {74}},\ \bibinfo {pages}
  {245113} (\bibinfo {year} {2006})}\BibitemShut {NoStop}%
\bibitem [{\citenamefont {{B}ulla}\ \emph {et~al.}(2008)\citenamefont
  {{B}ulla}, \citenamefont {{C}osti},\ and\ \citenamefont
  {{P}ruschke}}]{RevModPhys.80.395}%
  \BibitemOpen
  \bibfield  {author} {\bibinfo {author} {\bibfnamefont {R.}~\bibnamefont
  {{B}ulla}}, \bibinfo {author} {\bibfnamefont {T.~A.}\ \bibnamefont
  {{C}osti}}, \ and\ \bibinfo {author} {\bibfnamefont {T.}~\bibnamefont
  {{P}ruschke}},\ }\href@noop {} {\bibfield  {journal} {\bibinfo  {journal}
  {Rev. Mod. Phys.},\ }\textbf {\bibinfo {volume} {80}},\ \bibinfo {pages}
  {395} (\bibinfo {year} {2008})}\BibitemShut {NoStop}%
\bibitem [{\citenamefont {{B}ulla}\ \emph {et~al.}(2005)\citenamefont
  {{B}ulla}, \citenamefont {{L}ee}, \citenamefont {{T}ong},\ and\ \citenamefont
  {{V}ojta}}]{PhysRevB.71.045122}%
  \BibitemOpen
  \bibfield  {author} {\bibinfo {author} {\bibfnamefont {R.}~\bibnamefont
  {{B}ulla}}, \bibinfo {author} {\bibfnamefont {H.-J.}\ \bibnamefont {{L}ee}},
  \bibinfo {author} {\bibfnamefont {N.-H.}\ \bibnamefont {{T}ong}}, \ and\
  \bibinfo {author} {\bibfnamefont {M.}~\bibnamefont {{V}ojta}},\ }\href@noop
  {} {\bibfield  {journal} {\bibinfo  {journal} {Phys. Rev. B},\ }\textbf
  {\bibinfo {volume} {71}},\ \bibinfo {pages} {045122} (\bibinfo {year}
  {2005})}\BibitemShut {NoStop}%
\bibitem [{\citenamefont {{L}i}\ \emph {et~al.}(2005)\citenamefont {{L}i},
  \citenamefont {{L}e {H}ur},\ and\ \citenamefont
  {{H}ofstetter}}]{PhysRevLett.95.086406}%
  \BibitemOpen
  \bibfield  {author} {\bibinfo {author} {\bibfnamefont {M.-R.}\ \bibnamefont
  {{L}i}}, \bibinfo {author} {\bibfnamefont {K.}~\bibnamefont {{L}e {H}ur}}, \
  and\ \bibinfo {author} {\bibfnamefont {W.}~\bibnamefont {{H}ofstetter}},\
  }\href@noop {} {\bibfield  {journal} {\bibinfo  {journal} {Phys. Rev.
  Lett.},\ }\textbf {\bibinfo {volume} {95}},\ \bibinfo {pages} {086406}
  (\bibinfo {year} {2005})}\BibitemShut {NoStop}%
\bibitem [{\citenamefont {{L}e {H}ur}(2004)}]{PhysRevLett.92.196804}%
  \BibitemOpen
  \bibfield  {author} {\bibinfo {author} {\bibfnamefont {K.}~\bibnamefont {{L}e
  {H}ur}},\ }\Doi {10.1103/PhysRevLett.92.196804} {\bibfield  {journal}
  {\bibinfo  {journal} {Phys. Rev. Lett.},\ }\textbf {\bibinfo {volume} {92}},\
  \bibinfo {pages} {196804} (\bibinfo {year} {2004})}\BibitemShut {NoStop}%
\bibitem [{\citenamefont {{S}chmitteckert}(2004)}]{PhysRevB.70.121302}%
  \BibitemOpen
  \bibfield  {author} {\bibinfo {author} {\bibfnamefont {P.}~\bibnamefont
  {{S}chmitteckert}},\ }\Doi {10.1103/PhysRevB.70.121302} {\bibfield  {journal}
  {\bibinfo  {journal} {Phys. Rev. B},\ }\textbf {\bibinfo {volume} {70}},\
  \bibinfo {pages} {121302} (\bibinfo {year} {2004})}\BibitemShut {NoStop}%
\bibitem [{\citenamefont {{F}lorens}\ \emph {et~al.}(2011)\citenamefont
  {{F}lorens}, \citenamefont {{F}reyn}, \citenamefont {{V}enturelli},\ and\
  \citenamefont {{N}arayanan}}]{PhysRevB.84.155110}%
  \BibitemOpen
  \bibfield  {author} {\bibinfo {author} {\bibfnamefont {S.}~\bibnamefont
  {{F}lorens}}, \bibinfo {author} {\bibfnamefont {A.}~\bibnamefont {{F}reyn}},
  \bibinfo {author} {\bibfnamefont {D.}~\bibnamefont {{V}enturelli}}, \ and\
  \bibinfo {author} {\bibfnamefont {R.}~\bibnamefont {{N}arayanan}},\ }\Doi
  {10.1103/PhysRevB.84.155110} {\bibfield  {journal} {\bibinfo  {journal}
  {Phys. Rev. B},\ }\textbf {\bibinfo {volume} {84}},\ \bibinfo {pages}
  {155110} (\bibinfo {year} {2011})}\BibitemShut {NoStop}%
\bibitem [{\citenamefont {{K}eil}\ and\ \citenamefont
  {{S}choeller}(2001)}]{PhysRevB.63.180302}%
  \BibitemOpen
  \bibfield  {author} {\bibinfo {author} {\bibfnamefont {M.}~\bibnamefont
  {{K}eil}}\ and\ \bibinfo {author} {\bibfnamefont {H.}~\bibnamefont
  {{S}choeller}},\ }\Doi {10.1103/PhysRevB.63.180302} {\bibfield  {journal}
  {\bibinfo  {journal} {Phys. Rev. B},\ }\textbf {\bibinfo {volume} {63}},\
  \bibinfo {pages} {180302} (\bibinfo {year} {2001})}\BibitemShut {NoStop}%
\bibitem [{\citenamefont {{S}choeller}(2009)}]{epjst/e2009-00962-3}%
  \BibitemOpen
  \bibfield  {author} {\bibinfo {author} {\bibfnamefont {H.}~\bibnamefont
  {{S}choeller}},\ }\href {http://dx.doi.org/10.1140/epjst/e2009-00962-3}
  {\bibfield  {journal} {\bibinfo  {journal} {EPJ ST},\ }\textbf {\bibinfo
  {volume} {168}},\ \bibinfo {pages} {179} (\bibinfo {year}
  {2009})}\BibitemShut {NoStop}%
\bibitem [{\citenamefont {{P}letyukhov}\ \emph {et~al.}(2010)\citenamefont
  {{P}letyukhov}, \citenamefont {{S}churicht},\ and\ \citenamefont
  {{S}choeller}}]{PhysRevLett.104.106801}%
  \BibitemOpen
  \bibfield  {author} {\bibinfo {author} {\bibfnamefont {M.}~\bibnamefont
  {{P}letyukhov}}, \bibinfo {author} {\bibfnamefont {D.}~\bibnamefont
  {{S}churicht}}, \ and\ \bibinfo {author} {\bibfnamefont {H.}~\bibnamefont
  {{S}choeller}},\ }\Doi {10.1103/PhysRevLett.104.106801} {\bibfield  {journal}
  {\bibinfo  {journal} {Phys. Rev. Lett.},\ }\textbf {\bibinfo {volume}
  {104}},\ \bibinfo {pages} {106801} (\bibinfo {year} {2010})}\BibitemShut
  {NoStop}%
\bibitem [{\citenamefont {{A}ndergassen}\ \emph {et~al.}(2011)\citenamefont
  {{A}ndergassen}, \citenamefont {{P}letyukhov}, \citenamefont {{S}churicht},
  \citenamefont {{S}choeller},\ and\ \citenamefont
  {{B}orda}}]{andergassen_renormalization_2011}%
  \BibitemOpen
  \bibfield  {author} {\bibinfo {author} {\bibfnamefont {S.}~\bibnamefont
  {{A}ndergassen}}, \bibinfo {author} {\bibfnamefont {M.}~\bibnamefont
  {{P}letyukhov}}, \bibinfo {author} {\bibfnamefont {D.}~\bibnamefont
  {{S}churicht}}, \bibinfo {author} {\bibfnamefont {H.}~\bibnamefont
  {{S}choeller}}, \ and\ \bibinfo {author} {\bibfnamefont {L.}~\bibnamefont
  {{B}orda}},\ }\Doi {10.1103/PhysRevB.83.205103} {\bibfield  {journal}
  {\bibinfo  {journal} {Phys. Rev. B},\ }\textbf {\bibinfo {volume} {83}},\
  \bibinfo {pages} {205103} (\bibinfo {year} {2011})}\BibitemShut {NoStop}%
\bibitem [{\citenamefont {{E}ckstein}\ \emph {et~al.}(2009)\citenamefont
  {{E}ckstein}, \citenamefont {{H}ackl}, \citenamefont {{K}ehrein},
  \citenamefont {{K}ollar}, \citenamefont {{M}oeckel}, \citenamefont
  {{W}erner},\ and\ \citenamefont
  {{W}olf}}]{springerlink:10.1140/epjst/e2010-01219-x}%
  \BibitemOpen
  \bibfield  {author} {\bibinfo {author} {\bibfnamefont {M.}~\bibnamefont
  {{E}ckstein}}, \bibinfo {author} {\bibfnamefont {A.}~\bibnamefont {{H}ackl}},
  \bibinfo {author} {\bibfnamefont {S.}~\bibnamefont {{K}ehrein}}, \bibinfo
  {author} {\bibfnamefont {M.}~\bibnamefont {{K}ollar}}, \bibinfo {author}
  {\bibfnamefont {M.}~\bibnamefont {{M}oeckel}}, \bibinfo {author}
  {\bibfnamefont {P.}~\bibnamefont {{W}erner}}, \ and\ \bibinfo {author}
  {\bibfnamefont {F.}~\bibnamefont {{W}olf}},\ }\href
  {http://dx.doi.org/10.1140/epjst/e2010-01219-x} {\bibfield  {journal}
  {\bibinfo  {journal} {EPJ ST},\ }\textbf {\bibinfo {volume} {180}},\ \bibinfo
  {pages} {217} (\bibinfo {year} {2009})}\BibitemShut {NoStop}%
\bibitem [{\citenamefont {{H}ackl}\ and\ \citenamefont
  {{K}ehrein}(2008)}]{PhysRevB.78.092303}%
  \BibitemOpen
  \bibfield  {author} {\bibinfo {author} {\bibfnamefont {A.}~\bibnamefont
  {{H}ackl}}\ and\ \bibinfo {author} {\bibfnamefont {S.}~\bibnamefont
  {{K}ehrein}},\ }\href@noop {} {\bibfield  {journal} {\bibinfo  {journal}
  {Phys. Rev. B},\ }\textbf {\bibinfo {volume} {78}},\ \bibinfo {pages}
  {092303} (\bibinfo {year} {2008})}\BibitemShut {NoStop}%
\bibitem [{\citenamefont {{H}ackl}\ and\ \citenamefont
  {{K}ehrein}(2009)}]{0953-8984-21-1-015601}%
  \BibitemOpen
  \bibfield  {author} {\bibinfo {author} {\bibfnamefont {A.}~\bibnamefont
  {{H}ackl}}\ and\ \bibinfo {author} {\bibfnamefont {S.}~\bibnamefont
  {{K}ehrein}},\ }\href@noop {} {\bibfield  {journal} {\bibinfo  {journal} {J.
  Phys. Condens. Matter},\ }\textbf {\bibinfo {volume} {21}},\ \bibinfo {pages}
  {015601} (\bibinfo {year} {2009})}\BibitemShut {NoStop}%
\bibitem [{\citenamefont {{K}ennes}\ \emph
  {et~al.}(2012){\natexlab{a}}\citenamefont {{K}ennes}, \citenamefont
  {{J}akobs}, \citenamefont {{K}arrasch},\ and\ \citenamefont
  {{M}eden}}]{PhysRevB.85.085113}%
  \BibitemOpen
  \bibfield  {author} {\bibinfo {author} {\bibfnamefont {D.~M.}\ \bibnamefont
  {{K}ennes}}, \bibinfo {author} {\bibfnamefont {S.~G.}\ \bibnamefont
  {{J}akobs}}, \bibinfo {author} {\bibfnamefont {C.}~\bibnamefont
  {{K}arrasch}}, \ and\ \bibinfo {author} {\bibfnamefont {V.}~\bibnamefont
  {{M}eden}},\ }\Doi {10.1103/PhysRevB.85.085113} {\bibfield  {journal}
  {\bibinfo  {journal} {Phys. Rev. B},\ }\textbf {\bibinfo {volume} {85}},\
  \bibinfo {pages} {085113} (\bibinfo {year} {2012}{\natexlab{a}})}\BibitemShut
  {NoStop}%
\bibitem [{\citenamefont {{K}ennes}\ \emph
  {et~al.}(2012){\natexlab{b}}\citenamefont {{K}ennes}, \citenamefont
  {{K}ashuba}, \citenamefont {{P}letyukhov}, \citenamefont {{S}choeller},\ and\
  \citenamefont {{M}eden}}]{arXiv:1211.0293}%
  \BibitemOpen
  \bibfield  {author} {\bibinfo {author} {\bibfnamefont {D.~M.}\ \bibnamefont
  {{K}ennes}}, \bibinfo {author} {\bibfnamefont {O.}~\bibnamefont {{K}ashuba}},
  \bibinfo {author} {\bibfnamefont {M.}~\bibnamefont {{P}letyukhov}}, \bibinfo
  {author} {\bibfnamefont {H.}~\bibnamefont {{S}choeller}}, \ and\ \bibinfo
  {author} {\bibfnamefont {V.}~\bibnamefont {{M}eden}},\ }\href@noop {}
  {\bibfield  {journal} {\bibinfo  {journal} {arXiv:1211.0293}} (\bibinfo
  {year} {2012}{\natexlab{b}})}\BibitemShut {NoStop}%
\bibitem [{\citenamefont {{F}eynman}\ and\ \citenamefont {{V}ernon
  {J}r.}(1963)}]{feynmanvernon}%
  \BibitemOpen
  \bibfield  {author} {\bibinfo {author} {\bibfnamefont {R.}~\bibnamefont
  {{F}eynman}}\ and\ \bibinfo {author} {\bibfnamefont {F.}~\bibnamefont
  {{V}ernon {J}r.}},\ }\Doi {doi:10.1016/0003-4916(63)90068-X} {\bibfield
  {journal} {\bibinfo  {journal} {Ann. Phys. (NY)},\ }\textbf {\bibinfo
  {volume} {24}},\ \bibinfo {pages} {118} (\bibinfo {year} {1963})}\BibitemShut
  {NoStop}%
\bibitem [{\citenamefont {{L}esovik}\ \emph {et~al.}(2002)\citenamefont
  {{L}esovik}, \citenamefont {{L}ebedev},\ and\ \citenamefont
  {{I}mambekov}}]{imambek_jetp_02}%
  \BibitemOpen
  \bibfield  {author} {\bibinfo {author} {\bibfnamefont {G.~B.}\ \bibnamefont
  {{L}esovik}}, \bibinfo {author} {\bibfnamefont {A.~O.}\ \bibnamefont
  {{L}ebedev}}, \ and\ \bibinfo {author} {\bibfnamefont {A.~O.}\ \bibnamefont
  {{I}mambekov}},\ }\href@noop {} {\bibfield  {journal} {\bibinfo  {journal}
  {JETP Lett.},\ }\textbf {\bibinfo {volume} {75}},\ \bibinfo {pages} {474}
  (\bibinfo {year} {2002})}\BibitemShut {NoStop}%
\bibitem [{\citenamefont {{I}mambekov}\ \emph {et~al.}(2008)\citenamefont
  {{I}mambekov}, \citenamefont {{G}ritsev},\ and\ \citenamefont
  {{D}emler}}]{imambekov:063606}%
  \BibitemOpen
  \bibfield  {author} {\bibinfo {author} {\bibfnamefont {A.}~\bibnamefont
  {{I}mambekov}}, \bibinfo {author} {\bibfnamefont {V.}~\bibnamefont
  {{G}ritsev}}, \ and\ \bibinfo {author} {\bibfnamefont {E.}~\bibnamefont
  {{D}emler}},\ }\Doi {10.1103/PhysRevA.77.063606} {\bibfield  {journal}
  {\bibinfo  {journal} {Phys. Rev. A},\ }\textbf {\bibinfo {volume} {77}},\
  \bibinfo {eid} {063606} (\bibinfo {year} {2008})}\BibitemShut {NoStop}%
\bibitem [{\citenamefont {{I}mambekov}\ \emph {et~al.}(2006)\citenamefont
  {{I}mambekov}, \citenamefont {{G}ritsev},\ and\ \citenamefont
  {{D}emler}}]{ImambekovGritsevDemler-FundNoiseFermiProceed2006}%
  \BibitemOpen
  \bibfield  {author} {\bibinfo {author} {\bibfnamefont {A.}~\bibnamefont
  {{I}mambekov}}, \bibinfo {author} {\bibfnamefont {V.}~\bibnamefont
  {{G}ritsev}}, \ and\ \bibinfo {author} {\bibfnamefont {E.}~\bibnamefont
  {{D}emler}},\ }in\ \Doi {cond-mat/0703766} {\emph {\bibinfo {booktitle}
  {Proceedings of the 2006 Enrico Fermi Summer School on "Ultracold Fermi
  gases"}}},\ \bibinfo {editor} {edited by\ \bibinfo {editor} {\bibfnamefont
  {M.}~\bibnamefont {Inguscio}}, \bibinfo {editor} {\bibfnamefont
  {W.}~\bibnamefont {Ketterle}}, \ and\ \bibinfo {editor} {\bibfnamefont
  {C.}~\bibnamefont {Salomon}}}\ (\bibinfo {address} {Varenna, Italy},\
  \bibinfo {year} {2006})\BibitemShut {NoStop}%
\bibitem [{\citenamefont {{H}offerberth}\ \emph {et~al.}(2008)\citenamefont
  {{H}offerberth}, \citenamefont {{L}esanovsky}, \citenamefont {{S}chumm},
  \citenamefont {{I}mambekov}, \citenamefont {{G}ritsev}, \citenamefont
  {{D}emler},\ and\ \citenamefont {{S}chmiedmayer}}]{Hofferberth-NatPhys-2008}%
  \BibitemOpen
  \bibfield  {author} {\bibinfo {author} {\bibfnamefont {S.}~\bibnamefont
  {{H}offerberth}}, \bibinfo {author} {\bibfnamefont {I.}~\bibnamefont
  {{L}esanovsky}}, \bibinfo {author} {\bibfnamefont {T.}~\bibnamefont
  {{S}chumm}}, \bibinfo {author} {\bibfnamefont {A.}~\bibnamefont
  {{I}mambekov}}, \bibinfo {author} {\bibfnamefont {V.}~\bibnamefont
  {{G}ritsev}}, \bibinfo {author} {\bibfnamefont {E.}~\bibnamefont {{D}emler}},
  \ and\ \bibinfo {author} {\bibfnamefont {J.}~\bibnamefont {{S}chmiedmayer}},\
  }\href@noop {} {\bibfield  {journal} {\bibinfo  {journal} {Nat. Phys.},\
  }\textbf {\bibinfo {volume} {4}},\ \bibinfo {pages} {489} (\bibinfo {year}
  {2008})}\BibitemShut {NoStop}%
\bibitem [{\citenamefont {{O}rth}\ \emph
  {et~al.}(2010){\natexlab{b}}\citenamefont {{O}rth}, \citenamefont
  {{I}mambekov},\ and\ \citenamefont {{L}e {H}ur}}]{PhysRevA.82.032118}%
  \BibitemOpen
  \bibfield  {author} {\bibinfo {author} {\bibfnamefont {P.~P.}\ \bibnamefont
  {{O}rth}}, \bibinfo {author} {\bibfnamefont {A.}~\bibnamefont {{I}mambekov}},
  \ and\ \bibinfo {author} {\bibfnamefont {K.}~\bibnamefont {{L}e {H}ur}},\
  }\Doi {10.1103/PhysRevA.82.032118} {\bibfield  {journal} {\bibinfo  {journal}
  {Phys. Rev. A},\ }\textbf {\bibinfo {volume} {82}},\ \bibinfo {pages}
  {032118} (\bibinfo {year} {2010}{\natexlab{b}})}\BibitemShut {NoStop}%
\bibitem [{\citenamefont {{K}leinert}\ and\ \citenamefont
  {{S}habanov}(1995)}]{Kleinert1995224}%
  \BibitemOpen
  \bibfield  {author} {\bibinfo {author} {\bibfnamefont {H.}~\bibnamefont
  {{K}leinert}}\ and\ \bibinfo {author} {\bibfnamefont {S.~V.}\ \bibnamefont
  {{S}habanov}},\ }\href@noop {} {\bibfield  {journal} {\bibinfo  {journal}
  {Phys. Lett. A},\ }\textbf {\bibinfo {volume} {200}},\ \bibinfo {pages} {224}
  (\bibinfo {year} {1995})}\BibitemShut {NoStop}%
\bibitem [{\citenamefont {{S}tockburger}\ and\ \citenamefont
  {{M}ak}(1998)}]{PhysRevLett.80.2657}%
  \BibitemOpen
  \bibfield  {author} {\bibinfo {author} {\bibfnamefont {J.~T.}\ \bibnamefont
  {{S}tockburger}}\ and\ \bibinfo {author} {\bibfnamefont {C.~H.}\ \bibnamefont
  {{M}ak}},\ }\Doi {10.1103/PhysRevLett.80.2657} {\bibfield  {journal}
  {\bibinfo  {journal} {Phys. Rev. Lett.},\ }\textbf {\bibinfo {volume} {80}},\
  \bibinfo {pages} {2657} (\bibinfo {year} {1998})}\BibitemShut {NoStop}%
\bibitem [{\citenamefont {{S}trunz}\ \emph {et~al.}(1999)\citenamefont
  {{S}trunz}, \citenamefont {{D}i\'osi},\ and\ \citenamefont
  {{G}isin}}]{PhysRevLett.82.1801}%
  \BibitemOpen
  \bibfield  {author} {\bibinfo {author} {\bibfnamefont {W.~T.}\ \bibnamefont
  {{S}trunz}}, \bibinfo {author} {\bibfnamefont {L.}~\bibnamefont {{D}i\'osi}},
  \ and\ \bibinfo {author} {\bibfnamefont {N.}~\bibnamefont {{G}isin}},\
  }\href@noop {} {\bibfield  {journal} {\bibinfo  {journal} {Phys. Rev.
  Lett.},\ }\textbf {\bibinfo {volume} {82}},\ \bibinfo {pages} {1801}
  (\bibinfo {year} {1999})}\BibitemShut {NoStop}%
\bibitem [{\citenamefont {{S}tockburger}\ and\ \citenamefont
  {{G}rabert}(2002)}]{PhysRevLett.88.170407}%
  \BibitemOpen
  \bibfield  {author} {\bibinfo {author} {\bibfnamefont {J.~T.}\ \bibnamefont
  {{S}tockburger}}\ and\ \bibinfo {author} {\bibfnamefont {H.}~\bibnamefont
  {{G}rabert}},\ }\Doi {10.1103/PhysRevLett.88.170407} {\bibfield  {journal}
  {\bibinfo  {journal} {Phys. Rev. Lett.},\ }\textbf {\bibinfo {volume} {88}},\
  \bibinfo {pages} {170407} (\bibinfo {year} {2002})}\BibitemShut {NoStop}%
\bibitem [{\citenamefont {{S}tockburger}(2004)}]{Stockburger2004159}%
  \BibitemOpen
  \bibfield  {author} {\bibinfo {author} {\bibfnamefont {J.~T.}\ \bibnamefont
  {{S}tockburger}},\ }\href@noop {} {\bibfield  {journal} {\bibinfo  {journal}
  {Chemical Physics},\ }\textbf {\bibinfo {volume} {296}},\ \bibinfo {pages}
  {159} (\bibinfo {year} {2004})}\BibitemShut {NoStop}%
\bibitem [{\citenamefont {{E}gger}\ \emph {et~al.}(1997)\citenamefont
  {{E}gger}, \citenamefont {{G}rabert},\ and\ \citenamefont
  {{W}eiss}}]{egger_crossover_1997}%
  \BibitemOpen
  \bibfield  {author} {\bibinfo {author} {\bibfnamefont {R.}~\bibnamefont
  {{E}gger}}, \bibinfo {author} {\bibfnamefont {H.}~\bibnamefont {{G}rabert}},
  \ and\ \bibinfo {author} {\bibfnamefont {U.}~\bibnamefont {{W}eiss}},\ }\Doi
  {10.1103/PhysRevE.55.R3809} {\bibfield  {journal} {\bibinfo  {journal} {Phys.
  Rev. E},\ }\textbf {\bibinfo {volume} {55}},\ \bibinfo {pages} {R3809}
  (\bibinfo {year} {1997})}\BibitemShut {NoStop}%
\bibitem [{\citenamefont {{L}esage}\ and\ \citenamefont
  {{S}aleur}(1998)}]{PhysRevLett.80.4370}%
  \BibitemOpen
  \bibfield  {author} {\bibinfo {author} {\bibfnamefont {F.}~\bibnamefont
  {{L}esage}}\ and\ \bibinfo {author} {\bibfnamefont {H.}~\bibnamefont
  {{S}aleur}},\ }\Doi {10.1103/PhysRevLett.80.4370} {\bibfield  {journal}
  {\bibinfo  {journal} {Phys. Rev. Lett.},\ }\textbf {\bibinfo {volume} {80}},\
  \bibinfo {pages} {4370} (\bibinfo {year} {1998})}\BibitemShut {NoStop}%
\bibitem [{\citenamefont {{A}nders}\ and\ \citenamefont
  {{S}chiller}(2005)}]{PhysRevLett.95.196801}%
  \BibitemOpen
  \bibfield  {author} {\bibinfo {author} {\bibfnamefont {F.~B.}\ \bibnamefont
  {{A}nders}}\ and\ \bibinfo {author} {\bibfnamefont {A.}~\bibnamefont
  {{S}chiller}},\ }\href@noop {} {\bibfield  {journal} {\bibinfo  {journal}
  {Phys. Rev. Lett.},\ }\textbf {\bibinfo {volume} {95}},\ \bibinfo {pages}
  {196801} (\bibinfo {year} {2005})}\BibitemShut {NoStop}%
\bibitem [{\citenamefont {{E}gger}\ \emph {et~al.}(2000)\citenamefont
  {{E}gger}, \citenamefont {{M}\"uhlbacher},\ and\ \citenamefont
  {{M}ak}}]{PhysRevE.61.5961}%
  \BibitemOpen
  \bibfield  {author} {\bibinfo {author} {\bibfnamefont {R.}~\bibnamefont
  {{E}gger}}, \bibinfo {author} {\bibfnamefont {L.}~\bibnamefont
  {{M}\"uhlbacher}}, \ and\ \bibinfo {author} {\bibfnamefont {C.~H.}\
  \bibnamefont {{M}ak}},\ }\Doi {10.1103/PhysRevE.61.5961} {\bibfield
  {journal} {\bibinfo  {journal} {Phys. Rev. E},\ }\textbf {\bibinfo {volume}
  {61}},\ \bibinfo {pages} {5961} (\bibinfo {year} {2000})}\BibitemShut
  {NoStop}%
\bibitem [{\citenamefont {{D}ekker}(1987)}]{PhysRevA.35.1436}%
  \BibitemOpen
  \bibfield  {author} {\bibinfo {author} {\bibfnamefont {H.}~\bibnamefont
  {{D}ekker}},\ }\Doi {10.1103/PhysRevA.35.1436} {\bibfield  {journal}
  {\bibinfo  {journal} {Phys. Rev. A},\ }\textbf {\bibinfo {volume} {35}},\
  \bibinfo {pages} {1436} (\bibinfo {year} {1987})}\BibitemShut {NoStop}%
\bibitem [{\citenamefont {{A}slangul}\ \emph {et~al.}(1986)\citenamefont
  {{A}slangul}, \citenamefont {{P}ottier},\ and\ \citenamefont
  {{Saint-James}}}]{aslangul_spin-boson_1986}%
  \BibitemOpen
  \bibfield  {author} {\bibinfo {author} {\bibfnamefont {C.}~\bibnamefont
  {{A}slangul}}, \bibinfo {author} {\bibfnamefont {N.}~\bibnamefont
  {{P}ottier}}, \ and\ \bibinfo {author} {\bibfnamefont {D.}~\bibnamefont
  {{Saint-James}}},\ }\Doi {10.1051/jphys:0198600470100165700} {\bibfield
  {journal} {\bibinfo  {journal} {Journal de Physique},\ }\textbf {\bibinfo
  {volume} {47}},\ \bibinfo {pages} {5} (\bibinfo {year} {1986})}\BibitemShut
  {NoStop}%
\bibitem [{\citenamefont
  {{C}hild}(1974)}]{Child_molecular_collision_theory_book}%
  \BibitemOpen
  \bibfield  {author} {\bibinfo {author} {\bibfnamefont {M.~S.}\ \bibnamefont
  {{C}hild}},\ }\href@noop {} {\emph {\bibinfo {title} {{M}olecular {C}ollision
  {T}heory}}}\ (\bibinfo  {publisher} {Academic Press},\ \bibinfo {address}
  {London, U.K.},\ \bibinfo {year} {1974})\BibitemShut {NoStop}%
\bibitem [{\citenamefont {{W}ernsdorfer}\ and\ \citenamefont
  {{S}essoli}(1999)}]{W.Wernsdorfer04021999}%
  \BibitemOpen
  \bibfield  {author} {\bibinfo {author} {\bibfnamefont {W.}~\bibnamefont
  {{W}ernsdorfer}}\ and\ \bibinfo {author} {\bibfnamefont {R.}~\bibnamefont
  {{S}essoli}},\ }\href@noop {} {\bibfield  {journal} {\bibinfo  {journal}
  {Science},\ }\textbf {\bibinfo {volume} {284}},\ \bibinfo {pages} {133}
  (\bibinfo {year} {1999})}\BibitemShut {NoStop}%
\bibitem [{\citenamefont {{W}allraff}\ \emph {et~al.}(2004)\citenamefont
  {{W}allraff}, \citenamefont {{S}chuster}, \citenamefont {{B}lais},
  \citenamefont {{F}runzio}, \citenamefont {{H}uang}, \citenamefont {{M}ajer},
  \citenamefont {{K}umar}, \citenamefont {{G}irvin},\ and\ \citenamefont
  {{S}choelkopf}}]{Wallraff_Nature_2004}%
  \BibitemOpen
  \bibfield  {author} {\bibinfo {author} {\bibfnamefont {A.}~\bibnamefont
  {{W}allraff}}, \bibinfo {author} {\bibfnamefont {D.~I.}\ \bibnamefont
  {{S}chuster}}, \bibinfo {author} {\bibfnamefont {A.}~\bibnamefont {{B}lais}},
  \bibinfo {author} {\bibfnamefont {L.}~\bibnamefont {{F}runzio}}, \bibinfo
  {author} {\bibfnamefont {R.~S.}\ \bibnamefont {{H}uang}}, \bibinfo {author}
  {\bibfnamefont {J.}~\bibnamefont {{M}ajer}}, \bibinfo {author} {\bibfnamefont
  {S.}~\bibnamefont {{K}umar}}, \bibinfo {author} {\bibfnamefont {S.~M.}\
  \bibnamefont {{G}irvin}}, \ and\ \bibinfo {author} {\bibfnamefont {R.~J.}\
  \bibnamefont {{S}choelkopf}},\ }\href@noop {} {\bibfield  {journal} {\bibinfo
   {journal} {Nature (London)},\ }\textbf {\bibinfo {volume} {431}},\ \bibinfo
  {pages} {162} (\bibinfo {year} {2004})}\BibitemShut {NoStop}%
\bibitem [{\citenamefont {{S}illanp\"{a}\"{a}}\ \emph
  {et~al.}(2006)\citenamefont {{S}illanp\"{a}\"{a}}, \citenamefont
  {{L}ehtinen}, \citenamefont {{P}aila}, \citenamefont {{M}akhlin},\ and\
  \citenamefont {{H}akonen}}]{sillanpPRL}%
  \BibitemOpen
  \bibfield  {author} {\bibinfo {author} {\bibfnamefont {M.}~\bibnamefont
  {{S}illanp\"{a}\"{a}}}, \bibinfo {author} {\bibfnamefont {T.}~\bibnamefont
  {{L}ehtinen}}, \bibinfo {author} {\bibfnamefont {A.}~\bibnamefont {{P}aila}},
  \bibinfo {author} {\bibfnamefont {Y.}~\bibnamefont {{M}akhlin}}, \ and\
  \bibinfo {author} {\bibfnamefont {P.}~\bibnamefont {{H}akonen}},\ }\href@noop
  {} {\bibfield  {journal} {\bibinfo  {journal} {Phys. Rev. Lett.},\ }\textbf
  {\bibinfo {volume} {96}},\ \bibinfo {eid} {187002} (\bibinfo {year}
  {2006})}\BibitemShut {NoStop}%
\bibitem [{\citenamefont {{B}erns}\ \emph {et~al.}(2008)\citenamefont
  {{B}erns}, \citenamefont {{R}udner}, \citenamefont {{V}alenzuela},
  \citenamefont {{B}erggren}, \citenamefont {{O}liver}, \citenamefont
  {{L}evitov},\ and\ \citenamefont
  {{O}rlando}}]{OrlandoLevitov_AmplitSpectr_Nature_2008}%
  \BibitemOpen
  \bibfield  {author} {\bibinfo {author} {\bibfnamefont {D.~M.}\ \bibnamefont
  {{B}erns}}, \bibinfo {author} {\bibfnamefont {M.~S.}\ \bibnamefont
  {{R}udner}}, \bibinfo {author} {\bibfnamefont {S.~O.}\ \bibnamefont
  {{V}alenzuela}}, \bibinfo {author} {\bibfnamefont {K.~K.}\ \bibnamefont
  {{B}erggren}}, \bibinfo {author} {\bibfnamefont {W.~D.}\ \bibnamefont
  {{O}liver}}, \bibinfo {author} {\bibfnamefont {L.~S.}\ \bibnamefont
  {{L}evitov}}, \ and\ \bibinfo {author} {\bibfnamefont {T.~P.}\ \bibnamefont
  {{O}rlando}},\ }\href@noop {} {\bibfield  {journal} {\bibinfo  {journal}
  {Nature (London)},\ }\textbf {\bibinfo {volume} {455}},\ \bibinfo {pages}
  {51} (\bibinfo {year} {2008})}\BibitemShut {NoStop}%
\bibitem [{\citenamefont {{Z}ueco}\ \emph {et~al.}(2008)\citenamefont
  {{Z}ueco}, \citenamefont {{H}\"anggi},\ and\ \citenamefont
  {{K}ohler}}]{ZuecoHanggiKohler-LZInCavQED-NewJPhys-2008}%
  \BibitemOpen
  \bibfield  {author} {\bibinfo {author} {\bibfnamefont {D.}~\bibnamefont
  {{Z}ueco}}, \bibinfo {author} {\bibfnamefont {P.}~\bibnamefont {{H}\"anggi}},
  \ and\ \bibinfo {author} {\bibfnamefont {S.}~\bibnamefont {{K}ohler}},\
  }\href@noop {} {\bibfield  {journal} {\bibinfo  {journal} {New J. Phys.},\
  }\textbf {\bibinfo {volume} {10}},\ \bibinfo {pages} {115012} (\bibinfo
  {year} {2008})}\BibitemShut {NoStop}%
\bibitem [{\citenamefont {{Z}enesini}\ \emph {et~al.}(2009)\citenamefont
  {{Z}enesini}, \citenamefont {{L}ignier}, \citenamefont {{T}ayebirad},
  \citenamefont {{R}adogostowicz}, \citenamefont {{C}iampini}, \citenamefont
  {{M}annella}, \citenamefont {{W}imberger}, \citenamefont {{M}orsch},\ and\
  \citenamefont {{A}rimondo}}]{zenesini:090403}%
  \BibitemOpen
  \bibfield  {author} {\bibinfo {author} {\bibfnamefont {A.}~\bibnamefont
  {{Z}enesini}}, \bibinfo {author} {\bibfnamefont {H.}~\bibnamefont
  {{L}ignier}}, \bibinfo {author} {\bibfnamefont {G.}~\bibnamefont
  {{T}ayebirad}}, \bibinfo {author} {\bibfnamefont {J.}~\bibnamefont
  {{R}adogostowicz}}, \bibinfo {author} {\bibfnamefont {D.}~\bibnamefont
  {{C}iampini}}, \bibinfo {author} {\bibfnamefont {R.}~\bibnamefont
  {{M}annella}}, \bibinfo {author} {\bibfnamefont {S.}~\bibnamefont
  {{W}imberger}}, \bibinfo {author} {\bibfnamefont {O.}~\bibnamefont
  {{M}orsch}}, \ and\ \bibinfo {author} {\bibfnamefont {E.}~\bibnamefont
  {{A}rimondo}},\ }\href@noop {} {\bibfield  {journal} {\bibinfo  {journal}
  {Phys. Rev. Lett.},\ }\textbf {\bibinfo {volume} {103}},\ \bibinfo {eid}
  {090403} (\bibinfo {year} {2009})}\BibitemShut {NoStop}%
\bibitem [{\citenamefont {{C}hen}\ \emph {et~al.}(2011)\citenamefont {{C}hen},
  \citenamefont {{H}uber}, \citenamefont {{T}rotzky}, \citenamefont {{B}loch},\
  and\ \citenamefont {{A}ltman}}]{ChenBlochAltman-LZ1DBose-NatPhys-2011}%
  \BibitemOpen
  \bibfield  {author} {\bibinfo {author} {\bibfnamefont {Y.-A.}\ \bibnamefont
  {{C}hen}}, \bibinfo {author} {\bibfnamefont {S.~D.}\ \bibnamefont {{H}uber}},
  \bibinfo {author} {\bibfnamefont {S.}~\bibnamefont {{T}rotzky}}, \bibinfo
  {author} {\bibfnamefont {I.}~\bibnamefont {{B}loch}}, \ and\ \bibinfo
  {author} {\bibfnamefont {E.}~\bibnamefont {{A}ltman}},\ }\href@noop {}
  {\bibfield  {journal} {\bibinfo  {journal} {Nature Phys.},\ }\textbf
  {\bibinfo {volume} {7}},\ \bibinfo {pages} {61} (\bibinfo {year}
  {2011})}\BibitemShut {NoStop}%
\bibitem [{\citenamefont {{L}im}\ \emph {et~al.}(2012)\citenamefont {{L}im},
  \citenamefont {{F}uchs},\ and\ \citenamefont
  {{M}ontambaux}}]{PhysRevLett.108.175303}%
  \BibitemOpen
  \bibfield  {author} {\bibinfo {author} {\bibfnamefont {L.-K.}\ \bibnamefont
  {{L}im}}, \bibinfo {author} {\bibfnamefont {J.-N.}\ \bibnamefont {{F}uchs}},
  \ and\ \bibinfo {author} {\bibfnamefont {G.}~\bibnamefont {{M}ontambaux}},\
  }\Doi {10.1103/PhysRevLett.108.175303} {\bibfield  {journal} {\bibinfo
  {journal} {Phys. Rev. Lett.},\ }\textbf {\bibinfo {volume} {108}},\ \bibinfo
  {pages} {175303} (\bibinfo {year} {2012})}\BibitemShut {NoStop}%
\bibitem [{\citenamefont {{U}ehlinger}\ \emph {et~al.}(2012)\citenamefont
  {{U}ehlinger}, \citenamefont {{G}reif. {D}.}, \citenamefont {{J}otzu},
  \citenamefont {{T}arruell},\ and\ \citenamefont
  {{E}sslinger}}]{UehlingerEsslinger-LZDirac-arXiv-2012}%
  \BibitemOpen
  \bibfield  {author} {\bibinfo {author} {\bibfnamefont {T.}~\bibnamefont
  {{U}ehlinger}}, \bibinfo {author} {\bibnamefont {{G}reif. {D}.}}, \bibinfo
  {author} {\bibfnamefont {G.}~\bibnamefont {{J}otzu}}, \bibinfo {author}
  {\bibfnamefont {L.}~\bibnamefont {{T}arruell}}, \ and\ \bibinfo {author}
  {\bibfnamefont {T.}~\bibnamefont {{E}sslinger}},\ }\href@noop {} {\bibfield
  {journal} {\bibinfo  {journal} {arXiv:1210.0904}} (\bibinfo {year}
  {2012})}\BibitemShut {NoStop}%
\bibitem [{\citenamefont {{L}andau}(1932)}]{landau_lz}%
  \BibitemOpen
  \bibfield  {author} {\bibinfo {author} {\bibfnamefont {L.~D.}\ \bibnamefont
  {{L}andau}},\ }\href@noop {} {\bibfield  {journal} {\bibinfo  {journal}
  {Phys. Z. Sowjetunion},\ }\textbf {\bibinfo {volume} {2}},\ \bibinfo {pages}
  {46} (\bibinfo {year} {1932})}\BibitemShut {NoStop}%
\bibitem [{\citenamefont {{Z}ener}(1932)}]{zener_lz}%
  \BibitemOpen
  \bibfield  {author} {\bibinfo {author} {\bibfnamefont {C.}~\bibnamefont
  {{Z}ener}},\ }\href@noop {} {\bibfield  {journal} {\bibinfo  {journal} {Proc.
  R. Soc. London, Ser. A},\ }\textbf {\bibinfo {volume} {137}},\ \bibinfo
  {pages} {696} (\bibinfo {year} {1932})}\BibitemShut {NoStop}%
\bibitem [{\citenamefont {{S}t\"uckelberg}(1932)}]{stueckelberg_lz}%
  \BibitemOpen
  \bibfield  {author} {\bibinfo {author} {\bibfnamefont {E.~C.~G.}\
  \bibnamefont {{S}t\"uckelberg}},\ }\href@noop {} {\bibfield  {journal}
  {\bibinfo  {journal} {Helv. Phys. Acta},\ }\textbf {\bibinfo {volume} {5}},\
  \bibinfo {pages} {369} (\bibinfo {year} {1932})}\BibitemShut {NoStop}%
\bibitem [{\citenamefont {{M}ajorana}(1932)}]{majorana_lz}%
  \BibitemOpen
  \bibfield  {author} {\bibinfo {author} {\bibfnamefont {E.}~\bibnamefont
  {{M}ajorana}},\ }\href@noop {} {\bibfield  {journal} {\bibinfo  {journal}
  {Nuovo Cimento},\ }\textbf {\bibinfo {volume} {9}},\ \bibinfo {pages} {43}
  (\bibinfo {year} {1932})}\BibitemShut {NoStop}%
\bibitem [{\citenamefont {{A}o}\ and\ \citenamefont
  {{R}ammer}(1991)}]{PhysRevB.43.5397}%
  \BibitemOpen
  \bibfield  {author} {\bibinfo {author} {\bibfnamefont {P.}~\bibnamefont
  {{A}o}}\ and\ \bibinfo {author} {\bibfnamefont {J.}~\bibnamefont
  {{R}ammer}},\ }\href@noop {} {\bibfield  {journal} {\bibinfo  {journal}
  {Phys. Rev. B},\ }\textbf {\bibinfo {volume} {43}},\ \bibinfo {pages} {5397}
  (\bibinfo {year} {1991})}\BibitemShut {NoStop}%
\bibitem [{\citenamefont {{A}o}\ and\ \citenamefont
  {{R}ammer}(1989)}]{PhysRevLett.62.3004}%
  \BibitemOpen
  \bibfield  {author} {\bibinfo {author} {\bibfnamefont {P.}~\bibnamefont
  {{A}o}}\ and\ \bibinfo {author} {\bibfnamefont {J.}~\bibnamefont
  {{R}ammer}},\ }\href@noop {} {\bibfield  {journal} {\bibinfo  {journal}
  {Phys. Rev. Lett.},\ }\textbf {\bibinfo {volume} {62}},\ \bibinfo {pages}
  {3004} (\bibinfo {year} {1989})}\BibitemShut {NoStop}%
\bibitem [{\citenamefont {{K}ayanuma}\ and\ \citenamefont
  {{N}akayama}(1998)}]{PhysRevB.57.13099}%
  \BibitemOpen
  \bibfield  {author} {\bibinfo {author} {\bibfnamefont {Y.}~\bibnamefont
  {{K}ayanuma}}\ and\ \bibinfo {author} {\bibfnamefont {H.}~\bibnamefont
  {{N}akayama}},\ }\Doi {10.1103/PhysRevB.57.13099} {\bibfield  {journal}
  {\bibinfo  {journal} {Phys. Rev. B},\ }\textbf {\bibinfo {volume} {57}},\
  \bibinfo {pages} {13099} (\bibinfo {year} {1998})}\BibitemShut {NoStop}%
\bibitem [{\citenamefont {{P}okrovsky}\ and\ \citenamefont
  {{S}un}(2007)}]{PhysRevB.76.024310}%
  \BibitemOpen
  \bibfield  {author} {\bibinfo {author} {\bibfnamefont {V.~L.}\ \bibnamefont
  {{P}okrovsky}}\ and\ \bibinfo {author} {\bibfnamefont {D.}~\bibnamefont
  {{S}un}},\ }\Doi {10.1103/PhysRevB.76.024310} {\bibfield  {journal} {\bibinfo
   {journal} {Phys. Rev. B},\ }\textbf {\bibinfo {volume} {76}},\ \bibinfo
  {pages} {024310} (\bibinfo {year} {2007})}\BibitemShut {NoStop}%
\bibitem [{\citenamefont {{W}ubs}\ \emph {et~al.}(2006)\citenamefont {{W}ubs},
  \citenamefont {{S}aito}, \citenamefont {{K}ohler}, \citenamefont
  {{H}\"{a}nggi},\ and\ \citenamefont {{K}ayanuma}}]{wubs:200404}%
  \BibitemOpen
  \bibfield  {author} {\bibinfo {author} {\bibfnamefont {M.}~\bibnamefont
  {{W}ubs}}, \bibinfo {author} {\bibfnamefont {K.}~\bibnamefont {{S}aito}},
  \bibinfo {author} {\bibfnamefont {S.}~\bibnamefont {{K}ohler}}, \bibinfo
  {author} {\bibfnamefont {P.}~\bibnamefont {{H}\"{a}nggi}}, \ and\ \bibinfo
  {author} {\bibfnamefont {Y.}~\bibnamefont {{K}ayanuma}},\ }\href@noop {}
  {\bibfield  {journal} {\bibinfo  {journal} {Phys. Rev. Lett.},\ }\textbf
  {\bibinfo {volume} {97}},\ \bibinfo {eid} {200404} (\bibinfo {year}
  {2006})}\BibitemShut {NoStop}%
\bibitem [{\citenamefont {{S}aito}\ \emph {et~al.}(2007)\citenamefont
  {{S}aito}, \citenamefont {{W}ubs}, \citenamefont {{K}ohler}, \citenamefont
  {{K}ayanuma},\ and\ \citenamefont {{H}\"{a}nggi}}]{saito:214308}%
  \BibitemOpen
  \bibfield  {author} {\bibinfo {author} {\bibfnamefont {K.}~\bibnamefont
  {{S}aito}}, \bibinfo {author} {\bibfnamefont {M.}~\bibnamefont {{W}ubs}},
  \bibinfo {author} {\bibfnamefont {S.}~\bibnamefont {{K}ohler}}, \bibinfo
  {author} {\bibfnamefont {Y.}~\bibnamefont {{K}ayanuma}}, \ and\ \bibinfo
  {author} {\bibfnamefont {P.}~\bibnamefont {{H}\"{a}nggi}},\ }\href@noop {}
  {\bibfield  {journal} {\bibinfo  {journal} {Phys. Rev. B},\ }\textbf
  {\bibinfo {volume} {75}},\ \bibinfo {eid} {214308} (\bibinfo {year}
  {2007})}\BibitemShut {NoStop}%
\bibitem [{\citenamefont {{A}stafiev}\ \emph {et~al.}(2007)\citenamefont
  {{A}stafiev}, \citenamefont {{N}iskanen}, \citenamefont {{Y}amamoto},
  \citenamefont {{P}ashkin}, \citenamefont {{N}akamura},\ and\ \citenamefont
  {{T}sai}}]{Astafiev-SingleAtomLasing-Nature-2007}%
  \BibitemOpen
  \bibfield  {author} {\bibinfo {author} {\bibfnamefont {K.}~\bibnamefont
  {{A}stafiev}, \bibfnamefont {{O}.and~{I}nomata}}, \bibinfo {author}
  {\bibfnamefont {A.~O.}\ \bibnamefont {{N}iskanen}}, \bibinfo {author}
  {\bibfnamefont {T.}~\bibnamefont {{Y}amamoto}}, \bibinfo {author}
  {\bibfnamefont {Y.~A.}\ \bibnamefont {{P}ashkin}}, \bibinfo {author}
  {\bibfnamefont {Y.}~\bibnamefont {{N}akamura}}, \ and\ \bibinfo {author}
  {\bibfnamefont {J.~S.}\ \bibnamefont {{T}sai}},\ }\href@noop {} {\bibfield
  {journal} {\bibinfo  {journal} {Nature},\ }\textbf {\bibinfo {volume}
  {449}},\ \bibinfo {pages} {588} (\bibinfo {year} {2007})}\BibitemShut
  {NoStop}%
\bibitem [{\citenamefont {{Z}hirov}\ and\ \citenamefont
  {{S}hepelyansky}(2008)}]{PhysRevLett.100.014101}%
  \BibitemOpen
  \bibfield  {author} {\bibinfo {author} {\bibfnamefont {O.~V.}\ \bibnamefont
  {{Z}hirov}}\ and\ \bibinfo {author} {\bibfnamefont {D.~L.}\ \bibnamefont
  {{S}hepelyansky}},\ }\Doi {10.1103/PhysRevLett.100.014101} {\bibfield
  {journal} {\bibinfo  {journal} {Phys. Rev. Lett.},\ }\textbf {\bibinfo
  {volume} {100}},\ \bibinfo {pages} {014101} (\bibinfo {year}
  {2008})}\BibitemShut {NoStop}%
\bibitem [{\citenamefont {{Z}hirov}\ and\ \citenamefont
  {{S}hepelyansky}(2009)}]{PhysRevB.80.014519}%
  \BibitemOpen
  \bibfield  {author} {\bibinfo {author} {\bibfnamefont {O.~V.}\ \bibnamefont
  {{Z}hirov}}\ and\ \bibinfo {author} {\bibfnamefont {D.~L.}\ \bibnamefont
  {{S}hepelyansky}},\ }\Doi {10.1103/PhysRevB.80.014519} {\bibfield  {journal}
  {\bibinfo  {journal} {Phys. Rev. B},\ }\textbf {\bibinfo {volume} {80}},\
  \bibinfo {pages} {014519} (\bibinfo {year} {2009})}\BibitemShut {NoStop}%
\bibitem [{\citenamefont {{R}itsch}\ \emph {et~al.}(2012)\citenamefont
  {{R}itsch}, \citenamefont {{D}omokos}, \citenamefont {{B}rennecke},\ and\
  \citenamefont {{E}sslinger}}]{EsslingerRMP-arXiv2012}%
  \BibitemOpen
  \bibfield  {author} {\bibinfo {author} {\bibfnamefont {H.}~\bibnamefont
  {{R}itsch}}, \bibinfo {author} {\bibfnamefont {P.}~\bibnamefont {{D}omokos}},
  \bibinfo {author} {\bibfnamefont {F.}~\bibnamefont {{B}rennecke}}, \ and\
  \bibinfo {author} {\bibfnamefont {T.}~\bibnamefont {{E}sslinger}},\
  }\href@noop {} {\bibfield  {journal} {\bibinfo  {journal} {arXiv:1210.0013}}
  (\bibinfo {year} {2012})}\BibitemShut {NoStop}%
\bibitem [{\citenamefont {{S}assetti}\ and\ \citenamefont
  {{W}eiss}(1990){\natexlab{a}}}]{PhysRevA.41.5383}%
  \BibitemOpen
  \bibfield  {author} {\bibinfo {author} {\bibfnamefont {M.}~\bibnamefont
  {{S}assetti}}\ and\ \bibinfo {author} {\bibfnamefont {U.}~\bibnamefont
  {{W}eiss}},\ }\Doi {10.1103/PhysRevA.41.5383} {\bibfield  {journal} {\bibinfo
   {journal} {Phys. Rev. A},\ }\textbf {\bibinfo {volume} {41}},\ \bibinfo
  {pages} {5383} (\bibinfo {year} {1990}{\natexlab{a}})}\BibitemShut {NoStop}%
\bibitem [{\citenamefont {{S}mith}\ and\ \citenamefont
  {{C}aldeira}(1987)}]{PhysRevA.36.3509}%
  \BibitemOpen
  \bibfield  {author} {\bibinfo {author} {\bibfnamefont {C.~M.}\ \bibnamefont
  {{S}mith}}\ and\ \bibinfo {author} {\bibfnamefont {A.~O.}\ \bibnamefont
  {{C}aldeira}},\ }\Doi {10.1103/PhysRevA.36.3509} {\bibfield  {journal}
  {\bibinfo  {journal} {Phys. Rev. A},\ }\textbf {\bibinfo {volume} {36}},\
  \bibinfo {pages} {3509} (\bibinfo {year} {1987})}\BibitemShut {NoStop}%
\bibitem [{\citenamefont {{S}hiba}(1975)}]{PTP.54.967}%
  \BibitemOpen
  \bibfield  {author} {\bibinfo {author} {\bibfnamefont {H.}~\bibnamefont
  {{S}hiba}},\ }\Doi {10.1143/PTP.54.967} {\bibfield  {journal} {\bibinfo
  {journal} {Prog. Theor. Phys.},\ }\textbf {\bibinfo {volume} {54}},\ \bibinfo
  {pages} {967} (\bibinfo {year} {1975})}\BibitemShut {NoStop}%
\bibitem [{\citenamefont {{S}assetti}\ and\ \citenamefont
  {{W}eiss}(1990){\natexlab{b}}}]{sassetti_universality_1990}%
  \BibitemOpen
  \bibfield  {author} {\bibinfo {author} {\bibfnamefont {M.}~\bibnamefont
  {{S}assetti}}\ and\ \bibinfo {author} {\bibfnamefont {U.}~\bibnamefont
  {{W}eiss}},\ }\Doi {10.1103/PhysRevLett.65.2262} {\bibfield  {journal}
  {\bibinfo  {journal} {Phys. Rev. Lett.},\ }\textbf {\bibinfo {volume} {65}},\
  \bibinfo {pages} {2262} (\bibinfo {year} {1990}{\natexlab{b}})}\BibitemShut
  {NoStop}%
\bibitem [{\citenamefont {{C}osti}\ and\ \citenamefont
  {{K}ieffer}(1996)}]{PhysRevLett.76.1683}%
  \BibitemOpen
  \bibfield  {author} {\bibinfo {author} {\bibfnamefont {T.~A.}\ \bibnamefont
  {{C}osti}}\ and\ \bibinfo {author} {\bibfnamefont {C.}~\bibnamefont
  {{K}ieffer}},\ }\Doi {10.1103/PhysRevLett.76.1683} {\bibfield  {journal}
  {\bibinfo  {journal} {Phys. Rev. Lett.},\ }\textbf {\bibinfo {volume} {76}},\
  \bibinfo {pages} {1683} (\bibinfo {year} {1996})}\BibitemShut {NoStop}%
\bibitem [{\citenamefont {{C}osti}(1998)}]{PhysRevLett.80.1038}%
  \BibitemOpen
  \bibfield  {author} {\bibinfo {author} {\bibfnamefont {T.~A.}\ \bibnamefont
  {{C}osti}},\ }\Doi {10.1103/PhysRevLett.80.1038} {\bibfield  {journal}
  {\bibinfo  {journal} {Phys. Rev. Lett.},\ }\textbf {\bibinfo {volume} {80}},\
  \bibinfo {pages} {1038} (\bibinfo {year} {1998})}\BibitemShut {NoStop}%
\bibitem [{\citenamefont {{D}elbecq}\ \emph {et~al.}(2011)\citenamefont
  {{D}elbecq}, \citenamefont {{S}chmitt}, \citenamefont {{P}armentier},
  \citenamefont {{R}och}, \citenamefont {{V}iennot}, \citenamefont {{F}\`eve},
  \citenamefont {{H}uard}, \citenamefont {{M}ora}, \citenamefont {{C}ottet},\
  and\ \citenamefont {{K}ontos}}]{PhysRevLett.107.256804}%
  \BibitemOpen
  \bibfield  {author} {\bibinfo {author} {\bibfnamefont {M.~R.}\ \bibnamefont
  {{D}elbecq}}, \bibinfo {author} {\bibfnamefont {V.}~\bibnamefont
  {{S}chmitt}}, \bibinfo {author} {\bibfnamefont {F.~D.}\ \bibnamefont
  {{P}armentier}}, \bibinfo {author} {\bibfnamefont {N.}~\bibnamefont
  {{R}och}}, \bibinfo {author} {\bibfnamefont {J.~J.}\ \bibnamefont
  {{V}iennot}}, \bibinfo {author} {\bibfnamefont {G.}~\bibnamefont {{F}\`eve}},
  \bibinfo {author} {\bibfnamefont {B.}~\bibnamefont {{H}uard}}, \bibinfo
  {author} {\bibfnamefont {C.}~\bibnamefont {{M}ora}}, \bibinfo {author}
  {\bibfnamefont {A.}~\bibnamefont {{C}ottet}}, \ and\ \bibinfo {author}
  {\bibfnamefont {T.}~\bibnamefont {{K}ontos}},\ }\Doi
  {10.1103/PhysRevLett.107.256804} {\bibfield  {journal} {\bibinfo  {journal}
  {Phys. Rev. Lett.},\ }\textbf {\bibinfo {volume} {107}},\ \bibinfo {pages}
  {256804} (\bibinfo {year} {2011})}\BibitemShut {NoStop}%
\bibitem [{\citenamefont {{D}utt}\ \emph {et~al.}(2011)\citenamefont {{D}utt},
  \citenamefont {{K}och}, \citenamefont {{H}an},\ and\ \citenamefont {{L}e
  {H}ur}}]{DuttLeHur-AnnPhys-2011}%
  \BibitemOpen
  \bibfield  {author} {\bibinfo {author} {\bibfnamefont {P.}~\bibnamefont
  {{D}utt}}, \bibinfo {author} {\bibfnamefont {J.}~\bibnamefont {{K}och}},
  \bibinfo {author} {\bibfnamefont {J.~E.}\ \bibnamefont {{H}an}}, \ and\
  \bibinfo {author} {\bibfnamefont {K.}~\bibnamefont {{L}e {H}ur}},\
  }\href@noop {} {\bibfield  {journal} {\bibinfo  {journal} {Ann. Phys.},\
  }\textbf {\bibinfo {volume} {326}},\ \bibinfo {pages} {2963} (\bibinfo {year}
  {2011})}\BibitemShut {NoStop}%
\bibitem [{\citenamefont {{S}tone}\ and\ \citenamefont
  {{G}oldbart}(2009)}]{GoldbartStone-MathMethodsBook}%
  \BibitemOpen
  \bibfield  {author} {\bibinfo {author} {\bibfnamefont {M.}~\bibnamefont
  {{S}tone}}\ and\ \bibinfo {author} {\bibfnamefont {P.}~\bibnamefont
  {{G}oldbart}},\ }\href@noop {} {\emph {\bibinfo {title} {{M}athematics for
  physics. {A} guided tour for graduate students}}},\ \bibinfo {edition} {1st}\
  ed.\ (\bibinfo  {publisher} {Cambridge University Press},\ \bibinfo {address}
  {Cambridge, U.K.},\ \bibinfo {year} {2009})\BibitemShut {NoStop}%
\bibitem [{\citenamefont {{A}rfken}\ and\ \citenamefont
  {{W}eber}(2005)}]{ArfkenWeber-MathMethodsBook}%
  \BibitemOpen
  \bibfield  {author} {\bibinfo {author} {\bibfnamefont {G.~B.}\ \bibnamefont
  {{A}rfken}}\ and\ \bibinfo {author} {\bibfnamefont {H.~J.}\ \bibnamefont
  {{W}eber}},\ }\href@noop {} {\emph {\bibinfo {title} {{M}athematical methods
  for physicists}}},\ \bibinfo {edition} {6th}\ ed.\ (\bibinfo  {publisher}
  {Elsevier Academic Press},\ \bibinfo {address} {Burlington, MA},\ \bibinfo
  {year} {2005})\BibitemShut {NoStop}%
\end{thebibliography}

%

\end{document}